\documentclass[aps,pre,twocolumn,superscriptaddress,showpacs,amsfonts,10pt,%
tightenlines,floatfix]{revtex4-1}

\usepackage[utf8]{inputenc}
\usepackage{graphicx}
\usepackage{amsmath,amssymb}
\usepackage[usenames,dvipsnames]{xcolor}

\graphicspath{{./AbbildungenNeu/}}

\begin{document}

\title{Shallow shell theory of the buckling energy barrier: 
  From the  Pogorelov state to softening and imperfection sensitivity
  close to the buckling pressure}

\author{Lorenz Baumgarten}
\email[]{lbaumgarten@itp.uni-bremen.de}
\affiliation{Institute for Theoretical Physics,
University of Bremen, 28359 Bremen, Germany}
\affiliation{Physics Department, TU Dortmund University, 
44221 Dortmund, Germany}

\author{Jan Kierfeld}
\email[]{jan.kierfeld@tu-dortmund.de}

\affiliation{Physics Department, TU Dortmund University, 
44221 Dortmund, Germany}

\date{\today}

\begin{abstract}
We study the axisymmetric response of a complete spherical shell
under homogeneous compressive pressure $p$ to 
an additional point force. For a pressure $p$ 
below the classical critical buckling pressure $p_c$,
indentation by a  point force does not  lead to spontaneous buckling 
but an energy barrier has to be overcome. 
The states at the maximum of the energy barrier  represent
a subcritical branch of unstable stationary points, which are 
the transition states to a snap-through buckled state.
Starting from nonlinear shallow shell theory we obtain a closed
analytical expression for the energy barrier height, 
which facilitates its effective numerical evaluation as a function 
of pressure by continuation techniques.
We find  a clear crossover between two regimes: For 
$p/p_c\ll 1$ the post-buckling 
barrier state is a mirror-inverted Pogorelov dimple, and
for $(1-p/p_c)\ll 1$ the barrier state is a shallow dimple with
indentations smaller than shell thickness and exhibits
 extended oscillations, which are well described by linear
response.
We find  systematic expansions of the nonlinear shallow shell 
equations about the 
Pogorelov mirror-inverted dimple for $p/p_c\ll 1$
 and the linear response state for $(1-p/p_c)\ll 1$,  
which enable us to 
 derive asymptotic analytical results for the energy barrier landscape 
in both regimes. Upon 
approaching the buckling bifurcation at $p_c$ from below, we find
a softening of an ideal spherical shell.
 The stiffness for the linear response to point forces vanishes 
 $\propto (1-p/p_c)^{1/2}$; 
 the buckling energy barrier 
vanishes $\propto (1-p/p_c)^{3/2}$; and 
 the shell indentation in the  barrier state
 vanishes $\propto (1-p/p_c)^{1/2}$.
This makes shells sensitive to imperfections which can strongly reduce $p_c$
in an  avoided buckling bifurcation. We find the 
same softening scaling in the vicinity of the reduced critical 
buckling pressure also in the presence of imperfections. 
We can also show that the effect of axisymmetric 
imperfections on the buckling instability 
is identical to the effect of a point force that is preindenting 
the shell. 
In the Pogorelov limit, 
the energy barrier maximum diverges  $\propto (p/p_c)^{-3}$
and the corresponding indentation diverges $\propto (p/p_c)^{-2}$. 
Numerical prefactors for  proportionalities both in the softening 
and the Pogorelov regime
are calculated analytically.
This also enables us to obtain results for the 
critical unbuckling pressure and the Maxwell pressure. 
\end{abstract}


\maketitle

\section{Introduction}

When a complete spherical elastic shell is put under homogeneous 
mechanical compressive pressure, the spherical shape remains 
stable over a considerable pressure range until it finally 
collapses at the critical buckling pressure $p_c$. This pressure
has been known  for over one hundred years  since the work of Zoelly
\cite{Zoelly} and buckling is an  ubiquitous mode of 
failure for curved thin-walled shells with significant implications 
for all engineering applications \cite{Bushnell1981}.
Buckling represents a hysteretic bifurcation analogously to a
hysteretic 
first-order transition in a thermodynamic system 
because the buckled state is already metastable 
below $p_c$ \cite{Knoche2011}. Therefore, the shell 
can be ``pushed'' into a buckled state containing a single 
 axisymmetric dimple already below $p_c$ by applying  an additional 
localized point force. A threshold force is 
required to create a stable  dimple, and the required 
threshold value increases for decreasing $p$ further below $p_c$. 
 This corresponds to an 
energy barrier that has to be overcome by applying the additional 
point force before the spherical shell buckles. 
 This energy barrier has  been subject of a number of recent 
studies both for spherical 
\cite{Marthelot2017,Evkin2017,Hutchinson2017,Hutchinson2017b,Thompson2017,
Baumgarten2018,Hutchinson2018} and  
cylindrical \cite{Virot2017,Thompson2017,Hutchinson2018} shells.
Obviously, it is an important feature that governs the mechanical stability 
of spherical shell structures 
slightly below the buckling threshold with respect to 
localized point forces, but also with respect to thermal fluctuations 
\cite{Kosmrlj2017,Baumgarten2018}. 
It also plays a prominent role for the buckling behavior 
of a shell containing inhomogeneities or imperfections 
in the form of ``frozen-in'' normal displacements in the rest state 
of the shell \cite{Hutchinson1967,Koiter1969,Lee2016} or 
soft spots \cite{Paulose2012}; both are 
problems that we will also revisit. 
The energy barrier represents also an important feature of an 
spherical shell from a general theoretical 
point of view as  the barrier  vanishes  upon 
approaching the buckling bifurcation and how it vanishes   
 characterizes the critical behavior of  the buckling bifurcation. 
In the mechanics literature, the unstable barrier state 
is often referred to as post-buckling state \cite{Koiter1969}
as the shell already contains a dimple; the catastrophic nature
of the buckling instability is reflected in a  decreasing pressure 
$p<p_c$ of the barrier state, which leads to a snap-through 
buckling \cite{Knoche2011,Knoche2014o}. 
Many  quantitative analytical results on
buckling of spherical shells are based on the  
Pogorelov theory,  where the dimple is approximated as 
 a mirror-inverted  spherical cap-shaped indentation
 \cite{Pogorelov}.
Here, we present a  rigorous quantitative approach on the
energy barrier based on  systematic expansions of 
nonlinear shallow shell theory.
Expanding about the 
Pogorelov mirror-inverted dimple 
we find  analytical results for the energy barrier
in the Pogorelov limit  $p/p_c\ll 1$.
 This extends recent work of Gomez {\it et al.}
on the Pogorelov indentation in the absence of pressure \cite{Gomez2016}
and is conceptually similar to the boundary layer approach of 
Evkin {\it et al.} \cite{Evkin1989,Evkin2016,Evkin2017}.
We also derive  analytical results 
for the ``critical'' regime of pressures close to $p_c$
by an expansion about the
 linear response state for $(1-p/p_c)\ll 1$.
This enables us to characterize the softening of the 
capsule close to the critical pressure.

Elastic shells  are thin-walled elastic structures
with a curved reference shape. 
 Bending energy penalizes deviations in curvature from 
the spontaneous curvature of the reference shape,  and two-dimensional 
elastic energy penalizes stretching and shear 
deformations of the quasi-two-dimensional solid shell with 
respect to the reference shape, in which the 
capsule is stress free. 
Examples for elastic shells shells range from the micro- to the macroscale. 
On the microscale, artificial microcapsules enclosing a liquid
\cite{Artificial2,Artificial3,Neubauer2013,SeokKwon2013}
can be  described as elastic shells. 
On the macroscale, all thin-walled spherical structures in mechanical 
engineering (vessels, domelike structures, egg shells \cite{Vella2012}) 
 provide examples.  
 Red blood cells \cite{evans1980,Iglic1997,LimHW2002,RBC_Book}
 and shells of viruses \cite{Lidmar2003,Virus_Capsids2} 
have elastic properties similar to continuum elastic  shells but 
there are important differences, for example, regarding the 
reference shape and crystallinity.
Here, we consider elastic shells with a spherical reference shape
 with  radius $R_0$.
For red blood cells, the 
rest shape is, however, not spherical but
 an oblate spheroid \cite{LimHW2002,RBC_Book}.  
Because a sphere has minimal area for a given volume, any deformation 
of the spherical rest shape involves stretching, whereas 
red blood cells are known to undergo shape transformations  
at (even locally) conserved area \cite{evans1980,Iglic1997,LimHW2002,RBC_Book}.
Viruses are crystalline spherical shells consisting of
discrete protein building blocks.
Any triangulation of a sphere must contain at least 12 fivefold
disclinations. Continuum shell theory cannot account for such defects,
which give rise to faceted 
equilibrium shapes of large viruses, while 
sufficiently small viruses remain spherical \cite{Lidmar2003}.
The faceted equilibrium shape of large viruses is an 
important difference to spherical shells. 
In contrast to quasi-two-dimensional elastic shells,
vesicles are  quasi-two-dimensional fluid membranes made from lipid 
bilayers. Vesicles  also
have a bending and  stretching energy but lack a shear energy because
of their fluidity. They  show a distinct 
deformation 
behavior as compared to elastic capsules \cite{Seifert_Vesicles}.
In particular,  their response to additional
  point forces is different because of
 the lack of an elastic reference state. 
 For vesicles,   additional  point forces  lead to
 tube formation \cite{Bukman1996,Hochmuth1996} rather than the
 formation of a dimple;
such tubes
can also be stabilized by actin protrusions \cite{Mesarec2017}.
Only in its gel phase can a vesicle  acquire a shear modulus and 
exhibit elastic features similar to an elastic shell. 
Also, biological cells have  an elastic cortex
which can be modeled as an elastic shell if it is sufficiently thin 
\cite{Mietke2015}. If the cortex spans the entire cell,
the cell should be treated as a solid elastic sphere
\cite{Boulbitch1998,Mietke2015}. Moreover, active motor-induced 
stresses can modify the actin cortex elasticity \cite{Salbreux2017}.

If the capsule  material can be viewed as a thin shell of thickness
 $h$ ($\ll R_0$)  made from 
an isotropic and homogeneous elastic material with bulk Young's modulus 
$E$, the  shell 
 has a bending modulus $\kappa \propto Eh^3$ but a  two-dimensional 
Young's modulus $Y\propto Eh$ \cite{LandauLifshitz,Libai1998}. 
Therefore, bending deformations
are energetically preferred over  stretching
 or shear deformations for thin shells, as long as 
$R_0 \gg (\kappa/Y)^{1/2} \propto h$.
As a result, spherical elastic shells or capsules are very
resistant to compressive forces because there are 
no isometric, stretch- and shear-avoiding  deformations 
of a sphere. 
Only above the critical  pressure $p_c$ does a perfect spherical shell 
become unstable and  buckling  
occurs \cite{LandauLifshitz,Timoshenko1961}.

At $p_c$, the buckling instability is triggered by a short-wavelength mode,
which spreads over the whole sphere and leads to many 
small-amplitude dimples appearing on the sphere, 
as can be found in a linear stability analysis
\cite{Hutchinson1967,Koiter1969,Paulose2012}.
After this mode has developed, the shell can further 
lower its energy by increasing the amplitude, and 
   nonlinearities in the elastic theory finally  lead to 
coalescence of small dimples into a single dimple in the buckled 
energy minimum \cite{Baumgarten2018}.
Following Pogorelov \cite{Pogorelov}, the final dimple can be viewed as an 
approximative inverted spherical cap whose sharp edge at the rim   
is rounded to avoid infinite bending energies. 
Such a rounded spherical cap is an approximative 
isometry of the spherical rest shape.
For a fixed mechanical 
 pressure $p\ge p_c$, the dimple will actually snap through and grow 
 until opposite sides are in contact, whereas for osmotic pressure
 control or even volume control, a stable dimple shape is reached before 
 opposite sides come into contact \cite{Knoche2011,Knoche2014o}.
 A deep dimple can also  assume a polygonal shape 
 in a {\it secondary} buckling transition \cite{Quilliet2012,
   Knoche2014a,Knoche2014}.

Understanding the critical properties of the 
buckling instability is important both  from a 
structural mechanics perspective for macroscopic spherical shells 
and for many applications of spherical microcapsules. 
For ideal spherical shells the classical buckling pressure $p_{c}$
is known exactly. 
For a shell with rest radius $R_0$, bending rigidity $\kappa$, 
two-dimensional (2D)
Young's modulus $Y$, one finds \cite{Zoelly,Timoshenko1961,Ventsel}
\begin{equation}
  p_c  = 4 \frac{\sqrt{Y\kappa}}{R_0^2} =
  4\frac{Eh^2}{R_0^2\sqrt{12(1-\nu^2)}}  = 4 \frac{Y}{R_0}\gamma^{-1/2}.
\label{eq:pcb}
\end{equation} 
The second equality applies for thin shells of thickness $h$ 
made from an isotropic  elastic material with bulk Young modulus $E$ 
and Poisson ratio $\nu$, where $\kappa = Eh^3/12(1-\nu^2)$ and $Y = Eh$
\cite{Ventsel}. We also introduced 
 the F\"oppl-von K\'arm\'an number 
\begin{equation}
  \gamma \equiv \frac{YR_0^2}{\kappa} =
  12(1-\nu^2)\left(\frac{R_0}{h}\right)^{2},
\label{eq:FvK}
\end{equation} 
which is an inverse dimensionless bending
rigidity.  The ideal 
critical pressure $p_c$ is, however, not reached in experiments on macroscopic 
shells, because imperfections reduce the buckling pressure significantly
\cite{Hutchinson1967, Koiter1969}.

Buckling represents a hysteretic bifurcation analogous to a
hysteretic 
first order transition in a thermodynamic system;  metastable 
buckled states and a corresponding 
 unstable transition state  appear already subcritically for $p<p_c$ 
\cite{Knoche2011,Knoche2014o,Hutchinson2017b,Evkin2017,Baumgarten2018}. 
The buckled state with a single axisymmetric dimple 
becomes {\it energetically} favorable already for 
$p> p_{c1}$, above the so-called
{\it  Maxwell pressure}, which can be obtained from a 
Maxwell construction of equal energies  
\cite{Knoche2011,Knoche2014o,Hutchinson2017b,Evkin2017}
resulting in a parameter dependence 
$p_{c1} \sim  p_{c} \gamma^{-1/4}$ \cite{Knoche2014o}.
As a result, there is a rather wide pressure window  $p_{c} >p>p_{c1}$,
where buckling is energetically possible but  an 
energy barrier has to be overcome; the barrier state is  an unstable 
transition state.
One way to probe the energy barrier is by application of an 
additional point force, which ``pushes'' the shell into the buckled 
state
\cite{Marthelot2017,Evkin2017,Hutchinson2017,Hutchinson2017b,Thompson2017,Hutchinson2018},
see Fig.\ \ref{fig:barrier_snapshot}. 
If the dimple is created by a point force it is 
 axisymmetric about the force axis.  We  exclusively study the
axisymmetric situation in this paper. 
SURFACE EVOLVER simulations
have shown, however,  that the axisymmetric 
 dimple is also the relevant barrier state if the dimple is not 
forced into an axisymmetric shape by a point force \cite{Baumgarten2018}.

Below the Maxwell pressure, there is 
also a  critical {\it unbuckling pressure}
 $p_\text{cu}\sim 3p_{c1}/4$,
 below which no stable buckled shape exists  and 
which has the same parameter dependence as $p_{c1}$
\cite{Knoche2011,Knoche,Paulose2012,Evkin2017}.
This pressure is also called {\it minimum buckling load}
in the literature \cite{Karman1939,Koiter1969}.  
This gives the following general bifurcation scenario:
Buckled states and the unstable 
 barrier transition state  appear in a bifurcation 
at  $p=p_\text{cu}$. In the range   $p_\text{cu} <p<p_c$, three 
stationary shapes are present: Spherical 
and buckled states are (meta)stable and separated by the
 unstable barrier state. At $p=p_c$, the barrier state and the spherical 
state vanish in a second bifurcation.

Whereas the value of the critical buckling pressure $p_c$
 is known analytically, many aspects of  the  buckling bifurcation
 are unexplored, 
in particular with respect to the buckling 
energy barrier. 
One example is the properties of the subcritical 
 axisymmetric barrier state for $p$ close to $p_c$. They  characterize
 the  bifurcation at $p_c$ but have, so far, not been explored systematically.
A systematic numerical and analytical investigation in this regime 
is the focus of the present paper.
 Most of the known
 results for the barrier state and the energy barrier height have been 
derived in the limit $p\ll p_c$  either 
from numerical work starting from energy minimization
 \cite{Hutchinson2017,Hutchinson2017b,Baumgarten2018}  or 
 based on  the Pogorelov energy scaling of the buckled state, 
 which is only valid for relatively deep mirror-buckled 
indentations at $p\ll p_c$.
The scaling of the energy barrier height
$E_B\propto (p/p_c)^{-3}$  and 
the depth of the barrier indentation $z_B \propto (p/p_c)^{-2}$
can be  derived using this Pogorelov scaling
\cite{Paulose2012,Knoche2014o,Baumgarten2018}.
In the Pogorelov approach,
numerical prefactors in the scaling results can be  obtained 
from only an approximative  variational energy minimization
for the rounding of the sharp edge of  inverted 
 spherical cap shapes. 
Further progress has been made 
by Evkin and coworkers using a more systematic 
boundary layer formulation in shallow shell theory 
but still relying on   variational energy minimization
\cite{Evkin1989,Evkin2016,Evkin2017}.

In this paper, we start from the force equilibrium for axisymmetric states
and use the nonlinear shallow shell equations to systematically
derive properties of the buckling energy barrier.
First, we will present numerical results based on an 
exact and explicit expression for the energy barrier in axisymmetric
 nonlinear shallow shell theory. 
Then we will focus both on the Pogorelov limit $p\ll p_c$, where we 
systematically expand about a mirror-symmetric barrier state 
with a deep indentation,
and on the limit of compressive 
pressures below but close to  $p_c$, where the  barrier state 
is a very shallow dimple such that we can systematically 
expand about Reissner solutions of the  linearized shallow shell 
theory. In both limits, we derive the exact asymptotic behavior 
including numerical prefactors. This enables us to 
obtain a complete picture of the buckling energy landscape 
in both limits 
and shed light on the critical properties of the 
buckling bifurcation.  

In  numerical calculations, application of a point force 
allows us to slowly push the shell into a buckled state, to explore 
thereby the buckling energy landscape, to detect 
the barrier state as the unstable force-free transition state,
and to quantify the energy of the barrier state by measuring 
the work performed by the point force until the barrier is reached. 
Point forces are, however, also an important experimental tool 
to test shells \cite{Marthelot2017}. Of particular interest 
in applications is the initial linear response of a shell 
to point forces because many mechanical compression techniques
 (plate compression
\cite{carin_compression_2003,fery_mechanical_2007} or 
compression by microscopy tips \cite{Fery2004,fery_mechanical_2007}) 
are equivalent 
to  point force indentation in the initial small displacement regime.
We provide an expression for the linear stiffness of the elastic 
shell, which is valid for the entire pressure range and extends 
results for pressurized capsules with 
stretching pressures \cite{Vella2012b} 
 to compressive pressures up to the critical buckling pressure.
Knowledge of the linearized stiffness 
 can be used for measuring  elastic capsule properties 
and  capsule pressure \cite{Vella2012b}.

Within the same framework of  nonlinear shallow shell equations,
we finally  consider the effect of axisymmetric imperfections 
within the systematic expansion for shallow indentations.
This allows us to explore how 
softening of the shell close to $p_c$ makes
 shells sensitive to imperfections and 
 results in an  avoided buckling bifurcation.
We compare the effect of a point force that is preindenting 
the shell with the effects of  axisymmetric 
imperfections on the buckling instability
and find striking similarities.

\section{Nonlinear shallow shell theory}

\begin{figure*}
\begin{center}
\includegraphics[width=0.95\textwidth]{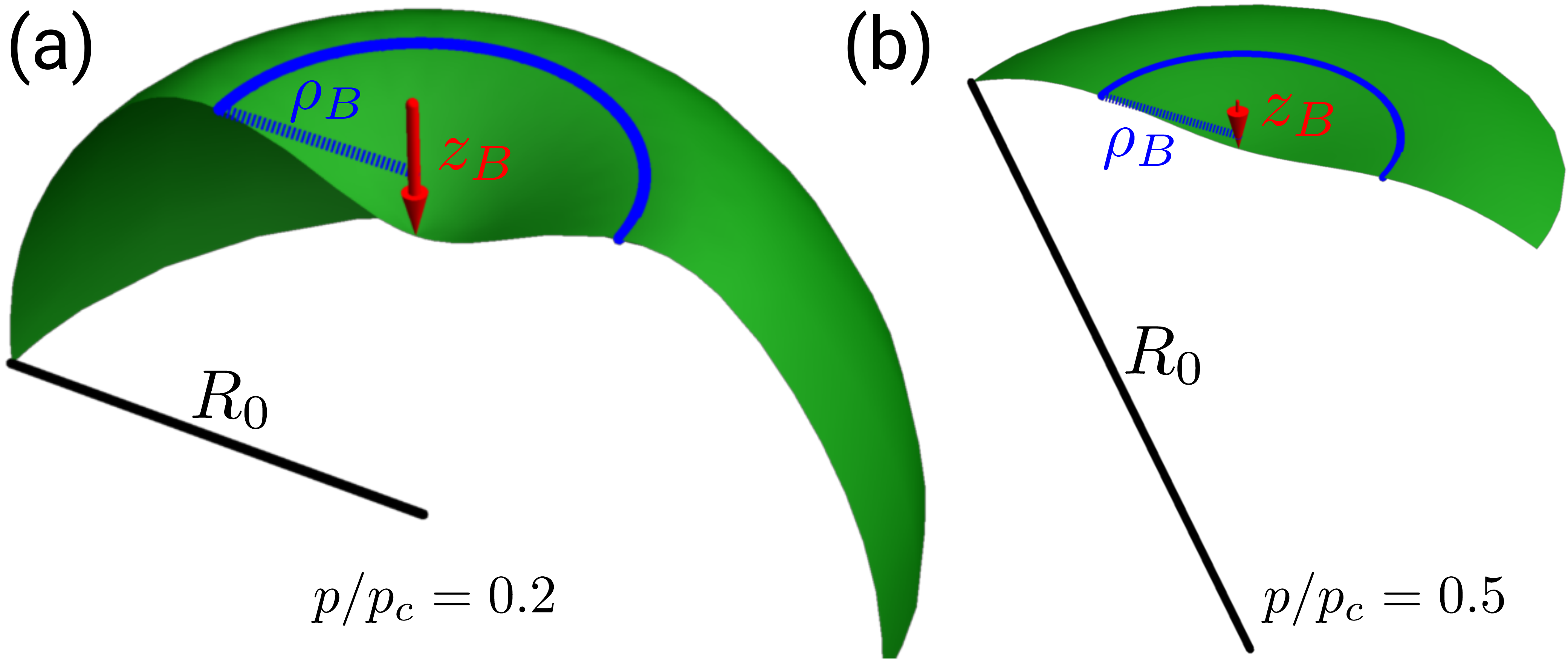}
\caption{
  Numerical results for the shape  
   of a buckled spherical shell with $\gamma=5000$
  and rest radius $R_0$ in the postbuckling barrier state for 
  pressures  $p/p_c=0.2$ (a) and $p/p_c=0.5$ (b) 
   according to nonlinear shallow shell equations 
   (only the relevant part of the sphere is shown).
  The red arrow indicates the normal displacement $z_B$ at the 
  pole, which is also the direction of the applied point force. 
 The blue ring indicates the width $\rho_B$ of the barrier state. 
}
\label{fig:barrier_snapshot}
\end{center}
\end{figure*}

We employ nonlinear 
shallow shell theory  for a thin spherical shell with equilibrium 
radius $R_0$ \cite{Ventsel,Libai1998}, 
which is subject to a homogeneous compressive  pressure $p$ {\it and}
an additional indenting point force $F$ normal to the surface
(see Fig.\ \ref{fig:barrier_snapshot}). 
We focus on  isotropic elastic materials; isotropic Hookean 
elasticity can describe the deformation behavior of 
most artificial microcapsules 
rather well \cite{Hegemann2018}.
Using polar coordinates $r$ and $\phi$ on the two-dimensional 
reference plane over which shallow shell configurations are 
parametrized with the point force acting at the pole, 
shallow shell theory gives 
two coupled equations for the normal displacement $w(r,\phi)$
(negative for inward displacement) and 
the Airy stress function $\Phi(r,\phi)$. 
For axisymmetric states, as they are enforced by the 
point force,  these functions
become independent of $\phi$, and we have two equations for $w(r)$ and 
the negative derivative of the Airy stress function $\psi(r)=-\Phi'(r)$,
which have been derived and are described  in detail in the literature
\cite{Ventsel,Hutchinson1967,Vella2012b,Paulose2012,Gomez2016} 
[see also Eqs.\ (\ref{eq:forcebalI}) and (\ref{eq:compI}) in 
Appendix \ref{app:shshell} with  $w_I=0$],
\begin{align}
  & \kappa \nabla^4 w
   + \frac{1}{R_0} \frac{1}{r} \partial_r (r\psi)
      - \frac{1}{r} \partial_r\left( \psi
   \partial_r w \right) 
 = -p - \frac{F}{2\pi} \frac{\delta(r)}{r},
  \label{eq:forcebal} \\
 &\frac{1}{Y} r  \partial_r \left[ 
     \frac{1}{r} \partial_r (r\psi) \right] 
  = \frac{r}{R_0} \partial_rw  - 
   \frac{1}{2} \left( \partial_r w\right)^2 
  \label{eq:comp}
\end{align}
(with $\nabla^2... =(\frac{1}{r}\partial_r)
+\partial_r^2)... 
 = \frac{1}{r}\partial_r r \partial_r ... $).
Positive $p$  corresponds to  a {\it compressive} pressure,
a positive $F$ corresponds to a compressive point force, and 
 the point force acts at $r=0$.
The first equation, (\ref{eq:forcebal}),  is the force balance in vertical 
direction, 
and the second equation, (\ref{eq:comp}), is
 the (integrated) compatibility of strains. 
The in-plane stresses are obtained
as $\sigma_{\phi\phi} = \psi'$ and $\sigma_{rr} = \psi/r$. 
We assume  thin shells $h/R_0 \ll 1$ and 
 shallow shells, i.e.,  small 
slopes  $|w'|\ll 1$ \cite{Ventsel}, in the above equations.

Equations (\ref{eq:forcebal}) and  (\ref{eq:comp}) 
have to be solved with boundary conditions 
$w(\infty)=w'(\infty) = 0$ and $\psi(\infty)=0$ (or $\psi'(\infty)=0$)
for $r\to \infty$;
at $r=0$, we require a given indentation $w(0)<0$, 
$w'(0)=0$ corresponding to the absence of kinks,
and $\lim_{r\to 0}(r\psi'(r)) - \nu \psi(0)=0$ corresponding to 
vanishing radial in-plane displacement to avoid 
tearing the shell. We 
prescribe the indentation $w(0)<0$ at the origin,  
solve Eq.\ (\ref{eq:forcebal}) in the domain $r>0$ 
where $F=0$,  and calculate 
the necessary force $F$ to induce this indentation only afterward 
from an integrated version of Eq.\ (\ref{eq:forcebal})
\cite{Vella2012b,Gomez2016}.

In the absence of a point force, the pressure $p$ puts the shell 
into a uniformly pre-compressed state  with $w(r) =w_0 <0$ and
 $\psi(r) = \psi_0(r)  = -pR_0r /2$ corresponding to stresses 
$\sigma_{rr} = \sigma_{\phi\phi} = \sigma_0 = -pR_0/2$. 
We consider changes with respect to this precompressed state and 
substitute
$w(r) \to w_0 + w(r)$ and $\psi(r) \to \psi_0(r) + \psi(r)$, such that 
Eq.\ (\ref{eq:forcebal}) becomes 
\begin{align}
  & \kappa \nabla^4 w
   + \frac{1}{R_0} \frac{1}{r} \partial_r (r\psi)
      -\sigma_0  \nabla^2 w
- \frac{1}{r} \partial_r \left( \psi
   \partial_r w \right) 
\nonumber\\
& ~~~ =  - \frac{F}{2\pi} \frac{\delta(r)}{r}
  \label{eq:forcebal2} 
\end{align}
while Eq.\ (\ref{eq:comp}) remains unchanged. 
The boundary conditions for $w(r)$ and $\psi(r)$ are 
unchanged by this substitution, and  we define the 
 indentation depth $z$ ($z>0$)
at the pole with respect to the precompressed state; i.e., 
we require  $w(0)=-z<0$ at $r=0$ after substitution. 
The additional term $+\sigma_0  \nabla^2 w$ in Eq.\ (\ref{eq:forcebal2})
induces a  tendency 
 of the precompressed state for oscillating
 $w$ fluctuations
 as it is the variation of an effective energy
 $ -\frac{1}{2}\sigma_0 \int d^2{\bf r} (\nabla w)^2$, which
 is lowered by oscillating   $w$ modes. 
This is the driving force for the classical 
instability with respect 
to oscillatory $w$ modes at the buckling pressure $p_c$. 

We introduce  dimensionless quantities 
\begin{equation}
\begin{split}
\bar{w}  &\equiv  w/(\kappa/Y)^{1/2},~~
\rho \equiv r/(\kappa R_0^2/Y)^{1/4},~
\\
\bar{\psi} &\equiv \psi/(\kappa^2Y/R_0^2)^{1/4},~~
\bar{E} \equiv E/2\pi(\kappa^{3}/YR_0^2)^{1/2};
\end{split}
\label{eq:non-dim}
\end{equation}
i.e., we measure normal displacements 
$w$ (and indentations $z$ at the pole)  in units of 
$(\kappa/Y)^{1/2} = R_0 \gamma^{-1/2}  = hk^{-2}$,  which is, 
apart from factors of  $k\equiv [12(1-\nu^2)]^{1/4}$, the shell thickness $h$
(the dimensionless radius is thus $\bar{R}_0= \gamma^{1/2}$),
  radial distances in units of the elastic length scale 
$l_{\rm el} =  (\kappa R_0^2/Y)^{1/4} = R_0 \gamma^{-1/4} = (hR_0)^{1/2}k^{-1}$ 
(the radial scale on which bending and stretching energy
 are balanced), which is also the unstable 
wave length at the buckling transition at $p_c$  \cite{Hutchinson1967},
 normal forces in units of 
$\kappa/R_0 = YR_0 \gamma^{-1}$, and energies in 
units of $2\pi(\kappa^{3}/YR_0^2)^{1/2} = 2\pi YR_0^2 \gamma^{-3/2}$.
The dimensionless shallow shell equations (\ref{eq:forcebal2}) 
and (\ref{eq:comp})  become 
\begin{align}
  &  \nabla_\rho^4 \bar{w}
   + \frac{1}{\rho} \partial_\rho (\rho\bar{\psi})
      +2\frac{p}{p_c} \nabla_\rho^2 \bar{w}
- \frac{1}{\rho} \partial_\rho \left( \bar{\psi} \partial_\rho\bar{w}\right) 
 =  - \frac{\bar{F}}{2\pi} \frac{\delta(\rho)}{\rho},
  \label{eq:forcebal3}\\
 & \rho  \partial_\rho \left[ 
     \frac{1}{\rho} \partial_\rho (\rho\bar{\psi}) \right] 
  = \rho \partial_\rho \bar{w}  - 
   \frac{1}{2} \left( \partial_\rho \bar{w}\right)^2 
  \label{eq:comp3}
\end{align}
with 
$\nabla_\rho^2... = \frac{1}{\rho}\partial_\rho(\rho \partial_\rho
...)$, 
and  $\bar{F} \equiv F \gamma/Y R_0$. 
We compare to other non-dimensionalization schemes of the problem in 
Table \ref{tab:dimensionless} in  Appendix \ref{app:dim}.
Shallow shell theory is applicable as long as $\partial_r w \ll 1$
  \cite{Ventsel}, which implies $\partial_\rho \bar{w} \ll \gamma^{1/4}$
  in dimensionless quantities.

\subsection{Exact analytical results}

Because $\bar{w}(\rho)$ decays exponentially for $\rho \gg 1$, we can 
obtain  
\begin{equation}
\bar{\psi} \sim -\bar{F}/2\pi \rho  ~~\mbox{for}~\rho\to \infty
\label{eq:psiasy}
\end{equation}
 from integrating 
 (\ref{eq:forcebal3}) over a circle of radius $\rho \to\infty$ on both sides,
 $\int_0^\rho d\tilde{\rho} \tilde\rho ...$, 
resulting in 
$\left. \tilde{\rho}\bar{\psi} \right|_0^\rho \sim \rho\bar{\psi} = 
   - \frac{\bar{F}}{2\pi}$ \cite{Vella2012b}. 
Equation (\ref{eq:psiasy}) also follows from  force balance in the 
point force direction \cite{Gomez2016}.

From the shallow shell equations (\ref{eq:forcebal3}) and (\ref{eq:comp3}), 
two exact relations can be obtained.
Multiplying by $\rho$ and  integrating 
 from $\rho$ to infinity on both sides of Eq.\ (\ref{eq:forcebal3})
and using (\ref{eq:psiasy}) at infinity
gives the first relation
\begin{align}
  & -\rho \partial_\rho (\nabla_\rho^2 \bar{w}) - \rho \bar{\psi} 
    + \bar{\psi} \partial_\rho \bar{w}  -2\frac{p}{p_c} \rho \partial_\rho
    \bar{w} = \frac{\bar{F}}{2\pi}.
\label{eq:exact1}
\end{align}
Dividing by $\rho$ and  integrating
 from $\rho$ to infinity on both sides of Eq.\  (\ref{eq:comp3}),
 multiplying by $\rho$ and integrating from $0$ to infinity 
on both sides,  
and  using (\ref{eq:psiasy}) at infinity and one partial integration on the 
right-hand side give the second relation
\begin{align}
 -\frac{\bar{F}}{2\pi} &= \int_0^{\infty} d\rho \rho \bar{w}
                         + \int_0^\infty d\rho \rho \frac{1}{4} (\partial_\rho
                         \bar{w})^2.
\label{eq:exact2}
\end{align}
Both of these  equations can be employed to  determine the point
force $\bar{F}$ for a given  
 indentation $\bar{z}$ and thus the 
force-indentation relation $\bar{F}=\bar{F}(\bar{z})$ numerically.

\begin{figure}
\begin{center}
\includegraphics[width=0.4\textwidth]{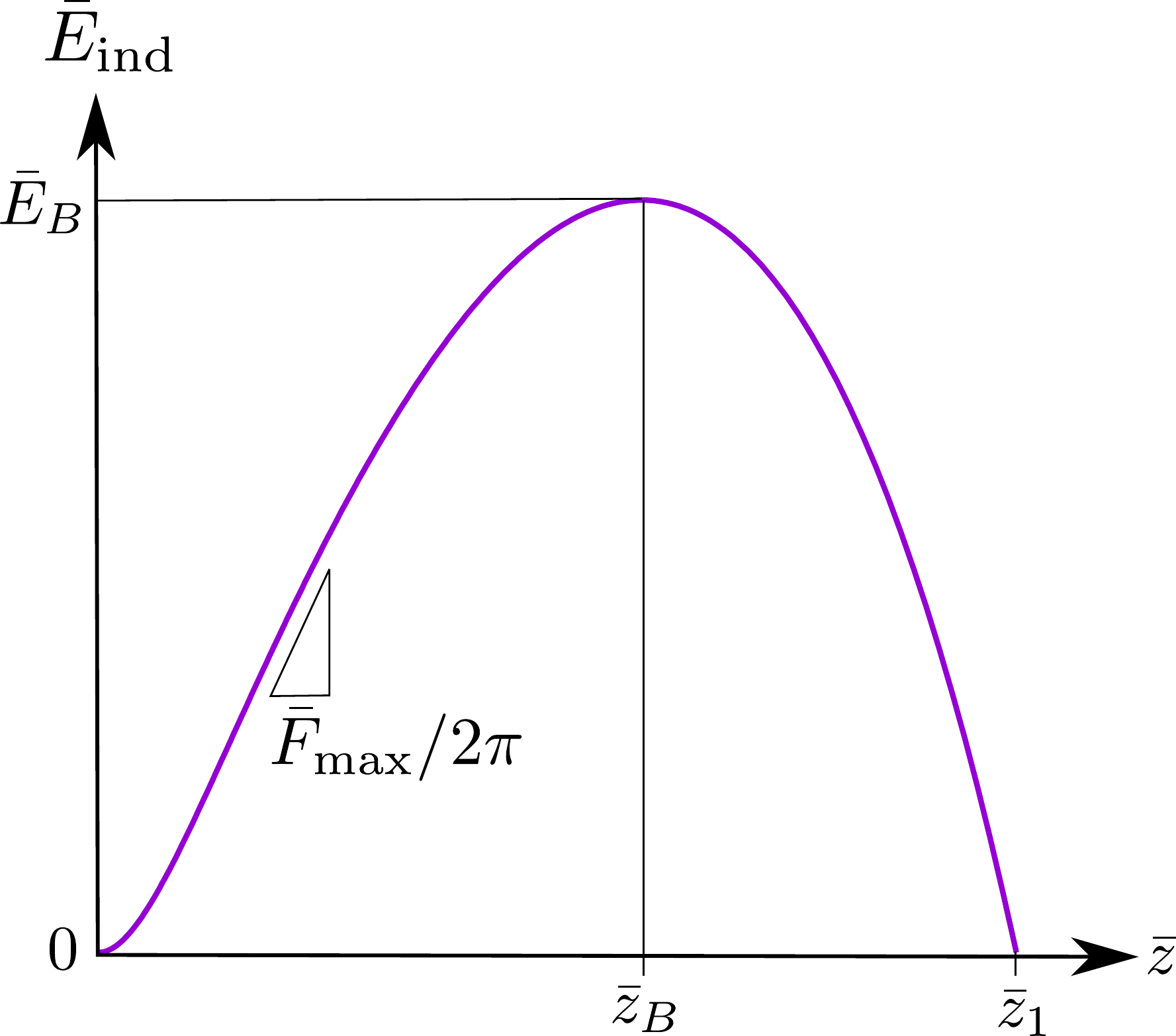}
\caption{
 Schematic energy landscape $\bar{E}_{\rm ind}$ as a function 
  of indentation depth  $\bar{z}$. The indentation at the 
energy barrier maximum is $\bar{z}_B$, the height of the energy barrier 
is $\bar{E}_B$. At depth $\bar{z}_1$, unindented and indented 
state have equal energies. 
}
\label{fig:landscape}
\end{center}
\end{figure}

The force-indentation relation can be integrated
to obtain the  indentation energy  $\bar{E}_{\rm ind}$
as a function of the indentation depth,
$\bar{E}_{\rm ind}(\bar{z}) = \frac{1}{2\pi} 
\int_0^{\bar{z}} \bar{F}(\tilde{\bar{z}}) d\tilde{\bar{z}}$.
We note that this is the total energy difference with respect to the 
precompressed spherical state (at pressure $p$) if an additional 
indentation of depth $\bar{z}$ is generated (by applying a 
point force $\bar{F}$). 
At the  barrier state, the indentation energy has a maximum
as a function of the indentation $\bar{z}$.
If we call the  indentation in the barrier state $\bar{z}_B$ (see Fig.\ 
\ref{fig:barrier_snapshot}), 
it fulfills  $\partial \bar{E}_{\rm ind}/\partial \bar{z}(\bar{z}_B) =0$ or 
 $\bar{F}(\bar{z}_B)=0$.
We introduce the barrier energy as 
 energy difference between barrier state and the 
precompressed spherical state, $\bar{E}_B = \bar{E}_{\rm ind}(\bar{z}_B)$. 
Vice versa, the force-indentation 
 relation $\bar{F}=\bar{F}(\bar{z})$ is obtained
from the  energy $\bar{E}_{\rm ind}(\bar{z})$ by 
minimizing the tilted indentation energy landscape 
$\bar{E}_{\rm ind}(\bar{z})- \bar{F}\bar{z}/2\pi$.
The slope of the energy landscape $\bar{E}_{\rm ind}(\bar{z})$ 
at $\bar{z}$
gives  the necessary point force $\bar{F}/2\pi$ to achieve 
an indentation $\bar{z}$. A 
 pushing compressive  point force is necessary to achieve 
indentations where the energy landscape is increasing; at the maximum
in the barrier state 
a force-free unstable equilibrium is achieved; at indentations where the 
energy landscape decreases, the shell can  only be stabilized by 
 a pulling point force.
Figure \ref{fig:landscape}
summarizes important features of a schematic buckling energy landscape.

The energy barrier, i.e., the difference in total energy 
$\Delta E_{\rm tot} = \Delta E_s + \Delta E_b +p\Delta V$ 
(the sum of stretching, bending, and 
mechanical pressure work) between barrier state and the 
precompressed spherical state  is given by 
the simple, explicit formula
\begin{align}
   \bar{E}_B &= \Delta \bar{E}_{\rm tot} = 
-\frac{1}{4} \int_0^{\infty}d\rho  \bar{\psi}
   (\partial_\rho \bar{w})^2,
\label{eq:Etot}
\end{align}
where  $\bar{\psi}$ and $\bar{w}$ are solutions of the 
 shallow shell equations (\ref{eq:forcebal3}) and (\ref{eq:comp3}) 
for $\bar{F}=0$. Equation (\ref{eq:Etot}) is derived 
 in Appendix \ref{app:barrierenergy}. 
This result allows direct numerical access to  the value 
of the energy  barrier.
Moreover, it will allow us to obtain analytical results 
both for the critical behavior
of the energy barrier close to $p_c$, i.e., for shallow barriers with 
small indentations $\bar{z}\ll 1$) and for $p\gg p_c$, i.e., 
for barrier states with deep indentations $\bar{z}\gg 1$, which 
are mirror-inverted Pogorelov dimples.
Equation (\ref{eq:Etot}) actually gives a positive energy because 
there is mainly compressive hoop stress ($\bar{\psi}<0$)  in the regions 
where $(\partial_\rho \bar{w})^2>0$ is large, i.e., at the 
rim of the indentation. For a Pogorelov 
dimple, this is exactly  the inner rim of the Pogorelov ridge
\cite{Knoche2014a,Knoche2014} [see also Fig.\  \ref{fig:barrier_conf}(c)].

The terms in the
second exact relation (\ref{eq:exact2}) are directly 
related to  the
dimensionless volume change
by indentation, $\overline{\Delta V} =
\Delta V/(\kappa R_0/Y)$, and  the 
dimensionless area change by indentation,
$\overline{\Delta A} = \Delta A/(\kappa/Y)$,
\begin{align}
    \overline{\Delta V}  &\approx 
                          2\pi \int_0^{\infty} d\rho \rho \bar{w}<0
 \label{eq:DeltaV} \\
 \overline{\Delta A}  &\approx 2\pi \int_0^\infty d\rho \rho
                      \left[ 2\bar{w}  +\frac{1}{2} (\partial_\rho \bar{w})^2
                        \right],
    \label{eq:DeltaA}
\end{align}
where we work in 
shallow shell approximation, i.e., assuming
$w/R \ll \partial_r w \ll 1$ or $\bar{w}\gamma^{-1/2}\ll
\partial_\rho \bar{w}\gamma^{-1/4}\ll 1$.
Therefore, relation (\ref{eq:exact2}) is equivalent to
a relation
\begin{equation}
  -\bar{F} = -\frac{\partial \bar{E}_{\rm ind}}
    {\partial \bar{z}} =  \frac{1}{2}\overline{\Delta A} 
  \label{eq:FVA}
\end{equation}
for the area change by point force indentation.
This implies that the area is decreased ($\Delta A<0$)  by a compressive
point force indentation or an increasing indentation energy up to
the barrier, whereas
it is increased  ($\Delta A>0$) for a decreasing indentation energy.
It also shows that the 
force-indentation relation
$\bar{F}(\bar{z}) = \partial\bar{E}_{\rm ind}/\partial \bar{z}(\bar{z})$
directly gives the  area change 
$\overline{\Delta A}(\bar{z})=-\bar{F}(\bar{z})/2$
as a function of indentation.  
Right 
  at the barrier state with $\bar{F}=0$,  the
  area change by indentation exactly vanishes,
 \begin{equation}
  0 = -\frac{\partial \bar{E}_{\rm ind}}
    {\partial \bar{z}}(\bar{z}_B)
   =  \frac{1}{2} \overline{\Delta A}(\bar{z}_B).
  \label{eq:FVAB}
\end{equation}
The mirror-inverted Pogorelov dimple exactly fulfills this requirement
by definition but this result not only
holds  in the Pogorelov limit $p/p_c\ll 1$
but for all pressures $p$.  In particular, it also holds
close to $p_c$, where
the barrier state does not resemble a Pogorelov dimple but
becomes shallow and oscillatory.
At the maximal point force $\bar{F}_{\rm max}$, which has to be applied
to overcome the energy barrier and which is the characteristic maximal  point 
force for structural stability below $p_c$ (see Fig.\ \ref{fig:landscape}),
the shell area is minimal, and $-\bar{F}_{\rm max} =
\frac{1}{2}\overline{\Delta A}_{\rm min}$.

After  nondimensionalization (\ref{eq:non-dim}), the 
 shallow shell equations (\ref{eq:forcebal3}) and (\ref{eq:comp3}) 
only depend on the parameters $p/p_c$ 
and $\bar{F}$, which is a function of the indentation depth 
$\bar{z}$.   Therefore, 
properties of the barrier state that can be directly obtained from 
solution of the dimensionless shallow shell equations, such as the 
dimensionless indentation $\bar{z}_B$,  
will only depend on $p/p_c$. 
Because the dimensionless energy barrier $\bar{E}_B$ can also be expressed 
directly by solutions of the shallow shell equations at $\bar{F}=0$  via 
Eq.\ (\ref{eq:Etot}),
also  $\bar{E}_B$ will only depend on $p/p_c$, 
see our main results  (\ref{eq:barrierPog}) and (\ref{eq:barrierlin}) below. 
In particular, $\bar{E}_B$ does not depend on the Poisson number $\nu$
in shallow shell theory.

\subsection{Numerical method}

We solve the nonlinear shell theory boundary problem
(\ref{eq:forcebal3}) and (\ref{eq:comp3}) 
numerically on a finite  domain $\rho_{\rm min}<\rho<\rho_{\rm max}$ 
($\rho_{\rm min} = 10^{-5}$, $\rho_{\rm max}=5000$)
using the MATLAB routine {\tt bvp4c}
with boundary conditions 
$\bar{w}(\rho_{\rm max})=\bar{w}'(\rho_{\rm max}) = 0$
 and $\bar{\psi}'(\rho_{\rm max})=-\bar{\psi}(\rho_{\rm max})/\rho_{\rm max}$ 
at ``infinity''; the last condition is crucial to enforce the 
correct asymptotics $\bar{\psi} \propto 1/\rho$  [see Eq.\ (\ref{eq:psiasy})].
At ``$\rho=0$'',  we use 
$\rho_{\rm min}\bar{\psi}'(\rho_{\rm min}) - \nu \bar{\psi}(\rho_{\rm min})=0$
for 
vanishing radial in-plane displacement  (with $\nu =1/3$), 
$\bar{w}'(\rho_{\rm min}) = 0$, and 
a prescribed indentation depth 
$\bar{w}(\rho_{\rm min}) = -\bar{z}<0$ instead of the  point force, 
which is absent in the domain $\rho>0$
 \cite{Vella2012b,Gomez2016}.

Inserting the numerical solution into Eqs.\ (\ref{eq:exact1}) 
(which holds pointwise for each $\rho$ but is used
after averaging over all $\rho$)
or (\ref{eq:exact2}) 
gives the value of the force $\bar{F}$ for the prescribed indentation 
depth  $\bar{z}$, which allows us to scan 
the force-indentation relation $\bar{F}=\bar{F}(\bar{z})$ by 
gradually increasing $\bar{z}$.  Knowledge of the entire 
force-indentation relation $\bar{F}(\bar{z})$ enables us  to calculate the 
energy barrier by numerical integration 
$\bar{E}_B = \frac{1}{2\pi} 
\int_0^{\bar{z}_B} \bar{F}(\tilde{\bar{z}}) d\tilde{\bar{z}}$ up to 
the barrier indentation $\bar{z}_B$ where the force vanishes, 
 $\bar{F}(\bar{z}_B)=0$.

While calculation of the  entire 
force-indentation relation  and numerical integration 
up to the barrier state where   $\bar{F}(\bar{z}_B)=0$
is an intuitive approach,
there is a much more efficient way to numerically 
calculate  the energy barrier:
The exact result (\ref{eq:Etot}) can be employed to 
evaluate the  energy barrier  directly 
 for a barrier state with $\bar{F}=0$.
To obtain the energy barrier as a function of $p$, we 
 continuate numerical  solutions of the shallow shell
equations (\ref{eq:forcebal3}) 
and (\ref{eq:comp3}) for the barrier states with $\bar{F}=0$
for small changes in $p$  and evaluate 
the energy barrier directly at each  barrier state 
using (\ref{eq:Etot}). This supersedes calculation of the entire 
force-indentation relation $\bar{F}(\bar{z})$ in order to calculate 
a single 
energy barrier value by numerical integration  of the 
force-indentation relation. 
We checked that we obtain numerically identical results with both methods.

\section{Linear response,  shell stiffness, 
   and softening close to buckling} 
\label{sec:lin}

Many mechanical compression tests
  such as plate compression
\cite{carin_compression_2003,fery_mechanical_2007} or 
compression by microscopy tips \cite{Fery2004,fery_mechanical_2007}
are equivalent 
to  point force indentation in the initial small displacement regime,
which can be described by linear response. 
We will rederive the linear stiffness of the shell  and
show that the result  for  a pressurized spherical shell with $p>0$
\cite{Vella2012b}  can be continued to 
 compressive pressures $0<p<p_c$.

Linearizing  Eqs.\ (\ref{eq:forcebal3}) 
and (\ref{eq:comp3}) gives the Reissner equations 
\cite{Reissner1946,Vella2012b,Paulose2012}
\begin{equation} 
\begin{split}
  &  \nabla_\rho^4 \bar{w}
   -\nabla_\rho^2 \bar{\Phi}
      + 2\frac{p}{p_c}  \nabla_\rho^2 \bar{w}
 =  - \frac{\bar{F}}{2\pi} \frac{\delta(\rho)}{\rho},
\\
 & \nabla_\rho^4 \bar{\Phi}
  = -\nabla_\rho^2 \bar{w}  
\end{split}
\label{eq:Reissner}
\end{equation}
 with the 
dimensionless Airy stress function $\bar{\Phi}$ 
($\bar{\psi} = - \partial_\rho\bar{\Phi}$).
In the domain $\rho>0$, where the $\delta$-function on the right-hand side 
 vanishes, 
these equations can be solved using the original ansatz  of Reissner
\cite{Reissner1946},
$f_\pm \equiv \bar{w} + \lambda_{\mp}\bar{\Phi}$,
which decouples equations to 
 $\nabla_\rho^4f_\pm - \lambda_\pm \nabla_\rho^2f_\pm  =0$
if 
$\lambda_\pm =  -p/p_c \pm i\left(1- (p/p_c)^2  \right)^{1/2}$
($\lambda_+\lambda_-=1$).
This  finally leads to solutions 
\begin{subequations}
\begin{align}
  \bar{w}_{\rm lin} &= \frac{\bar{z}}{\ln \lambda_+} 
\left(K_0(\lambda_+^{1/2}\rho) - K_0(\lambda_-^{1/2}\rho) \right),
\label{eq:barw}\\
  \bar{\Phi}_{\rm lin} &= \frac{\bar{z}}{\ln \lambda_+} 
\Bigg(\lambda_+K_0(\lambda_-^{1/2}\rho) -\lambda_- K_0(\lambda_+^{1/2}\rho) 
\nonumber\\
  &~~~~~   + (\lambda_+-\lambda_-)\ln\rho  \Bigg),
\label{eq:barPhi} \\
\bar{\psi}_{\rm lin} &= -\frac{\bar{z}}{\ln \lambda_+} 
\Bigg(\lambda_-^{1/2} K_1(\lambda_+^{1/2}\rho) 
   -\lambda_+^{1/2}K_1(\lambda_-^{1/2}\rho) +
\nonumber\\
  &~~~~~   + \frac{\lambda_+-\lambda_-}{\rho}  \Bigg)
\label{eq:barpsi}
\end{align}
\end{subequations}
satisfying all boundary conditions. 

\begin{figure}
\begin{center}
\includegraphics[width=0.45\textwidth]{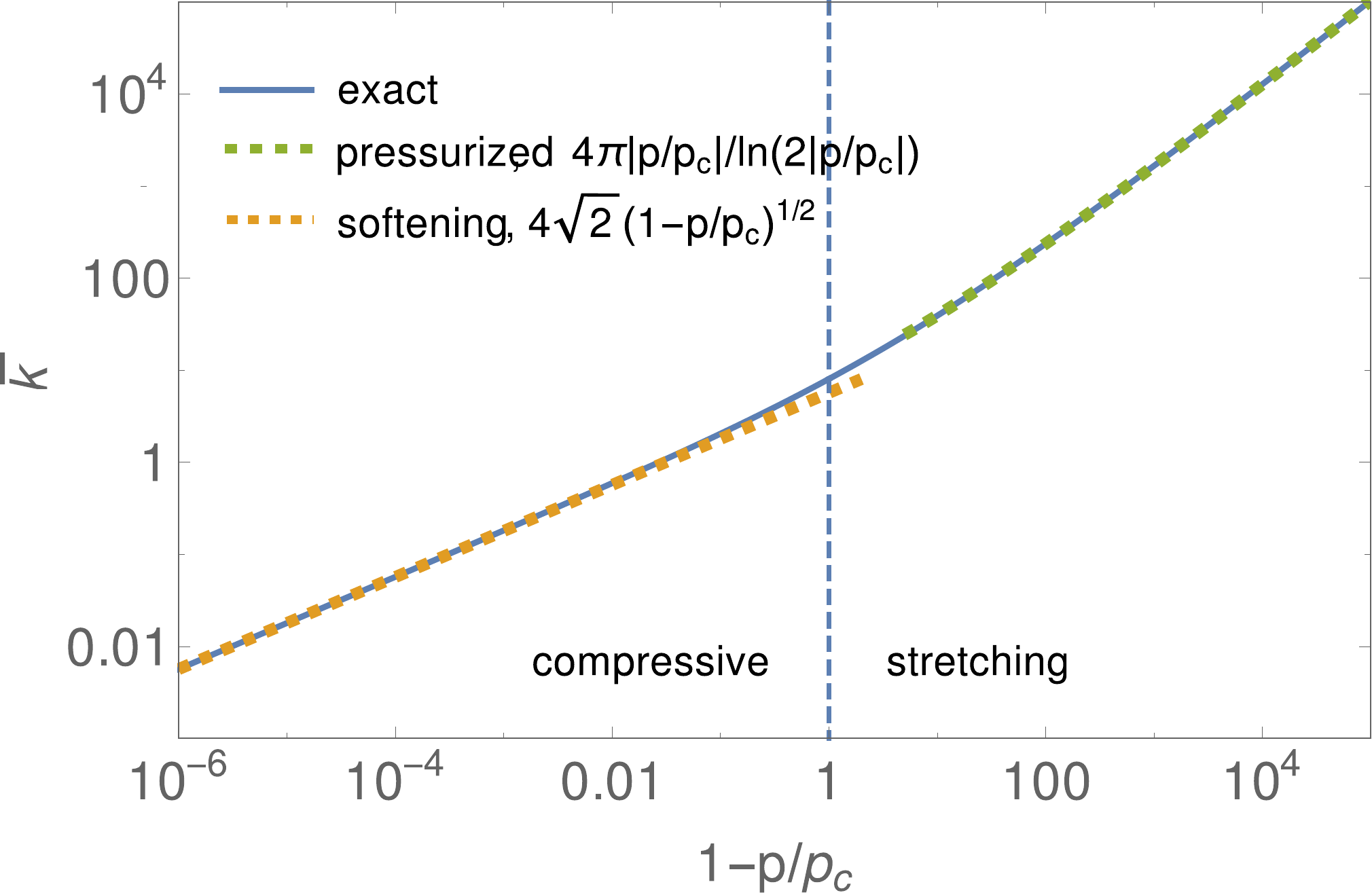}
\caption{
Double logarithmic plot of dimensionless
linear stiffness $\bar{k}$ 
as a function of pressure $1-p/p_c$ according to Eq.\ (\ref{eq:k})
both 
in the stretching ($1-p/p_c>1$) and compressive ($1-p/p_c<1$) regimes
(solid blue line). 
For $p=0$ (vertical dashed line), we have $\bar{k} = 8$.
Dashed lines are the asymptotic results for 
$\bar{k} \approx 4\pi |p/p_c|/\ln(2|p/p_c|)$ in the stretching
regime (right green line)
and $\bar{k}\propto  \left(1- p/p_c \right)^{1/2}$ in the compressive
  regime (left orange line); see Eq.\ (\ref{eq:klin}.
}
\label{fig:k}
\end{center}
\end{figure}

The force $\bar{F}$ for given $\bar{z}$ and thus 
the force-indentation relation in the linear approximation 
remains to be determined. 
It can be obtained from Eq.\ (\ref{eq:exact2}) 
by neglecting the last term, which is
  quadratic in $\bar{z}$,
\begin{align}
  -\frac{\bar{F}(\bar{z})}{2\pi} &= \int_0^{\infty} d\rho \rho \bar{w}_{\rm lin}
   = \bar{z} \frac{\lambda_--\lambda_+}{\ln \lambda_+} \nonumber\\
 &\approx - \frac{{2\sqrt{2}\bar{z}}}{\pi} \left(1- p/p_c  \right)^{1/2},
   \label{eq:Fwlin}
\end{align}
where 
the last approximation holds for $p\approx p_c$. 
Alternatively, we can inspect the asymptotics of the linear solution 
(\ref{eq:barpsi}) for $\rho \to \infty$,
\begin{align}
 \bar{\psi}_{\rm lin} &= - \partial_\rho\bar{\Phi}_{\rm lin}
  \approx  \bar{z} \frac{\lambda_--\lambda_+}{\ln \lambda_+} \frac{1}{\rho},
 \label{eq:Fwlin2}
\end{align}
which should be $\bar{\psi} \sim -\bar{F}/2\pi \rho$ according 
to (\ref{eq:psiasy}) or (\ref{eq:exact1}). 
Both (\ref{eq:Fwlin}) and (\ref{eq:Fwlin2}) lead to the same 
 dimensionless linear stiffness
\begin{align}
  \bar{k} = \frac{d\bar{F}}{d\bar{z}} 
  &=  \frac{4\pi i(1-\tau^2)^{1/2}}{\ln\left(\tau +
    i(1-\tau^2)^{1/2}\right)}
    = \frac{4\pi (1-(p/p_c)^2)^{1/2}}{\pi/2+\arcsin(p/p_c)}
  \label{eq:k}\\
   & \approx 4\sqrt{2} \left(1- p/p_c  \right)^{1/2},
 \label{eq:klin}
\end{align}
where $\tau \equiv -p/p_c$ and
with an arcsin branch $-\pi/2\le \arcsin x \le \pi/2$. 
Reverting the nondimensionalization we find the 
stiffness  $k =  Y \gamma^{-1/2}\bar{k}(p/p_c) 
=  (Y^{1/2}\kappa^{1/2}/R_0) \bar{k}(p/p_c)$.
In Ref.\ \cite{Vella2012b}, the same result has been 
obtained for  stretching pressures $p\le 0$ ($\tau\ge 0$).
We thus conclude that 
this result 
can be analytically  continued also to compressive pressures 
$0<p/p_c<1$. 
The stiffness (\ref{eq:klin}) {\it vanishes} as 
$\bar{k}\propto  \left(1- p/p_c  \right)^{1/2}$  close to $p_c$ 
corresponding to a softening of the capsule upon 
approaching the critical buckling pressure. 
Figure \ref{fig:k} clearly shows that the linear stiffness (\ref{eq:k}) 
is  monotonously decreasing with compressive pressure $p$ and 
exhibits essentially {\it two} scaling regimes, one for stretching pressures 
$-p/p_c \gg 1$, where $\bar{k} \approx 4\pi |p/p_c|/\ln(2|p/p_c|)$
\cite{Vella2012b} and the softening regime $\bar{k}\propto  \left(1- p/p_c
\right)^{1/2}$  close to $p_c$ according to (\ref{eq:klin}). 
The crossover between both regime happens around the pressure-free
case, where the Reissner results $\bar{k} = 8$ applies \cite{Reissner1946}.

The linear stiffness $k$ can be tested in various 
compression experiments in the initial 
small displacement regime. For microcapsules, 
most frequently used are plate compression
\cite{carin_compression_2003,fery_mechanical_2007,Wischnewski2018} or 
compression by microscopy tips \cite{Fery2004,fery_mechanical_2007}. 
In Ref.\ \cite{Wischnewski2018}, the result for the stiffness (\ref{eq:k})
could also be generalized if additional surface tensions are present,
which can arise, for example, from the shell-liquid interfaces or 
also as motor-induced 
active pressures if  biological 
cells are considered \cite{Mietke2015} and 
which effectively act as an additional stretching pressure. 
Our result for the linear softening of shells could be experimentally
tested in 
linear compression tests, where  an additional 
compressive pressure $0<p<p_c$ is applied.

The fact that $k>0$ for all $p<p_c$ implies that the barrier 
condition $\bar{F}=0$ can only be fulfilled 
for vanishing indentations at $p=p_c$; therefore, the barrier state 
is {\it not} directly accessible in the linear response regime,
and we will have to employ an additional expansion
around the linearized solutions. 

Close to $p_c$ the linearized solutions (\ref{eq:barw}) and (\ref{eq:barpsi})
approach
($\lambda_\pm^{1/2}\approx \pm i$)
 \begin{align}
  \bar{w}_{\rm lin} &\approx \frac{\bar{z}}{i\pi} 
\left(K_0(i\rho) - K_0(-i\rho) \right) = -\bar{z} J_0(\rho),
\label{eq:barw1}\\
 \bar{\psi}_{\rm lin} &\approx -\bar{w}_{\rm lin}' \approx  -\bar{z} J_1(\rho),
\label{eq:barpsi1}
\end{align}
where $J_\nu(x)$ and $K_\nu(x)$ are Bessel functions. 
The normal displacement thus exhibits
  extended oscillations with a period $\Delta \rho \approx 2\pi$ 
corresponding to $\Delta r= 2\pi l_{\rm el}$. This is reminiscent 
of the appearance of an unstable wavelength  
$\lambda_c =   2\pi l_\text{el}$ at the buckling threshold $p=p_c$ 
 in the absence of an additional point force which localizes
the dimple \cite{Hutchinson1967}.

\section{Numerical results for the  barrier state}

\begin{figure*}
\begin{center}
\includegraphics[width=0.95\textwidth]{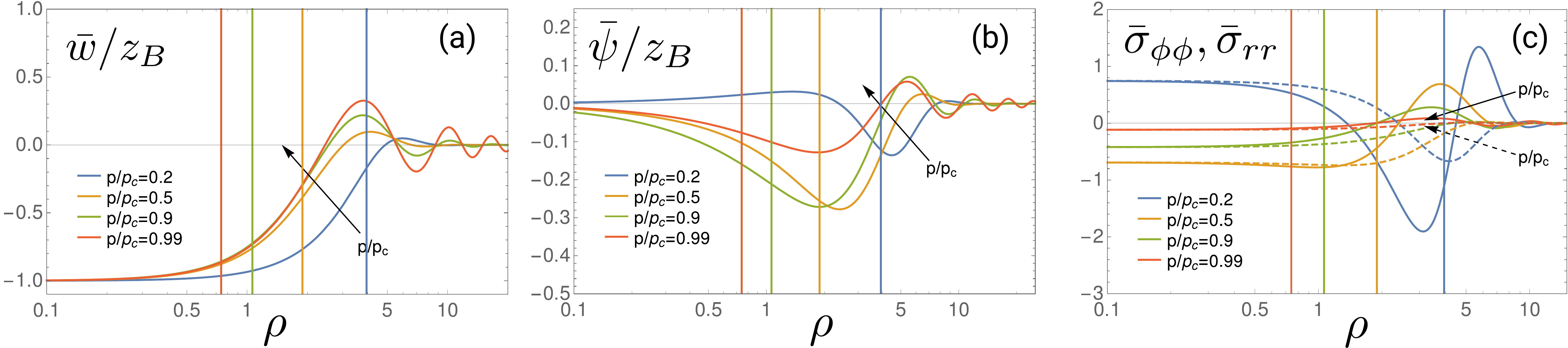}
\caption{Numerical results for the 
   barrier state of a buckled spherical shell 
   (see also shapes in Fig.\ \ref{fig:barrier_snapshot}).
   Arrows indicate increasing pressure.
   (a) Normalized dimensionless
       normal displacement $\bar{w}(\rho)/\bar{z}_B$, 
   (b) normalized dimensionless stress function  $\bar{\psi}(\rho)/\bar{z}_B$, 
   and (c) dimensionless
     hoop stress  $\bar{\sigma}_{\phi\phi} = \partial_\rho\psi$  (solid lines)
    and radial  stress $\bar{\sigma}_{rr} = \psi/\rho$ (dashed lines); 
 vertical lines  indicate the indentation width
 $\rho_B =(2\overline{\Delta V}_B/\pi\bar{z}_B)^{1/2}$.
}
\label{fig:barrier_conf}
\end{center}
\end{figure*}

Figure \ref{fig:barrier_snapshot} shows numerical results 
for the shell configuration in the barrier state, and 
Fig.\  \ref{fig:barrier_conf}  shows
the normalized displacement $\bar{w}(\rho)/\bar{z}_B$,
 the stress function, and stress distribution along 
the shell for various pressures.

In the following, we present numerical results for the 
energy barrier $\bar{E}_B$ and  the corresponding pole  
indentation $\bar{z}_B$ at the barrier state (the indentation, 
where $\bar{F}(\bar{z}_B)=0$) as a function of pressure.
The scaling of these quantities
with pressure starting from $p$ close to $p_c$ down to $p/p_c \ll 1$ 
always shows a clear crossover between  {\it two} scaling regimes. 
One scaling regime governs the softening behavior close to $p_c$
and characterizes the critical properties and exponents 
of the buckling instability; the other scaling regime 
for $p/p_c \ll 1$ corresponds to a barrier state, 
which is  a well-developed
mirror-inverted Pogorelov dimple. 
The crossover between both regimes takes place at $\bar{z}_B \sim 1$
corresponding to $z_B \sim h$ or pressures $p/p_c \sim 1/2$.   

For $p \ll p_c$, we find the typical Pogorelov scaling 
for the energy barrier.
Here, the indentation at the barrier state 
 is deep ($\bar{z}_B\gg 1$) and typically an inverted spherical cap
which is localized to $\rho < \rho_B \sim \bar{z}_B^{1/2}$;
see also Fig.\ \ref{fig:barrier_conf}(a).
For $p\ll p_c$, a deep indentation by a point force is necessary 
to carry the shell into the snap-through buckled state. 
The Pogorelov dimple consists of a mirror inverted spherical cap
 whose sharp edge at the rim becomes 
 rounded to avoid infinite bending energies \cite{Pogorelov}. 
This rounding happens over a boundary layer of width 
$\xi \sim  R_0 \gamma^{-1/4}k = (hR_0)^{1/2}\sim l_{\rm el} k$ 
\cite{Knoche2014a,Knoche2014}
or, in dimensionless units, 
$\bar{\xi}\sim k \sim O(1)$. For $\bar{z}_B < 1$ or 
$\rho_B<1$ (corresponding to larger pressures $p/p_c > 1/2$), 
the Pogorelov dimple at the barrier state becomes too shallow to fully 
develop this boundary layer, and a crossover to the softening regime 
happens  close to $p_c$.

Close to $p_c$, not only the linear stiffness $k$ vanishes.
Also the energy barrier, which protects the 
unbuckled state from spontaneous buckling,  and the corresponding 
indentation $\bar{z}_B$  at the barrier state must vanish
at $p_c$  in order to connect smoothly  to an unstable
energy landscape with  $\partial_z E_{\rm tot}(z=0) <0$ 
corresponding to a  spontaneous buckling instability for $p>p_c$. 
In this regime, the indentation $\bar{w}(\rho)$ in the barrier state 
is very shallow [see Fig.\ \ref{fig:barrier_conf}(a)]
and exhibits extended 
oscillations on the typical length scale $\Delta \rho \sim 1$ 
corresponding
to $\Delta r\sim l_{\rm el}$ reminiscent of the linearized theory. 
If  $p$ is already close to $p_c$, a small 
additional localized indentation by a point force is sufficient to 
carry the shell over the energy barrier into the snap-through buckled state.

After presenting the numerical results along with some 
scaling arguments we will derive exact analytical results for 
the asymptotics of 
the barrier energy  $\bar{E}_B$ and the barrier 
indentation $\bar{z}_B$ in both limits close to $p_c$ and 
for $p\ll p_c$ in the following sections.

\subsection{Indentation in the barrier state}

\begin{figure}
\begin{center}
\includegraphics[width=0.4\textwidth]{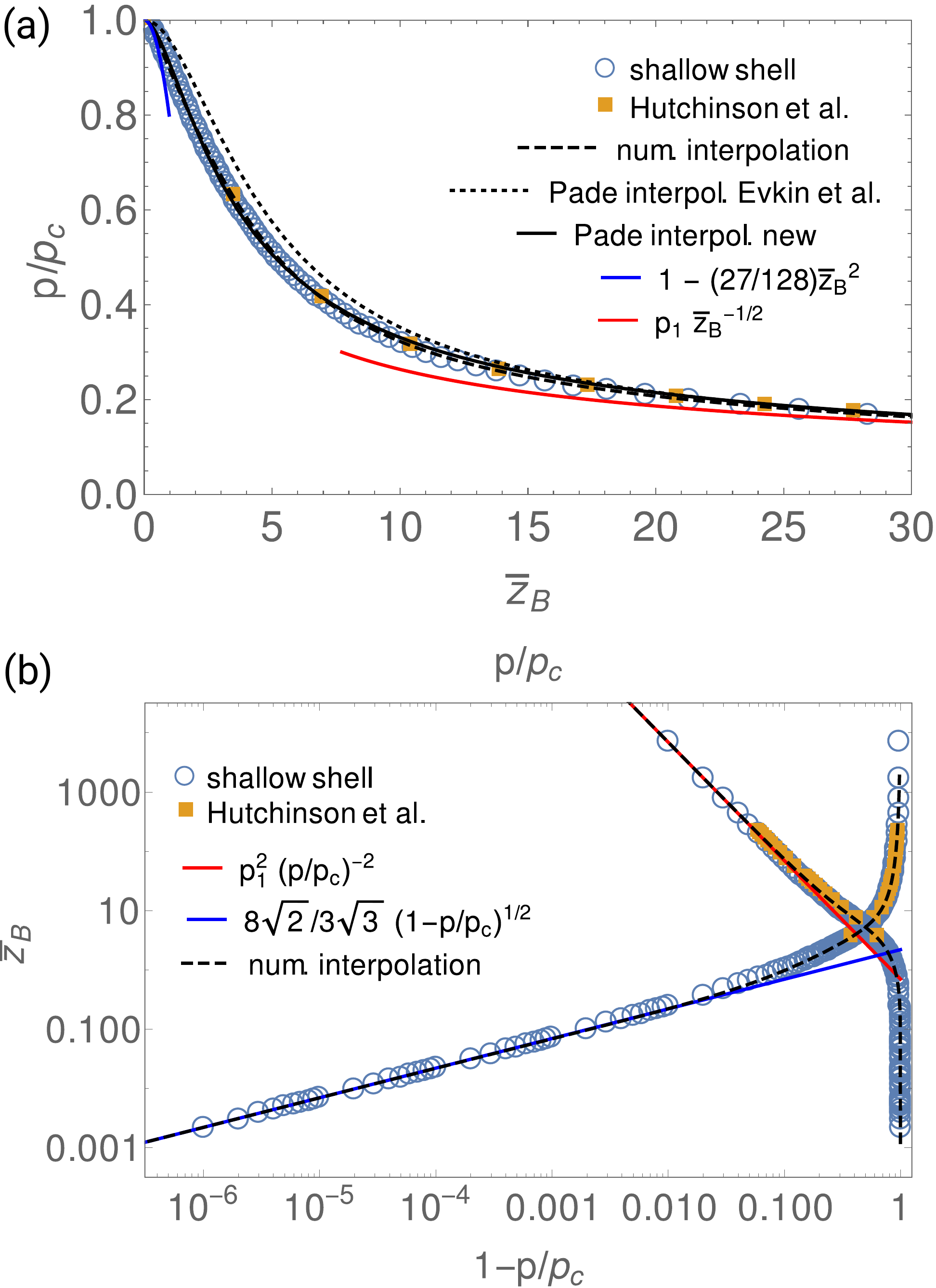}
\caption{
 Numerical shallow shell 
  results for relation between  indentation $\bar{z}_B$  at the barrier state 
     and pressure $p/p_c$.
(a) Pressure $p/p_c$ as a function of $\bar{z}_B$ 
 together with the asymptotic analytical results (\ref{eq:pzPog}) 
(lower solid red line, $p_1 \simeq 0.8337$)
and  (\ref{eq:zBlin2}) (upper solid blue line).
(b) Double logarithmic plot of $\bar{z}_B$ as a function of
$p/p_c$ (upper curve and upper horizontal scale)
together with the analytical result (\ref{eq:zBPog}) (upper
solid red line
 and as a function of  $1-p/p_c$  (lower curve and lower horizontal scale)
 together with the analytical result (\ref{eq:zBlin}) (lower
 solid blue line).
In both panels (a) and (b), we also show  the 
interpolation formula  (\ref{eq:zBinterpol}) (black dashed line)
 and  the data for the function $p/p_c=f(\xi)$ versus 
$\bar{z}_B=\sqrt{12}\xi$
from Hutchinson {\it et al.} \cite{Hutchinson2017b,Hutchinson2018};
see Table \ref{tab:dimensionless} in Appendix \ref{app:dim}.
In panel (a), we also show Pad{\'e} interpolations from  Evkin {\it et al.} 
\cite{Evkin2016} (black dotted line) and as derived below 
[see Eq.\ (\ref{eq:Pade}), black solid line].
}
\label{fig:zp}
\end{center}
\end{figure}

Close to $p_c$ the indentation in the barrier state 
becomes small $\bar{z}_B\ll 1$, 
such that it resembles the oscillating 
 linearized solutions (\ref{eq:barw}) and
(\ref{eq:barPhi}). For $p\gg p_c$, 
on the other hand, also the  barrier state is a mirror-inverted 
Pogorelov dimple with $\bar{z}_B\gg 1$. Many of its scaling properties 
can be explained based on the Pogorelov approach in this regime
\cite{Baumgarten2018}. Figure \ref{fig:zp} shows   
numerical shallow shell   results for the relation between $\bar{z}_B$ 
and pressure $p/p_c$.

\begin{figure*}
\begin{center}
\includegraphics[width=0.99\textwidth]{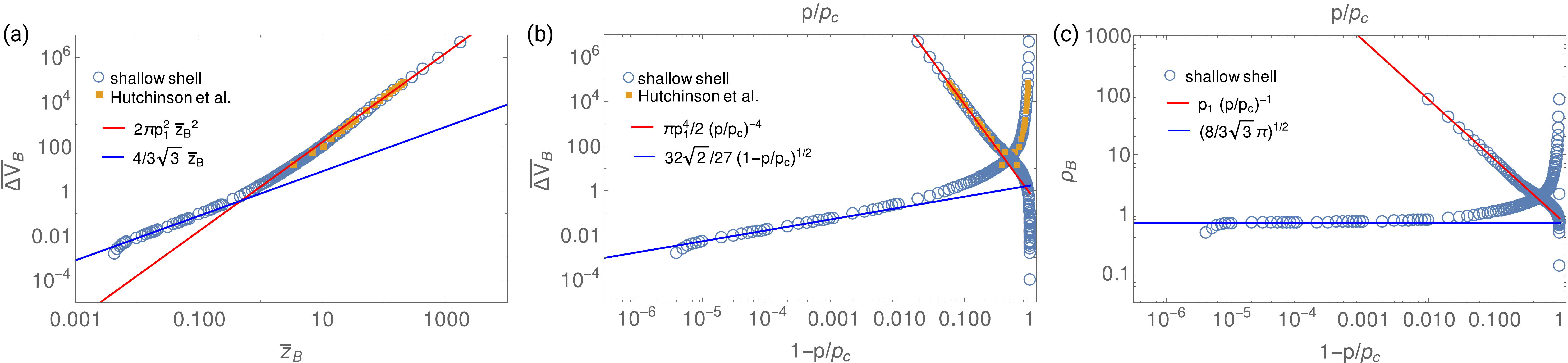}
\caption{
Numerical shallow shell results for the  indentation volume 
   $\overline{\Delta V}_B$ and the 
 effective indentation width  
 $\rho_B\equiv (2\overline{\Delta V}_B/\pi\bar{z}_B)^{1/2}$ 
at the barrier state.
 (a)  Dimensionless indentation volume 
   $\overline{\Delta V}_B$ as a function of indentation depth 
   $\bar{z}_B$ as compared to analytical results (\ref{eq:VindPog})
   (blue line fitting shallow indentations $\bar{z}_B\ll 1$)
   and  (\ref{eq:Vindlin}) (red line, $p_1 \simeq 0.8337$, fitting deep
   indentations $\bar{z}_B\gg 1$).
(b) $\overline{\Delta V}_B/2\pi$ 
 as a function of pressure $p/p_c$ (upper curves 
and upper horizontal scale)
 together with analytical result (\ref{eq:VindPog}) (upper red line)
 and  as a function of $1-p/p_c$  (lower curves and lower horizontal scale)
 together with analytical result (\ref{eq:Vindlin}) (lower blue line).
We also show the data from Hutchinson {\it et al.} 
 \cite{Hutchinson2017b,Hutchinson2018}
for the function $\overline{\Delta V}_B = 24\pi h(\xi)$ 
 plotted as a function of 
$\bar{z}_B=\sqrt{12}\xi$ (a) or as a function of $p/p_c=f(\xi)$ and
$1-p/p_c$ (b); see Table \ref{tab:dimensionless}.
(c)
  Effective indentation width  
 $\rho_B$ 
 at the barrier state as a function of pressure $p/p_c$ and  $1-p/p_c$
 together with the analytical results (\ref{eq:rhoBPog})
 (upper red line) and
  (\ref{eq:rhoBlin}) (lower blue line), respectively.
}
\label{fig:Vind}
\end{center}
\end{figure*}

Close to $p_c$, the indentation $\bar{z}_B$ 
at the barrier [$\bar{F}(\bar{z}_B)=0$] becomes small, $\bar{z}_B\ll 1$.
Numerically, we find for $\bar{z}_B$ as a 
function of pressure a crossover between just two scaling regimes,
\begin{align}
\bar{z}_B  &\propto 
\begin{cases} (1-p/p_c)^{1/2} &\mbox{for}~ p \approx p_c
\\
  (p/p_c)^{-2} &\mbox{for}~ p \ll p_c
\end{cases},
\label{eq:zp}
\end{align}
with a clear crossover at $\bar{z}_B \sim 1$; see Fig.\ \ref{fig:zp}(b).
The indentation $\bar{z}_B$ at the barrier is monotonously 
decreasing as a function of $p$, which shows that 
increasingly deep indentations are necessary to reach the 
metastable barrier beyond which the shell will spontaneously 
fall into the snap-through buckled state. 
 
Both scaling results in the limits $p$ close to $p_c$ and $p\ll p_c$ 
are non-trivial results, which we will rationalize in the course of this 
paper. In Fig.\ \ref{fig:zp}, we compare with the exact asymptotics 
including numerical prefactors that will be calculated in the 
following sections and find excellent agreement. 
We also see that the numerical data given in
Refs.\ \cite{Hutchinson2017b,Hutchinson2018} is in 
excellent agreement but
does not cover the asymptotics for $p$ close to $p_c$. 
 The  scaling of the 
depth of the barrier state for $p\ll p_c$ has been obtained previously 
in Ref.\ \cite{Paulose2012}
based on the Pogorelov energy estimate
and in Refs.\ \cite{Evkin1989,Evkin2016} using a boundary 
layer approach with variational energy minimization 
that turns out to be equivalent to the 
systematic expansion that we will employ below.

The dimensionless
indentation volume $\overline{\Delta V}_B
= - \overline{\Delta V}(\bar{z}_B)>0$ [see Eq.\  (\ref{eq:DeltaV})]
at the barrier state shows a characteristic 
dependence on the indentation $\bar{z}_B$ 
at the barrier:
\begin{align}
\overline{\Delta V}_B  =  &\propto 
\begin{cases}  \bar{z}_B &\mbox{for}~ \bar{z}_B \ll 1
\\
  \bar{z}_B^2 &\mbox{for}~ \bar{z}_B \gg 1
\end{cases},
\label{eq:Vind}
\end{align}
see Fig.\ \ref{fig:Vind}(a), 
with  a clear crossover at
$\bar{z}_B \sim 1$ ($z \sim h$) between shallow and 
deep indentations.
 Figures \ref{fig:Vind}(a) and (b) also show  that the numerical data from
Refs.\ \cite{Hutchinson2017b,Hutchinson2018} are in 
excellent agreement. 
When we combine  (\ref{eq:zp}) and (\ref{eq:Vind}), the pressure 
dependence of the indentation volume follows as 
\begin{align}
\overline{\Delta V}_B   &\propto 
\begin{cases} (1-p/p_c)^{1/2} &\mbox{for}~ p\approx p_c
\\
  (p/p_c)^{-4} &\mbox{for}~ p \ll p_c
\end{cases},
\label{eq:Vindp}
\end{align}
in agreement with the numerical results in Fig.\ \ref{fig:Vind}(b).

From the indentation volume $\overline{\Delta V}_B $
and the indentation depth 
$\bar{z}_B$ at the barrier state, we can define an effective
width $\rho_B$ of the indentation as 
\begin{align}
  \rho_B &\equiv (2\overline{\Delta V}_B /\pi\bar{z}_B)^{1/2} &
           \!\!\!\!   \propto  
\begin{cases}  {\rm const} &\!\mbox{for}~ p\approx p_c
\\
 \bar{z}_B^{1/2} \propto (p/p_c)^{-1}  &\!\mbox{for}~  p\ll p_c
\end{cases};
\label{eq:rhoB}
\end{align}
see Fig.\ \ref{fig:Vind}(c).
We choose the numerical prefactor  in the 
definition of $\rho_B$ such that a mirror-inverted Pogorelov dimple
with  $\bar{w}(\rho) = -\bar{z}_B+ \rho^2$ 
and $\overline{\Delta V}_B  =  \pi \bar{z}_B^2/2$
has $\rho_B = \bar{z}_B^{1/2}$ in accordance with $\bar{w}(\rho_B)=0$.
This is exactly the behavior of $\overline{\Delta V}_B$ and $\rho_B$ 
 for  $\bar{z}_B \gg 1$ or $p\ll p_c$.
The effective indentation width $\rho_B$ 
 remains remarkably constant $\simeq 0.70$ 
for pressures $p$ close to $p_c$
corresponding to an indentation width $r_B \sim l_{\rm el}$.
This behavior will have interesting consequences for the 
buckling behavior of small  soft spots. 
The depth $\bar{z}_B$ is vanishing for  $p$ close to $p_c$ such that 
the indentation at the barrier becomes not only increasingly shallow 
but also increasingly broad with 
a  width-to-depth ratio $r_B/z_B \propto \gamma^{1/4}  (1-p/p_c)^{-1/2}$
[note the different dimensionless units for  $r_B$ and $z_B$ in Eq.\ 
  (\ref{eq:non-dim})].
For $p\ll p_c$, the width of the barrier state increases
with decreasing pressure such that the indentation becomes 
not only increasingly deep but also increasingly 
 narrow with a depth-to-width ratio
 $\propto \gamma^{1/4}(p/p_c)^{-1}$.

For a Pogorelov dimple, the scaling  
$\overline{\Delta V}_B\sim (p/p_c)^{-4}$ 
has been shown in Ref.\ \cite{Baumgarten2018} based on the 
Pogorelov energy estimate for the elastic energy of a mirror-inverted 
dimple. 
Together with the geometric result 
$\overline{\Delta V}_B= \pi \bar{z}_B^2/2$ for mirror-inverted dimples,
this rationalizes the numerically observed scaling (\ref{eq:zp})
of the indentation for  $p\ll p_c$.  
We will present a strict 
derivation in the framework of nonlinear shallow shell
theory below.

Because the  indentation $\bar{z}_B$ 
at the barrier remains small 
close to $p_c$, the solution $\bar{w}(\rho)$ resembles 
the linear approximation (\ref{eq:barw}) in this regime.
We can use the exact relation (\ref{eq:exact2})  
at  $\bar{F}=0$  and see that  the indentation volume 
[the first term on the right-hand side of  relation (\ref{eq:exact2})]
must be given by  the second 
term on the right-hand side which is of second order in $\bar{z}$.
Our numerics confirm  that this term 
 can  still be obtained using 
  the linearized solution (\ref{eq:barw}) 
 to a good approximation, 
\begin{align}
 -\frac{\overline{\Delta V}}{2\pi} &=  
   \frac{1}{4}   \int_0^\infty d\rho \rho (\partial_\rho \bar{w})^2
    \approx
   \frac{1}{4}  \int_0^\infty d\rho \rho  (\partial_\rho \bar{w}_{\rm lin})^2
   \nonumber\\
    &\approx
   -\frac{\bar{z}^2}{4\ln^2 \lambda_+} 
  \left( 1- \ln\lambda_+ \frac{\lambda_++\lambda_-}{\lambda_+-\lambda_-}\right)
 \nonumber\\
  &\approx  \frac{\bar{z}^2}{4\sqrt{2}\pi}  \left(1- p/p_c  \right)^{-1/2}
\label{eq:intwlin}
\end{align}
for $\bar{z}\ll 1$.
The numerical results (\ref{eq:Vind}), (\ref{eq:Vindp}),
 and (\ref{eq:zp})  suggest, on the other hand, 
 that 
\begin{equation}
-\frac{\overline{\Delta V}}{2\pi} = - \int_0^{\infty} d\rho \rho \bar{w} 
   =  {\rm const}\,   \bar{z}
\end{equation}
 still holds at the barrier, i.e., that the term is still 
 linear in $\bar{z}$ to a good approximation and the indentation 
 extends over $\rho = O(1)$.
The indentation volume does, however, not contain 
a factor $(1-p/p_c)^{1/2}$ as in the linearized solution [see 
 Eq.\ (\ref{eq:Fwlin})]. Nonlinear corrections are  affecting 
 the shape of the indentation at the barrier such that the cancellation of 
 oscillating contributions that governs the linearized result 
 (\ref{eq:Fwlin}) no longer happens but  the indentation 
 still extends over $\rho = O(1)$ as in the linearized solution.
Equating with Eq.\ (\ref{eq:intwlin}) at the barrier gives 
\begin{align*}
    \bar{z}_B  &\propto  \left(1- p/p_c \right)^{1/2},
\end{align*}     
 which explains  the numerically observed scaling (\ref{eq:zp})
 of the indentation close to $p_c$. 
We will present a strict derivation in the framework 
of nonlinear shallow shell
theory below.

\begin{figure}
\begin{center}
\includegraphics[width=0.4\textwidth]{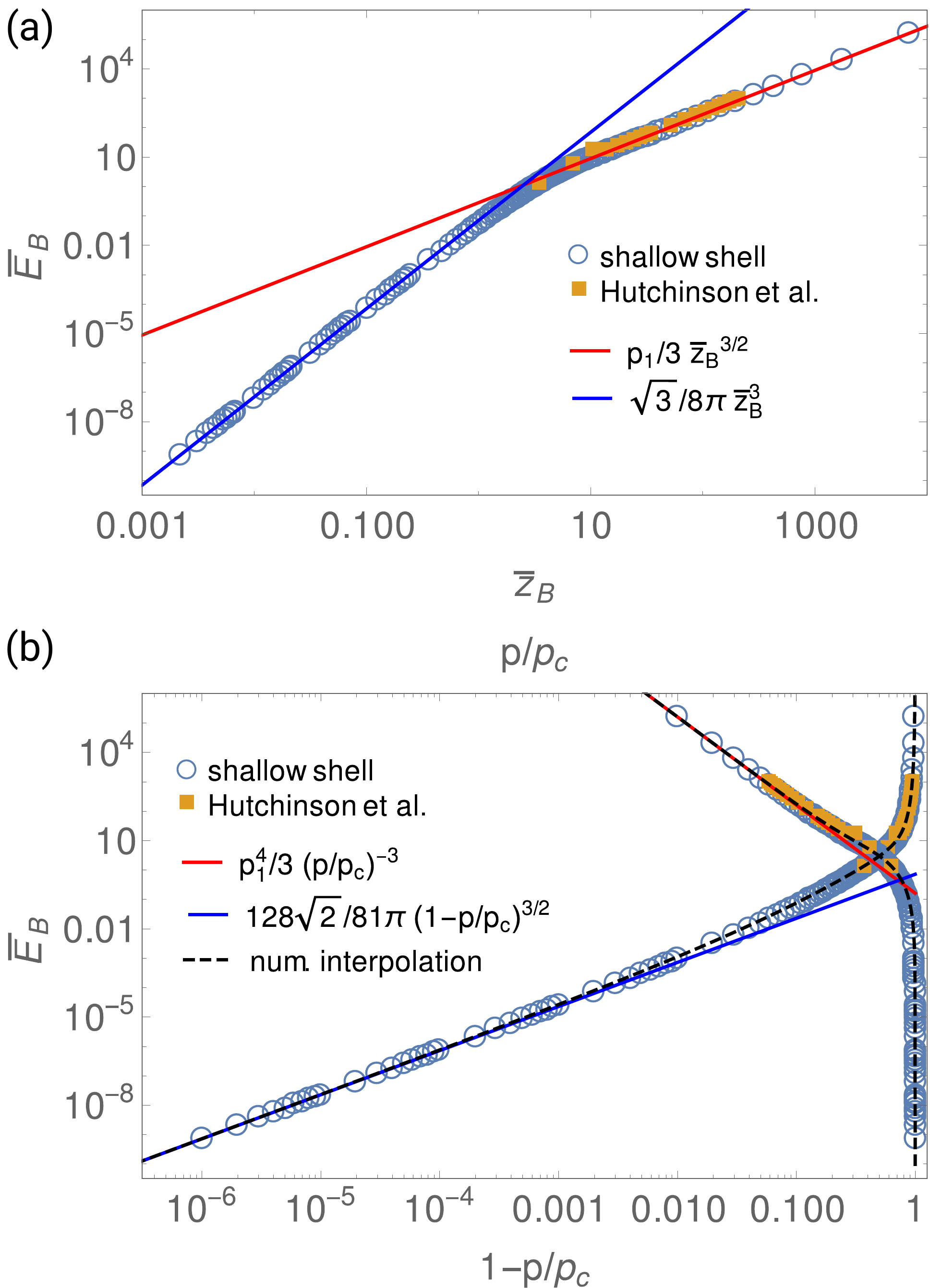}
\caption{
 Numerical shallow shell results for the   energy barrier $\bar{E}_B$.
(a) Double logarithmic plot of $\bar{E}_B$
  as a function of barrier indentation $\bar{z}_B$
  with the analytical results (\ref{eq:EBzPog}) (red line, $p_1 \simeq
  0.8337$, fitting deep indentations $\bar{z}_B\gg 1$)
  and (\ref{eq:EBzlin}) (blue line
  fitting shallow indentations $\bar{z}_B\ll 1$). 
(b) Double logarithmic plot of  $\bar{E}_B$
as a function of pressure $p/p_c$ (upper curves and upper horizontal scale)
 together with the analytical result (\ref{eq:EBpPog}) 
(upper red line)
 and as a function of  $1-p/p_c$  (lower curves and lower horizontal scale)
together with the analytical result (\ref{eq:EBplin}) (lower blue line). 
Also shown are   the  data from Hutchinson {\it et al.} 
 \cite{Hutchinson2017b,Hutchinson2018} and the 
interpolation formula  (\ref{eq:interpol}) (black dashed line in panel (b)).
}
\label{fig:EB}
\end{center}
\end{figure}

\subsection{Energy barrier}

Now we address the energy barrier itself. 
Numerically, we find for the energy barrier height
 \begin{align}
\bar{E}_B &\propto
  \begin{cases}
    \bar{z}_B^3 &\mbox{for}~  \bar{z}_B \ll 1\\
   \bar{z}_B^{3/2} &\mbox{for}~ \bar{z}_B \gg 1
 \end{cases},
\label{eq:EBz}
\end{align}
again with clearly two regimes and 
a crossover at $\bar{z}_B \sim 1$, see Fig.\ \ref{fig:EB}(a).
For small indentations $\bar{z} \ll 1$, the linear regime 
with 
$\bar{\psi}, \bar{w} \propto \bar{z}$ is a good approximation up to 
the barrier, and the typical 
radial extent of the indentation is $\rho= O(1)$, resulting in 
$\bar{E}_B \propto  \bar{z}_B^3$ according to (\ref{eq:Etot}). 
For deep indentations $\bar{z} \gg 1$, the characteristic behavior of a 
mirror-inverted Pogorelov dimple is  $\partial_\rho \bar{w} \sim
 \bar{z}^{1/2}$ and $\bar{\psi} \sim  \bar{z}^{1/2}$   
over a width  $\Delta \rho = O(1)$ at $\rho \sim \bar{z}^{1/2}$ 
(in the absence of pressure)  \cite{Gomez2016}, 
which results in $\bar{E}_B \propto  \bar{z}_B^{3/2}$ 
according to (\ref{eq:Etot}). 
This means both scaling limits in (\ref{eq:EBz}) 
can be rationalized by nonlinear shallow shell theory. 
 In Fig.\ \ref{fig:EB}(a), we compare with the exact asymptotics 
including numerical prefactors that will be calculated rigorously 
in the following sections.

\begin{figure}
\begin{center}
\includegraphics[width=0.4\textwidth]{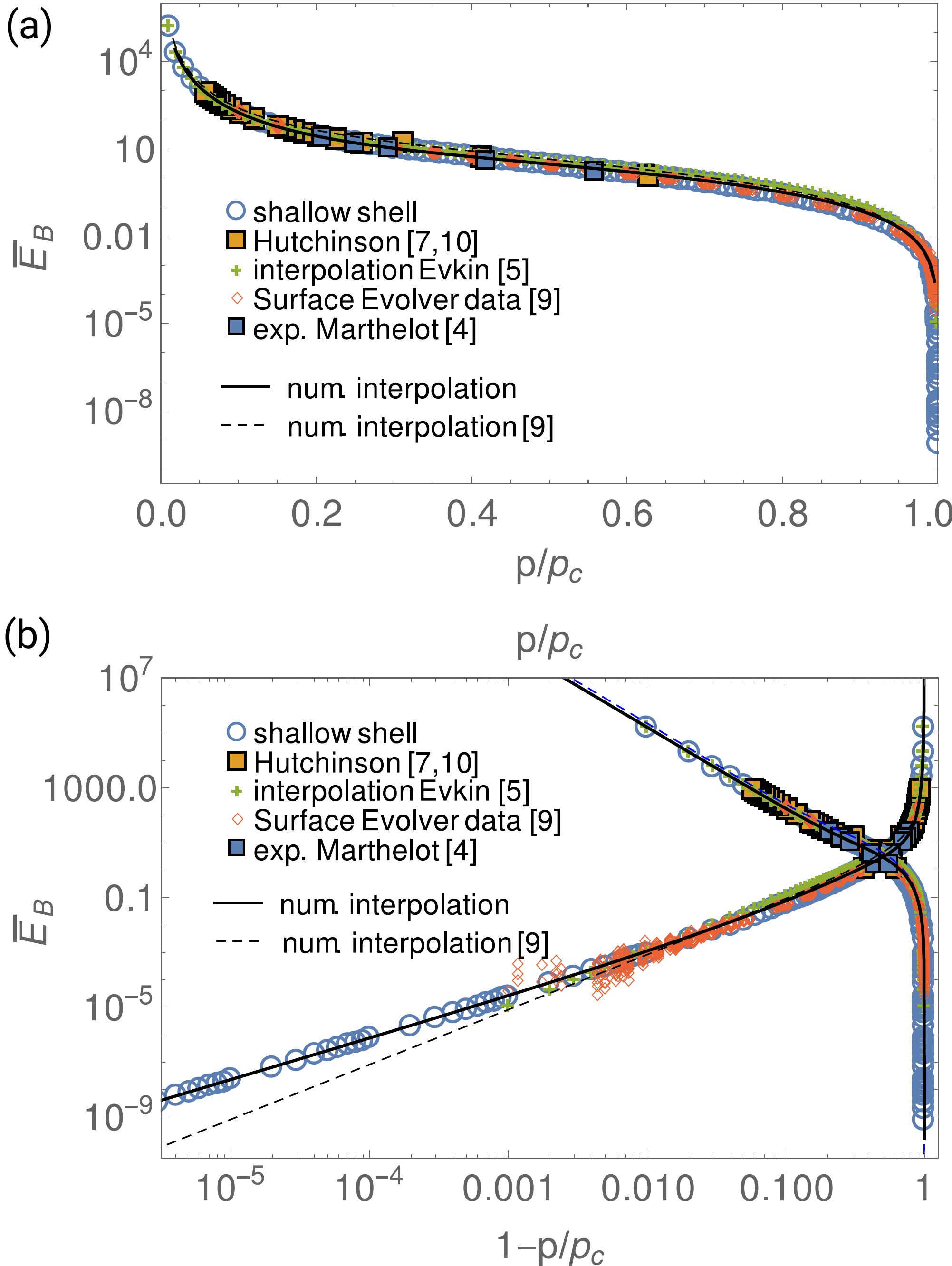}
\caption{
 Comparison of our numerical shallow shell results for the 
energy barrier  $\bar{E}_B$ as a function of $p/p_c$  or $1-p/p_c$ 
[(a) logarithmic, (b) double-logarithmic]
  to  different analytical, numerical, and experimental 
 energy barrier results  from the literature:
 numerical data  from Hutchinson {\it et al.}
   \cite{Hutchinson2017b,Hutchinson2018} (orange big squares),  
  energy barrier from integrating an 
    analytical interpolation of the force-indentation relation 
   from  Evkin {\it et al.}
  \cite{Evkin2017} (green crosses), 
  experimental data from  Marthelot  {\it et al.} \cite{Marthelot2017}
   (blue small squares), 
  and  SURFACE EVOLVER data (red diamonds) and an approximative 
    numerical interpolation formula from Ref.\ \cite{Baumgarten2018}
    (dashed line).
We also compare to the 
   new interpolation formula (\ref{eq:interpol}) (solid line). 
Clearly, shallow shell theory is correct through the whole range of pressures.
}
\label{fig:compareEB}
\end{center}
\end{figure}

Together with (\ref{eq:zp}), this 
results in a pressure dependence
\begin{align}
\bar{E}_B &\propto
  \begin{cases}
    (1-p/p_c)^{3/2} &\mbox{for}~  p\approx p_c\\
    (p/p_c)^{-3} &\mbox{for}~  p \ll p_c
 \end{cases},
\label{eq:EBp}
\end{align}
in agreement with the numerical results in Fig.\ \ref{fig:EB}(b).
The scaling  $\bar{E}_B \propto (p/p_c)^{-3}$ for deep indentations 
has been obtained before in Refs.\ 
\cite{Knoche2014o,Evkin2016,Baumgarten2018} (and implicitly also 
in Ref.\ \cite{Paulose2012})
based on the arguments of Pogorelov for the energy cost of a 
buckling indentation.  
The scaling $\bar{E}_B \propto (1-p/p_c)^{3/2}$  governs the softening 
of the shell close to the buckling
pressure and is a new result that corrects the conjecture 
$\bar{E}_B \propto (1-p/p_c)^{2}$ that has been obtained based on 
numerical data from SURFACE EVOLVER simulations 
in Ref.\ \cite{Baumgarten2018}. The SURFACE EVOLVER is, however, 
not well suited to investigate very shallow dimples as they
occur close to $p_c$. Shallow shell theory and the numerical 
continuation approach give much better
results in this regime, which extend over several decades  of the 
small parameter $1-p/p_c$ and reveal the actual exponent $3/2$. 
In Fig.\ \ref{fig:compareEB}, we compare our numerical results from 
shallow shell theory to the SURFACE EVOLVER simulation,  to 
 numerical data from Hutchinson and coworkers from 
Refs.\ \cite{Hutchinson2017b,Hutchinson2018} from moderate rotation theory, 
 to an analytical interpolation formula from  Evkin {\it et al.}
  \cite{Evkin2017}, and to 
   experimental data from  Marthelot  {\it et al.} \cite{Marthelot2017}.
We find excellent agreement and see that only the present numerical 
approach accesses the asymptotics for $p$ close to $p_c$. 

\section{Shallow shell theory for the Pogorelov barrier state}

In this section, we derive several analytical results for the 
energy barrier state  from nonlinear shallow shell 
theory in the  Pogorelov limit $p\ll p_c$ corresponding to a deep 
indentation $\bar{z}_B \gg 1$ in the barrier state:
\begin{subequations}
\begin{align}
  \bar{z}_B &= p_1^2\,  (p/p_c)^{-2} + O[(p/p_c)^{0}],
\label{eq:zBPog}\\
  \overline{\Delta V}_B &\approx \frac{\pi}{2}\bar{z}_B^2  \approx 
    \frac{\pi p_1^4}{2}\, (p/p_c)^{-4},
\label{eq:VindPog}\\
 \rho_B &\approx p_1 \, (p/p_c)^{-1},
\label{eq:rhoBPog}\\
  \bar{E}_B &=\frac{p_1}{3} \, \bar{z}_B^{3/2} +  O(\bar{z}_B),
\label{eq:EBzPog}\\
\bar{E}_B &=\frac{p_1^4}{3}\, (p/p_c)^{-3} +  O[(p/p_c)^{-2}]
  ~~\mbox{with}~
\label{eq:EBpPog}\\
  p_1 &\simeq 0.83370854.
\label{eq:p1}
\end{align}
\label{eq:barrierPog}
\end{subequations}
Thus  we derive all  Pogorelov scaling exponents
[see Refs.\ \cite{Knoche2014o,Evkin2017,Baumgarten2018}
and  Eqs.\  (\ref{eq:zp}), (\ref{eq:Vind}), (\ref{eq:Vindp}), (\ref{eq:rhoB}),
 (\ref{eq:EBz}), and (\ref{eq:EBp})]
from nonlinear shallow shell theory 
and also obtain exact numerical prefactors. 
The  number $p_1$  can be written as analytic expression
in terms of an integral over a
 solution of simple differential equations and is  numerically
easily accessible [see Eq.\ (\ref{eq:p1_2}) below]. 
The prefactors  accurately agree with the asymptotic numerical 
results; see 
Figs.\ \ref{fig:zp}, \ref{fig:Vind}, and \ref{fig:EB}.
We will further show that the total  indentation energy landscape 
in the presence of pressure is given by 
\begin{align}
 \bar{E}_{\rm ind}(\bar{z}) &=
       \bar{E}_{\rm ind,p=0}(\bar{z})  +\overline{p\Delta V}(\bar{z}) 
\nonumber\\
  &= 
     \frac{4p_1}{3} \bar{z}^{3/2} -   \frac{p}{p_c} \bar{z}^2;
\label{eq:landscapePog}
\end{align}
i.e., the pressure dependence is only via the mechanical work and the
elastic part of the indentation energy $\bar{E}_{\rm ind}(\bar{z})$ is 
independent of pressure. 
For deep  dimples with $\bar{z}\gg 1$ and $p\ll p_c$, 
where Eq.\ (\ref{eq:landscapePog}) is valid, 
an additional  pressure represents only a  small  perturbation 
which is apparently 
not modifying the elastic energy in leading order.

In the regime 
 $p\ll p_c$, where the barrier state is a deep mirror-inverted
dimple with $\bar{z}_B \gg 1$, we can start from the following 
mirror-inverted  solution of shallow shell equations at the 
barrier $\bar{F}=0$:
\begin{equation}
\begin{split}
\bar{w}(\rho) &= 
\begin{cases}  
  -\bar{z}_B+ \rho^2 &\mbox{for}~~ \rho< \bar{z}_B^{1/2}\\
     0 &\mbox{for}~~ \rho > \bar{z}_B^{1/2}
\end{cases},
\\
\bar{\psi}(\rho) &= 
 \begin{cases}  4 (p/p_c) \rho &\mbox{for}~~ \rho< \bar{z}_B^{1/2}\\
     0 &\mbox{for}~~ \rho > \bar{z}_B^{1/2}
\end{cases},
\end{split}
\label{eq:barwpsiPog}
\end{equation}
which is an exact solution everywhere except right at the 
rim of the dimple at $\rho= \bar{z}_B^{1/2}$, where it exhibits 
discontinuities.
Following Ref.\ \cite{Gomez2016} (where the case 
$p=0$ and $\bar{F}>0$ was considered), we smooth these discontinuities by 
an ansatz ($x\equiv \rho - \bar{z}_B^{1/2}$)
\begin{subequations}
\begin{align}
\partial_\rho\bar{w}(\rho) &= f(x) + 
  \begin{cases}  2(\bar{z}_B^{1/2}+x)  &\mbox{for}~~ x<0\\
     0 &\mbox{for}~~ x>0
  \end{cases},
\label{eq:barwA}\\
\bar{\psi}(\rho) &= \chi(x) + 
  \begin{cases}  4(p/p_c)(\bar{z}_B^{1/2}+x)  &\mbox{for}~~ x<0\\
     0 &\mbox{for}~~ x>0
  \end{cases},
\label{eq:barpsiA}
\end{align}
\label{eq:barwpsiA}
\end{subequations}
where the functions $f(x)$ and $\chi(x)$ have discontinuities 
at $x=0$ in order to lead to smooth functions $\bar{w}$ and $\bar{\psi}$.
This ansatz is conceptually similar to the boundary layer approach 
of Evkin {\it et al.} \cite{Evkin1989,Evkin2016}. 

As in Ref.\  \cite{Gomez2016}, we determine 
$f$, $\chi$, and $p$ in an expansion in inverse powers of $\bar{z}_B$.
Gomez  {\it et al.} considered the Pogorelov dimple created by a point 
force $\bar{F}>0$ in the absence of pressure and calculated 
the force-indentation curve, i.e., the point force necessary to maintain 
a given indentation $\bar{z}$. Here, we consider a
 metastable 
Pogorelov barrier state with $\bar{F}=0$ with a given 
indentation $\bar{z}_B$ and calculate the pressure 
$p$ necessary to maintain such a state. 
A major difference between both cases is the behavior of $\psi$ 
in the inner region of the dimple. Because $\bar{\psi} \sim -\bar{F}/2\pi \rho$ 
for a mirror-inverted Pogorelov state close to the ridge 
(the inner side of the ridge is compressed),  this divergence of $\bar{\psi}$ 
for small $\rho$ demands for the existence of four
additional scaling 
regions in the interior of the dimple ($\rho < \sqrt{\bar{z}}$)
\cite{Gomez2016}, resulting in the existence of a total of seven 
scaling regions. 
For the barrier state, on the other hand,
 we have $\bar{F}=0$ and this divergence of $\bar{\psi}$ is
absent.
Therefore, the solution (\ref{eq:barwpsiPog}) 
is valid in the entire region $\rho < \sqrt{\bar{z}_B}$ apart from 
the immediate ridge region and no additional regimes are present.
In a sense, the original Pogorelov picture with three regions -- 
mirror-buckled inside, Pogorelov ridge and undeformed 
outside -- is recovered for the barrier state. 
Moreover, it is the existence of the additional inner regions
that calls for an expansion of $f$, $\chi$, and $p$  in powers 
of $\bar{z}^{-1/4}$. If these regions are absent, the scaling 
in (\ref{eq:barwpsiA}) actually suggests that an 
expansion in powers of $\bar{z}_B^{-1/2}$ is sufficient:
\begin{equation}
\begin{split}
    f(x) &= \bar{z}_B^{1/2}f_0 + f_1 + \bar{z}_B^{-1/2} f_2 + ...,
\\
    \chi(x) &= \bar{z}_B^{1/2}\chi_0 + \chi_1 + \bar{z}_B^{-1/2} \chi_2 + ...,
\\
   p/p_c &= p_0 + \bar{z}_B^{-1/2} p_1 + ... .
\end{split}
\label{eq:expansion}
\end{equation}
In order to assure a continuous solution, $f(x)$ and $\chi(x)$
have to fulfill the following jump conditions at $x=0$:
\begin{equation}
\begin{split}
  \left. f\right|_{0-}^{0+} &= 2 \bar{z}_B^{1/2}~,~~
   \left. f'\right|_{0-}^{0+} = 2,
\\
 \left. \chi\right|_{0-}^{0+} &= 4\frac{p}{p_c} \bar{z}_B^{1/2}~,~~
   \left. \chi'\right|_{0-}^{0+} = 4\frac{p}{p_c}.
\end{split}
\label{eq:jump}
\end{equation}
Moreover, $f(x)$ and $\chi(x)$ vanish exponentially for $x \to \pm\infty$. 
We expect $f(x)$ and $\chi(x)$ to decay exponentially on the dimensionless 
 length set by the width of the Pogorelov rim 
$\bar{\xi}\sim O(1)$.
We note that Evkin {\it et al.} \cite{Evkin1989,Evkin2016} use a 
conceptually 
similar boundary layer approach which is based on essentially 
the same  expansion parameter 
$\varepsilon \sim \bar{z}_B^{-1/2}$.

\subsection{Leading order}

\begin{figure}
\begin{center}
\includegraphics[width=0.4\textwidth]{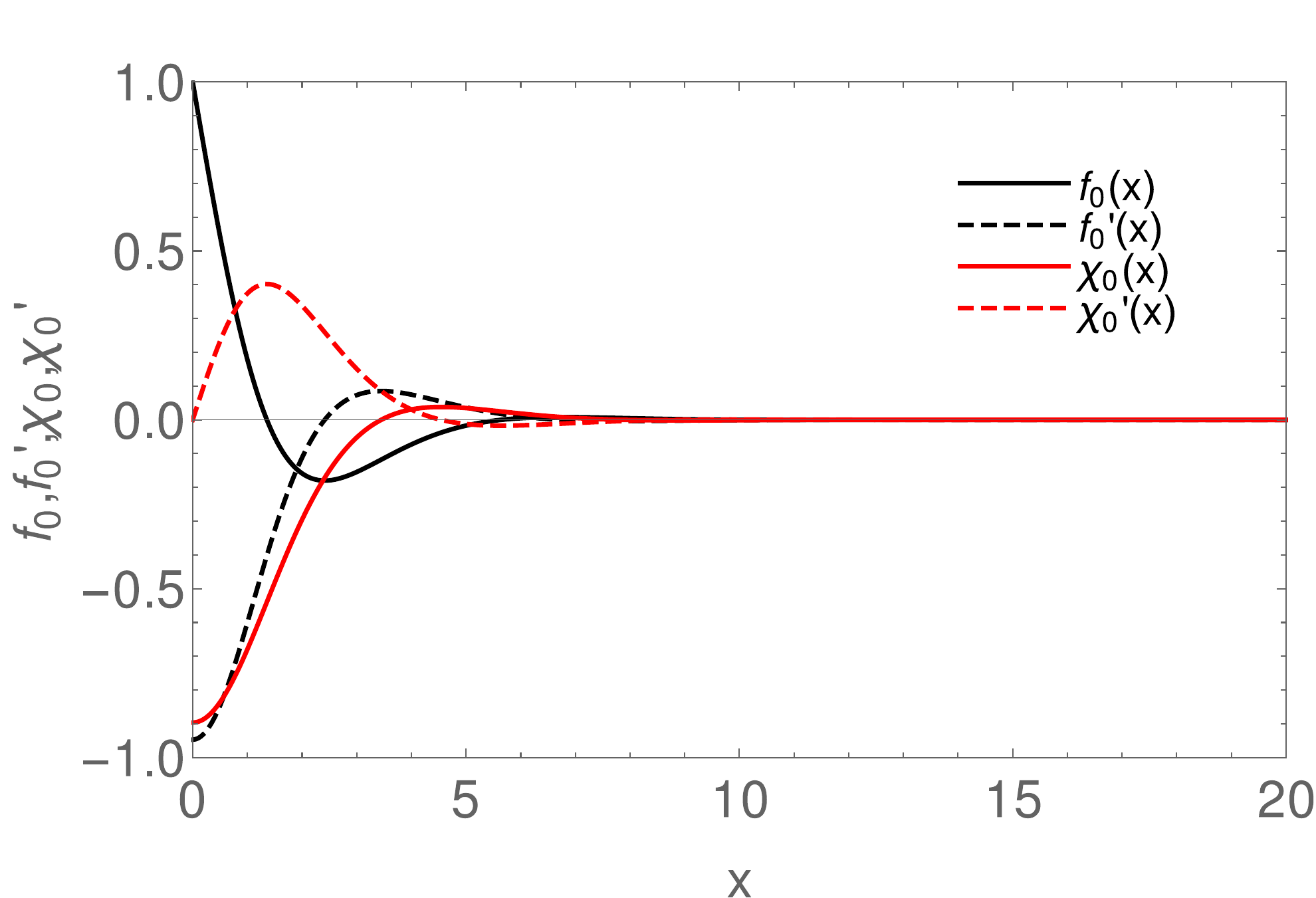}
\caption{
Leading-order functions $f_0(x)$ and $\chi_0(x)$ (solid black and red
  lines) and their derivatives (dashed black and red  lines)
for $x >0$ and $p_0=0$;
on the domain $x<0$ functions are obtained by symmetry, 
$f_0(x)$ and $\chi_0'(x)$ are odd, and $\chi_0(x)$ and $f_0'(x)$ are even. 
All functions decay exponentially on a length scale of order unity.
}
\label{fig:f0chi0}
\end{center}
\end{figure}

Inserting the  expansion (\ref{eq:expansion}) and the ansatz 
(\ref{eq:barwpsiA})
into the integrated force balance (\ref{eq:exact1}) for $\bar{w}$ and 
the compatibility condition (\ref{eq:comp3}) for $\bar{\psi}$,
we obtain in order $\bar{z}_B$ the following 
differential equations for $f_0(x)$ and $\chi_0(x)$
\footnote{
These equations are exactly identical to 
  the leading-order Eqs.\ (4.8) and (4.9) 
 from Ref.\ \cite{Gomez2016} with 
  $2p_0=F_0/2\pi$.
}:
\begin{equation}
\begin{split}
    f_0'' + {\rm sgn} x\,(\chi_0 + 2p_0 f_0) 
    &= f_0 \chi_0,
\\
  \chi_0'' - {\rm sgn} x\,f_0 &= -\frac{1}{2}f_0^2.
\end{split}
\label{eq:f0chi0}
\end{equation}
The first  equation is multiplied by $f_0'$, the 
second by $\chi_0'$, and then both equations are subtracted and 
integrated once (with a vanishing integration constant because of 
boundary conditions at infinity) to  give a first integral,
\begin{align}
   \frac{1}{2}\left(f_0'^2 - \chi_0'^2 - f_0^2\chi_0\right)
  + {\rm sgn}x\, f_0\chi_0 &= - {\rm sgn}x\, p_0 f_0^2,
\label{eq:firstI}
\end{align}
which 
holds both in $x>0$ and $x<0$ (but not right at $x=0$).
Subtracting this relation for $x=0-$ from the relation at $x=0+$ 
and employing the jump conditions at order $\bar{z}_B^{1/2}$,
\begin{equation}
\begin{split}
  \left. f_0\right|_{0-}^{0+} &= 2~,~~
   \left. f_0'\right|_{0-}^{0+} = 0,
\\
 \left. \chi_0\right|_{0-}^{0+} &= 4p_0 ~,~~
   \left. \chi_0'\right|_{0-}^{0+} = 0,
\end{split}
\label{eq:jump0}
\end{equation}
 we finally 
obtain the relation $2p_0 [f_0(0-)+2]^2=0$, from which we
conclude 
\begin{equation*}
  p_0 =0.
\end{equation*}
For $p_0=0$, Eqs.\ (\ref{eq:f0chi0}) are symmetric such that 
\begin{equation*}
  f_0(x)~{\rm odd},~~\chi_0(x)~{\rm even}.
\end{equation*}
This symmetry together with the discontinuities (\ref{eq:jump0}) 
also requires $f_0(0+)=1$. A numerical solution of 
Eqs.\ (\ref{eq:f0chi0}) for $x>0$ using the MATLAB routine {\tt bvp4c}
is shown in Fig.\ \ref{fig:f0chi0}. The boundary conditions are 
$f_0(x_{\rm min})=1$, $\chi_0'(x_{\rm min})=0$,
 and $f_0(x_{\rm max})=0$, $\chi_0(x_{\rm max})=0$
(using $x_{\rm min}=10^{-7}$ and $x_{\rm max}=1000$). 
For $p_0=0$, the above Eqs.\ (\ref{eq:f0chi0}) and boundary conditions
(\ref{eq:jump0}) become parameter free. Therefore, solutions 
 fall off exponentially on a parameter-independent length 
scale $O(1)$ (see Fig.\ \ref{fig:f0chi0}), 
which corresponds  to the  width of the Pogorelov rim 
$\bar{\xi}\sim O(1)$ as argued above.

\subsection{First  order}

In order $\bar{z}_B^{1/2}$, we obtain for $f_1(x)$ and $\chi_1(x)$ 
\begin{equation}
\begin{split}
&f_1'' +  ({\rm sgn} x -f_0) \chi_1  - \chi_0 f_1
\\
&= 
   -f_0' - xf_0'' + {\rm sgn}x\, x\chi_0 - 2 p_1 f_0  {\rm sgn}x,
\\
 & \chi_1'' +  (f_0- {\rm sgn}x)  f_1
\\
 &=  -\chi_0' - x\chi_0'' + {\rm sgn}x\, xf_0,
\end{split}
\label{eq:f1chi1}
\end{equation}
where we used already $p_0=0$.
This inhomogeneous differential equation has to be solved 
with the jump conditions 
\begin{equation}
\begin{split}
  \left. f_1\right|_{0-}^{0+} &= 0~,~~
   \left. f_1'\right|_{0-}^{0+} = 2,
\\
 \left. \chi_1\right|_{0-}^{0+} &= 4p_1 ~,~~
   \left. \chi_1'\right|_{0-}^{0+} = 4p_0
\end{split}
\label{eq:jump1}
\end{equation}
at order $\bar{z}_B^0$.

Using Eq.\ (\ref{eq:f0chi0}),
 the inhomogeneous equation (\ref{eq:f1chi1}) can be written as 
\footnote{
This equation is exactly identical to the second-order 
Eq.\ (6.3) from Ref.\ \cite{Gomez2016} with 
  $2p_1=F_2/2\pi$.
}
\begin{equation}
  \hat{L} \begin{pmatrix} f_1 \\ \chi_1 \end{pmatrix}  = 
 \begin{pmatrix}  
 -f_0' - xf_0\chi_0 - 2 p_1 f_0  {\rm sgn}x
 \\ 
-\chi_0'  + \frac{1}{2} x f_0^2 
\end{pmatrix}
\label{eq:f1chi1L}
\end{equation}
with  a linear differential operator 
\begin{equation*}
  \hat{L} \begin{pmatrix} f_1 \\ \chi_1 \end{pmatrix} \equiv 
   \begin{pmatrix}  
   f_1'' + ({\rm sgn} x -f_0) \chi_1  - \chi_0 f_1
 \\
  \chi_1'' -  ({\rm sgn}x-f_0)  f_1
\end{pmatrix}.
\end{equation*}
For the adjoint operator $\hat{L}^+$ [with respect to the scalar product 
$\langle (a,  b), (c,d) \rangle \equiv
\int_{-\infty}^{\infty} dx (a(x)c(x) + b(x)d(x)) $],
we can show that $\hat{L} (a,b)=0$ is equivalent to 
$\hat{L}^+(a,-b)=0$; i.e., homogeneous solutions of the 
 problem  (\ref{eq:f1chi1}) are, apart from a minus sign, also homogeneous 
solutions of the adjoint  problem. This holds, however, only for continuous
functions $a(x)$ and $b(x)$. The functions $f_i$ and $\chi_i$ are 
discontinuous at $x=0$ [see Eqs.\ (\ref{eq:jump0}) and (\ref{eq:jump1})],
and we have to  carefully check boundary contributions. 

Nevertheless, we will make use of the fact that 
 one solution of the homogeneous problem  
can be explicitly constructed, 
\begin{equation*}
  \hat{L}    \begin{pmatrix} f_0' \\ \chi_0' \end{pmatrix} =0
 ~~\mbox{for}~ x\neq 0, 
\end{equation*}
as can be checked by taking one derivative in the first-order equation 
(\ref{eq:f0chi0}). 
This suggests that a Fredholm solvability condition for the inhomogeneous 
problem (\ref{eq:f1chi1L}) 
can be derived by forming the scalar product 
$ \langle (f_0' , -\chi_0'), \hat{L} (f_1, \chi_1)\rangle$ on both sides.
On the left-hand side, this should give zero apart from boundary terms from 
$x=0$. We find 
\begin{align*}
   \left\langle   \begin{pmatrix} f_0' \\ -\chi_0' \end{pmatrix},
    \hat{L} \begin{pmatrix} f_1 \\ \chi_1 \end{pmatrix} \right\rangle &=  
    \left.   -f_0'f_1'   \right|_{0-}^{0+} -  
 \left.   \chi_0''\chi_1'   \right|_{0-}^{0+}
 \\
 & = -2 f_0'(0) - 2p_1.
\end{align*}
Forming the scalar product also with the right-hand side
(the inhomogeneity)  of (\ref{eq:f1chi1L}), 
using the symmetry of $f_0$ and $\chi_0$ and 
 the first integral (\ref{eq:firstI}) for $p_0=0$, 
 integrating by parts,
and using  Eq.\ (\ref{eq:f0chi0}) for $p_0=0$
 finally gives the following solvability condition 
for $p_1$:
\begin{align}
  p_1 &= \frac{1}{4} \int_0^\infty dx 
 \left( \chi_0'^2 - f_0'^2 + 2\chi_0  \right)
    = - \frac{3}{5} \int_0^\infty dx \chi_0
\nonumber\\
  &\simeq 0.83370854,
\label{eq:p1_2}
\end{align}
where the right-hand side has been evaluated numerically using the 
solutions shown in Fig.\ \ref{fig:f0chi0}. 
The last equality of integrals 
is obtained by using  Eq.\ (\ref{eq:f0chi0}) for $p_0=0$
after partial integration and 
the first integral (\ref{eq:firstI}) for $p_0=0$.
Together with $p_0=0$, we have 
\begin{equation}
\begin{split}
    p/p_c &=p_1 \bar{z}_B^{-1/2}
             +O(\bar{z}_B^{-1}),
\\ 
   \bar{z}_B &=  p_1^2 \, (p/p_c)^{-2}  + O((p/p_c)^{-1})
\end{split}
\label{eq:pzPog}
\end{equation}  
with  $p_1^2 \simeq   0.69506993$, which is Eq.\ (\ref{eq:zBPog}).
Further corrections $p_2$, $p_3$, etc., 
can be calculated by extending this scheme to higher orders. 
Based on the symmetry properties that 
$f_0(x)$ and $\chi_1(x)$ are odd and $\chi_0(x)$ and $f_1(x)$ are 
even, we can show in the next order that 
\begin{equation}
 p_2=0.
\label{eq:p2}
\end{equation}
This means that the leading non-vanishing 
corrections in Eq.\ (\ref{eq:pzPog}) 
are actually of higher order: the leading correction to $p/p_c$ 
is $O(\bar{z}_B^{-3/2})$ and the leading correction to $\bar{z}_B$
in eq.\  (\ref{eq:zBPog})
is $O((p/p_c)^{0})$. 
This is supported by our numerics.

The result (\ref{eq:pzPog}) is  also in agreement 
with the work of Evkin {\it et al.} \cite{Evkin1989,Evkin2016}.
It can be shown that the boundary layer approach of Ref.\ 
\cite{Evkin2016} is equivalent to our expansion (\ref{eq:expansion})
in powers of $\bar{z}_B^{-1/2}$  (with the identification
$f_0 = 2w_1'$ and $\chi_0 = - 2\phi_1'$ in the notation of Ref.\ 
\cite{Evkin2016}). 
Evkin  {\it et al.} use a variational approach rather than 
Fredholm integrability conditions to obtain the expansion 
coefficients $p_1$ and $p_2$.
Using the variational approach they obtain
$p/p_c  = (3J_0/8)  \bar{z}_B^{-1/2} + O(\bar{z}_B^{-3/2})$
(see also Table \ref{tab:dimensionless}) with 
a numerical constant $J_0$ for which 
we can show  the exact equality
\begin{equation} 
 J_0 = 2\int_0^\infty (\chi_0'^2+f_0'^2) = 8p_1/3,
\label{eq:J0}
\end{equation}
establishing the  equivalence with Eq.\ (\ref{eq:pzPog});
 the missing correction
in order   $O(\bar{z}_B^{-1})$ corresponds to  $p_2=0$. 
Our numerical evaluation of $p_1$ gives a slightly
different $J_0\simeq 2.22322$ as compared to  $J_0\simeq 2.23$ given 
 in  Ref.\  \cite{Evkin2016}.
We also note that the existence of only three spatial 
scaling regions for $\bar{\psi}$ and $\bar{w}$ as a function of 
$\rho$ in the barrier state cannot be justified systematically 
in the approach of Evkin  {\it et al.}

We now return to the 
 remarkable coincidence between our calculation for the barrier state
($\bar{F}=0$, $p>0$) and the complementary 
calculation by Gomez  {\it et al.}
\cite{Gomez2016} for the force-indentation curve in the absence of pressure
($p=0$, $\bar{F}>0$). 
Both governing differential equations (\ref{eq:f0chi0}) for 
the leading-order corrections $f_0$, $\chi_0$ 
  and  (\ref{eq:f1chi1L}) for the first-order corrections $f_1$, $\chi_1$
for the Pogorelov barrier state 
are {\it identical} to the corresponding 
equations of Gomez  {\it et al.} for the Pogorelov dimple with a 
point force in the absence of 
pressure.
We can show that both results are exactly consistent
if the elastic part of the indentation energy $\bar{E}_{\rm ind}(\bar{z})$ is 
{\it independent of the pressure} [apart from terms of order $(p/p_c)^2$]
in the Pogorelov regime $\bar{z}\gg 1$. This means that the 
Pogorelov dimple energy is actually independent of a 
precompression of the spherical shape by a pressure $p$, 
which is often tacitly assumed
(for example in Refs.\ \cite{Evkin2016,Baumgarten2018}).  
To show this consistency, we integrate
the result from Ref.\ \cite{Gomez2016} for the force-indentation 
relation in the Pogorelov limit for $p=0$,
$\bar{F} = F_2 \bar{z}^{1/2}+O(1)$ 
with $F_2/2\pi \simeq 1.6674$,
to obtain the $p=0$ indentation energy, 
\begin{align}
   \bar{E}_{\rm ind,p=0}(\bar{z})  &=   \frac{F_2}{3\pi} \bar{z}^{3/2}
     +O(\bar{z}^{1}).
\label{eq:Eind} 
\end{align}
If this is the elastic part of the indentation energy 
 independent of pressure $p$ (apart from pressure dependence in 
higher order terms),
the only effect of an applied pressure is 
to  add the mechanical pressure work 
to the total indentation energy,
 \begin{align*}
   \overline{p\Delta V} &= 4\frac{p}{p_c} \int d\rho \rho \bar{w}
      = \frac{2}{\pi} \frac{p}{p_c}  \overline{\Delta V}
 \approx - \frac{p}{p_c} \bar{z}^2,
\label{eq:pV} 
\end{align*}
which is the  leading-order result 
for a mirror-inverted dimple $\bar{w}(\rho) = -\bar{z}+ \rho^2$
with  $\overline{\Delta V} =  -\pi \bar{z}^2/2$
[corrections should be $O(\bar{z}^1)$ if an expansion analogous to 
(\ref{eq:barwA}) applies with an odd function $f(x)$].
 The barrier state with $\bar{F}=0$ then corresponds
to an energy extremum of 
$\bar{E}_{\rm ind}(\bar{z}) = 
  \bar{E}_{\rm ind,p=0}(\bar{z})  +\overline{p\Delta V}(\bar{z})$ 
with respect to variation 
of the indentation $\bar{z}$. 
This leads to our above result (\ref{eq:pzPog}), 
\begin{align*}
  p/p_c &= \frac{F_2}{4\pi} \bar{z}_B^{-1/2} +  O(\bar{z}_B^{-1}),
\end{align*}
if  $p_1 = F_2/4\pi$. This is indeed fulfilled 
because we can show the exact equality
\begin{equation}
  F_2/2\pi  = \frac{1}{2}\int_0^\infty (f_0'^2-\chi_0'^2-\chi_0) = 
     2p_1
\label{eq:F2}
\end{equation}
from Eq.\ (\ref{eq:p1_2})
(this equality is also exactly fulfilled on the level of the second order 
Eq.\  (\ref{eq:f1chi1L}) \cite{Note2}).
Our above finding $p_2=0$ suggests that the next non-vanishing 
term  is actually smaller than $O(\bar{z}_B^{-1})$.
This is in accordance with speculations in Ref.\ \cite{Gomez2016}
that the leading non-vanishing correction in Eq.\ (\ref{eq:Eind})
 is smaller than $O(\bar{z}^{1})$. 
Using Eq.\ (\ref{eq:J0}), we also obtain the relation
 $F_2 = 3\pi J_0/2$, which shows that  the 
 force-indentation  relation at $p=0$ from  Ref.\  \cite{Gomez2016},
$\bar{F} = F_2 \bar{z}^{1/2}+O(1)$, is exactly 
identical to the force-indentation relation that has been 
obtained before in 
Refs.\ \cite{Evkin2005,Evkin2016}.

\subsection{Energy barrier and force-indentation relation}

We can use the exact result (\ref{eq:Etot}), where we insert the 
ansatz  (\ref{eq:barwpsiA})  to find the leading-order
result for the energy barrier, which turns out to be of order 
$O(\bar{z}_B^{3/2}) = O((p/p_c)\bar{z}_B^{2})$, 
\begin{align*}
 \bar{E}_B &= -\frac{1}{4} \int_0^{\infty}d\rho  
  \bar{\psi} (\partial_\rho\bar{w})^2
    \nonumber\\
    &= -\frac{p}{p_c} \bar{z}_B^2 - \frac{1}{2} \bar{z}_B^{3/2}
      \int_0^\infty dx \chi_0\left( 2- 2f_0+ f_0^2\right)
  + O(\bar{z}_B^{1})
    \nonumber\\
    &= \frac{p_1}{3} \bar{z}_B^{3/2} + O(\bar{z}_B^{1})
\end{align*}
with  $p_1/3 \simeq  0.27793199$, which 
 leads to  Eqs.\ (\ref{eq:EBzPog}) and (\ref{eq:EBpPog}). 
To show the last equality we use $p_0=0$ in the expansion of $p$, 
 the first integral (\ref{eq:firstI}) 
for $p_0=0$, and the last equality in (\ref{eq:p1_2}). 
Our numerics show that the next non-vanishing 
terms in Eqs.\ (\ref{eq:EBzPog}) and (\ref{eq:EBpPog}) are
 actually $O(\bar{z}_B^{1/2})$ rather than 
 $O(\bar{z}_B)$ and $O((p/p_c)^{-1})$ rather
 than $O((p/p_c)^{-2})$, respectively.
It is not possible to establish this result analytically 
[analogously to $p_2=0$, see Eq.\ (\ref{eq:p2})]
from symmetry considerations only. 
Evkin {\it et al.}  find 
$\bar{E}_B \approx (3/16)^3J_0^4  (p/p_c)^{-3}$ \cite{Evkin2016}
(see also Table \ref{tab:dimensionless} in  Appendix \ref{app:dim}),
which agrees  with  our result 
(\ref{eq:EBpPog}) including the identical 
 numerical prefactor  $(3/16)^3J_0^4 =p_1^4/3$ [see Eq.\ (\ref{eq:J0})]. 
Again, our result for the energy barrier
 is also consistent with the result (\ref{eq:Eind}) 
of Gomez {\it et al.}
for the indentation energy if it is  independent of pressure $p$
 [apart from terms of order $(p/p_c)^2$]
because 
\begin{align*}
  \bar{E}_B &= \bar{E}_{\rm ind,p=0}(\bar{z}_B) +\overline{p\Delta V}(\bar{z}_B)
=  \frac{F_2}{12\pi}  \bar{z}_B^{3/2}
\end{align*}
 is exactly our above result 
for  $p_1 = F_2/4\pi$; see Eq.\ (\ref{eq:F2}).

We conclude that the  total indentation energy landscape
is given by 
$\bar{E}_{\rm ind}(\bar{z}) = 
  \bar{E}_{\rm ind,p=0}(\bar{z})  +\overline{p\Delta V}(\bar{z})$
in the Pogorelov limit $\bar{z}\gg 1$,
which confirms Eq.\   (\ref{eq:landscapePog}).
Differentiating  the  energy landscape
gives the force-indentation relation 
    $\bar{F}(\bar{z})/2\pi =  d\bar{E}_{\rm ind}/d\bar{z}$ 
in the presence of a pressure $p$, 
\begin{align}
   \frac{\bar{F}(\bar{z})}{2\pi} 
   &= 2p_1 \bar{z}^{1/2} -  2 \frac{p}{p_c} \bar{z}.
\label{eq:FzPog}
\end{align}
This generalizes the $p=0$ result of Ref.\ \cite{Gomez2016} and
 is valid for $\bar{z}\gg 1$ and $p\ll p_c$. 
It is also identical with the force-indentation relation in the presence 
of pressure which was 
conjectured in Ref.\ \cite{Evkin2016},
tacitly assuming that the 
Pogorelov dimple energy is independent of the  pressure $p$. 
We can now also obtain 
the maximal force needed to overcome the buckling barrier,
\begin{align*}
\frac{\bar{F}_{\rm max}}{2\pi} &= \frac{1}{2} p_1^2\,  (p/p_c)^{-2} = 
    \frac{3}{2} \frac{\bar{E}_B}{\bar{z}_B},
\end{align*}
which is the characteristic maximal  point 
force for structural stability below $p_c$.

\section{Shallow shell theory for the shallow barrier 
state close to the buckling pressure}
\label{sec:theory2}

In this section, we derive several analytical results
 for the buckling energy landscape  for $p$ close to $p_c$.
The total indentation energy landscape is
\begin{align}
 \bar{E}_{\rm ind} &=
\frac{\sqrt{2}}{\pi} (1-p/p_c)^{1/2}  \, \bar{z}^{2} 
      -\frac{\sqrt{3}}{4\pi} \bar{z}^3
+ O(\bar{z}^4)
\label{eq:landscapelin}
\end{align}
 for  
shallow indentations with $\bar{z} \ll 1$.
By maximizing with respect to $\bar{z}$ at $\bar{F}=0$,  we obtain 
several analytical results for the 
energy barrier state for $p$ close to $p_c$
 corresponding to 
shallow barrier states with $\bar{z}_B \ll 1$:
\begin{subequations}
\begin{align}
  \bar{z}_B &= \frac{8\sqrt{2}}{3\sqrt{3}}\, (1-p/p_c)^{1/2} 
       +  O(1-p/p_c),
\label{eq:zBlin}\\  
 \overline{\Delta V}_B  &\approx \frac{4}{3\sqrt{3}}\bar{z}_B  \approx 
      \frac{32\sqrt{2}}{27}\, (1-p/p_c)^{1/2},
\label{eq:Vindlin}\\ 
\rho_B &\approx \left(\frac{8}{3\sqrt{3}\pi}\right)^{1/2}  \simeq 0.70,
\label{eq:rhoBlin}\\
\bar{E}_B &= \frac{\sqrt{3}}{8\pi} \bar{z}_B^3  + O(\bar{z}_B^4),
\label{eq:EBzlin}\\
\bar{E}_B &= \frac{128\sqrt{2}}{81\pi}\, (1-p/p_c)^{3/2} 
 + O((1-p/p_c)^2).
\label{eq:EBplin}
\end{align}
\label{eq:barrierlin}
\end{subequations}
Thus,  we can derive all  critical properties 
of the buckling transition, i.e.,  all relevant scaling 
exponents for barrier indentation and barrier energy 
 close to the bifurcation in accordance with the 
numerical results 
[see Eqs.\ (\ref{eq:zp}), (\ref{eq:Vind}), (\ref{eq:Vindp}), (\ref{eq:rhoB}),
 (\ref{eq:EBz}), and (\ref{eq:EBp})]
from nonlinear shallow shell theory.
Energy barrier height 
 and barrier indentation  vanish as $\bar{E}_B\propto (1-p/p_c)^{3/2}$
and  $\bar{z}_B\propto (1-p/p_c)^{1/2}$, respectively, which gives 
rise to softening of the shell close to $p_c$.
We  also obtain exact numerical prefactors, which accurately 
agree with the asymptotic numerical results as
Figs.\ \ref{fig:zp}, \ref{fig:Vind}, and \ref{fig:EB} show.

For $p$ close to $p_c$ 
 the barrier state is a very shallow
dimple with $\bar{z}_B \ll 1$, and we can expand 
about the linear solution (\ref{eq:barw}) and (\ref{eq:barpsi}),
\begin{equation}
\begin{split}
    \bar{w}(\rho) &= \bar{z}\bar{w}_{\rm lin,0} + \bar{z}^2\bar{w}_1 +...,
\\
\bar{\psi}(\rho) &= \bar{z}\bar{\psi}_{\rm lin,0}
+ \bar{z}^2\bar{\psi}_1 +...,
\\
   \frac{\bar{F}}{2\pi} &= \bar{z} F_0 + \bar{z}^2 F_1 + ...,
\end{split}
\label{eq:expansionlin}
\end{equation}
where we define $\bar{w}_{\rm lin,0}\equiv \bar{w}_{\rm lin}/\bar{z}$
and $\bar{\psi}_{\rm lin,0}\equiv \bar{\psi}_{\rm lin}/\bar{z}$ as
normalized linear displacement and stress function.
As $\bar{w}_{\rm lin}$ and $\bar{\psi}_{\rm lin}$ from Eqs.\ (\ref{eq:barw}) 
and (\ref{eq:barpsi}) fulfill the correct boundary conditions,
$\bar{w}_1$, $\bar{w}_1'$,  $\bar{\psi}_1$, and $\bar{\psi}_1'$  
must vanish at $\rho=0$ and $\rho\to \infty$.
We note that we perform an expansion for the full problem with 
$p\neq 0$ close to $p_c$ and also $\bar{F} \neq 0$; i.e., 
we do not only aim at the $\bar{F}=0$ barrier state as 
for the Pogorelov limit in the previous section.

\subsection{Leading order}

We insert the  expansion (\ref{eq:expansionlin})
into the Eq.\ (\ref{eq:exact1}) for $\bar{w}$ and 
the compatibility condition (\ref{eq:comp3}) for $\bar{\psi}$.
To leading  linear order $\bar{z}$ 
the expansion (\ref{eq:expansionlin})  gives the linearized solutions
by construction, which motivates the form of the linear 
term in the expansion (\ref{eq:expansionlin}).
They fulfill an inhomogeneous linear differential equation, 
which is equivalent to the Reissner
equations (\ref{eq:Reissner}),
\begin{align}
  & \hat{M} 
\begin{pmatrix} \bar{w}_{\rm lin,0}' \\ \bar{\psi}_{\rm lin,0} \end{pmatrix}
+\begin{pmatrix} 2(1-p/p_c)\bar{w}_{\rm lin,0}' \\ 0 \end{pmatrix}
  =  \begin{pmatrix}   F_0
 \\
  0 
\end{pmatrix} 
~~\mbox{with}
\nonumber\\
 &\hat{M} \begin{pmatrix} a \\ b \end{pmatrix} 
   \equiv   \begin{pmatrix}  
   \frac{1}{\rho} a  -a' - \rho a'' -\rho b -2\rho a 
 \\
    -\frac{1}{\rho} b  +b' + \rho b'' -\rho a
\end{pmatrix}.
\label{eq:wlin0psilin0}
\end{align}
We wrote these equations using a linear 
 operator $\hat{M}$ which is self-adjoint with respect to the 
scalar product $\langle (a,  b), (c,d) \rangle \equiv
\int_{0}^{\infty} d\rho  (a(\rho)c(\rho) + b(\rho)d(\rho))$.

We will 
 perform the expansion  (\ref{eq:expansionlin}) 
by employing  the linearized solutions 
in the limit  $p\to p_c$, where  
\begin{align}
   \bar{w}_{0} &\equiv  \lim_{p\to p_c} \bar{w}_{\rm lin,0}
      =   -J_0(\rho),
\label{eq:w0}
\\
   \bar{\psi}_0  &\equiv  \lim_{p\to p_c} \bar{\psi}_{\rm lin,0}
     = -J_1(\rho) = -\bar{w}_0'
\nonumber
\end{align}
[see Eqs.\ (\ref{eq:barw1}) and  (\ref{eq:barpsi1})]. 
The functions $\bar{w}_0'$ and $\bar{\psi}_0$ 
 provide  a solution of the 
linearized problem  at $p=p_c$ where also $\bar{F}=0$
[because the stiffness vanishes for $p=p_c$ as shown 
in Sec.\ \ref{sec:lin},  see (\ref{eq:Fwlin})]:
\begin{align}
   \hat{M} \begin{pmatrix} \bar{w}_0' \\ \bar{\psi}_0 \end{pmatrix} 
    &= 0.
\label{eq:homlin}
\end{align}
We can use this homogeneous solution  to  obtain 
 a  Fredholm solvability condition for the inhomogeneous 
problem (\ref{eq:wlin0psilin0})  by scalar multiplication with 
$(\bar{w}_0' , \bar{\psi}_0)$ on both sides resulting in 
\begin{align*}
   2(1-p/p_c) \int_0^\infty d\rho \rho \bar{w}_0' \bar{w}_{\rm lin,0}' &=
        F_0 \int_0^\infty d\rho \bar{w}_0'.
 \end{align*}
Using 
\begin{align} 
   \int_0^\infty d\rho \bar{w}_0'  &= 1,
  \label{eq:intw01}
\\
  \int_0^\infty d\rho \rho \bar{w}_0' \bar{w}_{\rm lin,0}' &\approx 
    \frac{\sqrt{2}}{\pi} (1-p/p_c)^{-1/2}
\label{eq:intwo2}
\end{align}
[where the last integral  is  performed in the limit $p\to p_c$,
see also Eq.\ (\ref{eq:intwlin})],
we rediscover our above result (\ref{eq:k}) for the linear 
stiffness of the shell,
\begin{equation*}
   \bar{k} = \lim_{\bar{z}\to 0} \frac{\bar{F}}{\bar{z}}=
       2\pi F_0 = 4\sqrt{2}  (1-p/p_c)^{1/2}.
\end{equation*}

\subsection{First  order}

In the next order $\bar{z}^2$, we obtain 
the inhomogeneous equation 
\begin{align*} 
\hat{M}   \begin{pmatrix} \bar{w}_{1}' \\ \bar{\psi}_{1} \end{pmatrix}
+ \begin{pmatrix} 2(1-p/p_c)\bar{w}_{1}' \\ 0 \end{pmatrix}
  &=  \begin{pmatrix}  
 - \bar{\psi}_{\rm lin,0}\bar{w}_{\rm lin,0}'+ F_1
 \\
    -\frac{1}{2} \bar{w}_{\rm lin,0}'^2
\end{pmatrix}.
\end{align*}
Because $F_1$ will not vanish in the limit $p\to p_c$, 
we can perform this limit explicitly 
and obtain 
\begin{align} 
 \hat{M}  
    \begin{pmatrix} \bar{w}_{1}' \\ \bar{\psi}_{1} \end{pmatrix}
  &=  \begin{pmatrix}  
 - \bar{\psi}_{0}\bar{w}_{0}'+ F_1
  \\
    -\frac{1}{2} \bar{w}_{0}'^2
\end{pmatrix}.
\label{eq:w1psi12}
\end{align}
Again, we use the homogeneous solution (\ref{eq:homlin}) to 
obtain  a  Fredholm solvability condition
 by scalar multiplication  with 
$(\bar{w}_0' , \bar{\psi}_0)$ on both sides.
This gives [using again the integral  (\ref{eq:intw01})]
\begin{align}
 0 &= -\frac{3}{2}\int_0^\infty d\rho \bar{\psi}_{0}\bar{w}_{0}'^2
    +F_1
 =   \frac{3}{2}  \int_0^\infty d\rho J_1^3(\rho)  +F_1.
\label{eq:intw0b}
 \end{align}
Evaluating the last integral,   we finally obtain 
\begin{equation*}
F_1 = -\frac{3\sqrt{3}}{4\pi}.
\end{equation*}

\subsection{Energy barrier and force-indentation relation}

From our results for $F_0$ and $F_1$, we find the 
 force-indentation relation 
\begin{align}
  \frac{\bar{F}(\bar{z})}{2\pi} 
    &=  \frac{2\sqrt{2}}{\pi} (1-p/p_c)^{1/2} \bar{z} 
    - \frac{3\sqrt{3}}{4\pi}  \bar{z}^2  + O(\bar{z}^3)
\label{eq:Fz}
\end{align}
in the presence of pressure $p$ 
for shallow dimples $\bar{z}\ll 1$ and for $p$ close to $p_c$.
The force-indentation relation is related by 
$\bar{F}(\bar{z})/2\pi =  d\bar{E}_{\rm ind}/d\bar{z}$  
to the indentation energy landscape.
As opposed to the Pogorelov limit $\bar{z}\gg 1$ and $p\ll p_c$ [see
Eq.\ (\ref{eq:landscapePog})], the pressure 
does not only enter via the mechanical work term 
$\overline{p\Delta V}(\bar{z})$ for shallow dimples. 
[$\overline{p\Delta V} =  (2/\pi) (p/p_c)  \overline{\Delta V} \propto 
 -(p/p_c) \bar{z}$  would result in a constant contribution in 
the force-indentation relation (\ref{eq:Fz})].
For shallow dimples  close to $p_c$, the softening of the shell 
profoundly modifies the indentation energy already in leading order. 
Structural stability is governed by 
the maximal force needed to overcome the barrier,
\begin{align}
\frac{\bar{F}_{\rm max}}{2\pi}  &= \frac{8}{3\sqrt{3}\pi} (1-p/p_c) =
    \frac{3}{2} \frac{\bar{E}_B}{\bar{z}_B},
\label{eq:Fmaxlin}
\end{align}
which becomes small close to $p_c$, reflecting the softening of the 
shell. 

For $\bar{F}=0$, we obtain the relation between pressure and indentation 
in the post-buckling barrier state (\ref{eq:zBlin}), 
which can also be written as 
\begin{equation}
  p/p_c = 1 -  \frac{27}{128} \bar{z}_B^2.
\label{eq:zBlin2}
\end{equation}
This is the  same  asymptotic form as found by Evkin {\it et al.} 
\cite{Evkin2016}. Based on the incorrect assumption of 
zero  curvature at the pole in the barrier state [as 
the normal displacement profiles in Fig.\ \ref{fig:barrier_conf}(a) 
clearly show]
they find
$p/p_c \simeq 1-0.048\bar{z}_B^2$ with 
 a numerical prefactor that differs significantly 
from our result $27/128\simeq 0.211$. This leads to significant 
deviations of their Pad{\'e} interpolation of  $p/p_c$ 
from the numerical data [see Fig.\ \ref{fig:zp}(a)], 
whereas Eq.\ (\ref{eq:zBlin2}) is in excellent agreement 
with the numerics over several decades of the 
small parameter $1-p/p_c$ (see Fig.\ \ref{fig:zp}). 
An early result of Thompson 
for axisymmetric post-buckling shapes (in the absence of a point force) 
 does not agree and features a
 linear  term $O(\bar{z}_B)$ \cite{Thompson1964,Hutchinson1967}.

Integrating $\bar{E}_{\rm ind} = \frac{1}{2\pi}
\int_0^{\bar{z}} \bar{F}(\tilde{\bar{z}}) d\tilde{\bar{z}}$,
we find the total indentation  energy landscape 
(\ref{eq:landscapelin}) and all results for the energy 
barrier state. 
The energy barrier can also be calculated directly from
 the exact result (\ref{eq:Etot}), where we obtain in leading order
\begin{align*}
 \bar{E}_B &= -\frac{1}{4} \int_0^{\infty}d\rho  \bar{\psi} \bar{w}'^2
\nonumber\\
    &\approx  -\bar{z}^3 
      \frac{1}{4} \int_0^{\infty}d\rho  \bar{\psi}_0 \bar{w}_0'^2
     = \bar{z}_B^3 \frac{1}{4}  \int_0^\infty d\rho J_1^3(\rho)
     = \frac{\sqrt{3}}{8\pi}\bar{z}_B^3, 
\end{align*}
in agreement with  (\ref{eq:EBzlin}). 
The indentation volume (\ref{eq:Vindlin}) at the barrier state 
is  found by 
using the barrier indentation (\ref{eq:zBlin}) 
in the exact relation (\ref{eq:exact2}) or 
 relation (\ref{eq:intwlin}).

\section{Energy landscape and barrier }

\subsection{Energy landscape}

\begin{figure}
\begin{center}
\includegraphics[width=0.4\textwidth]{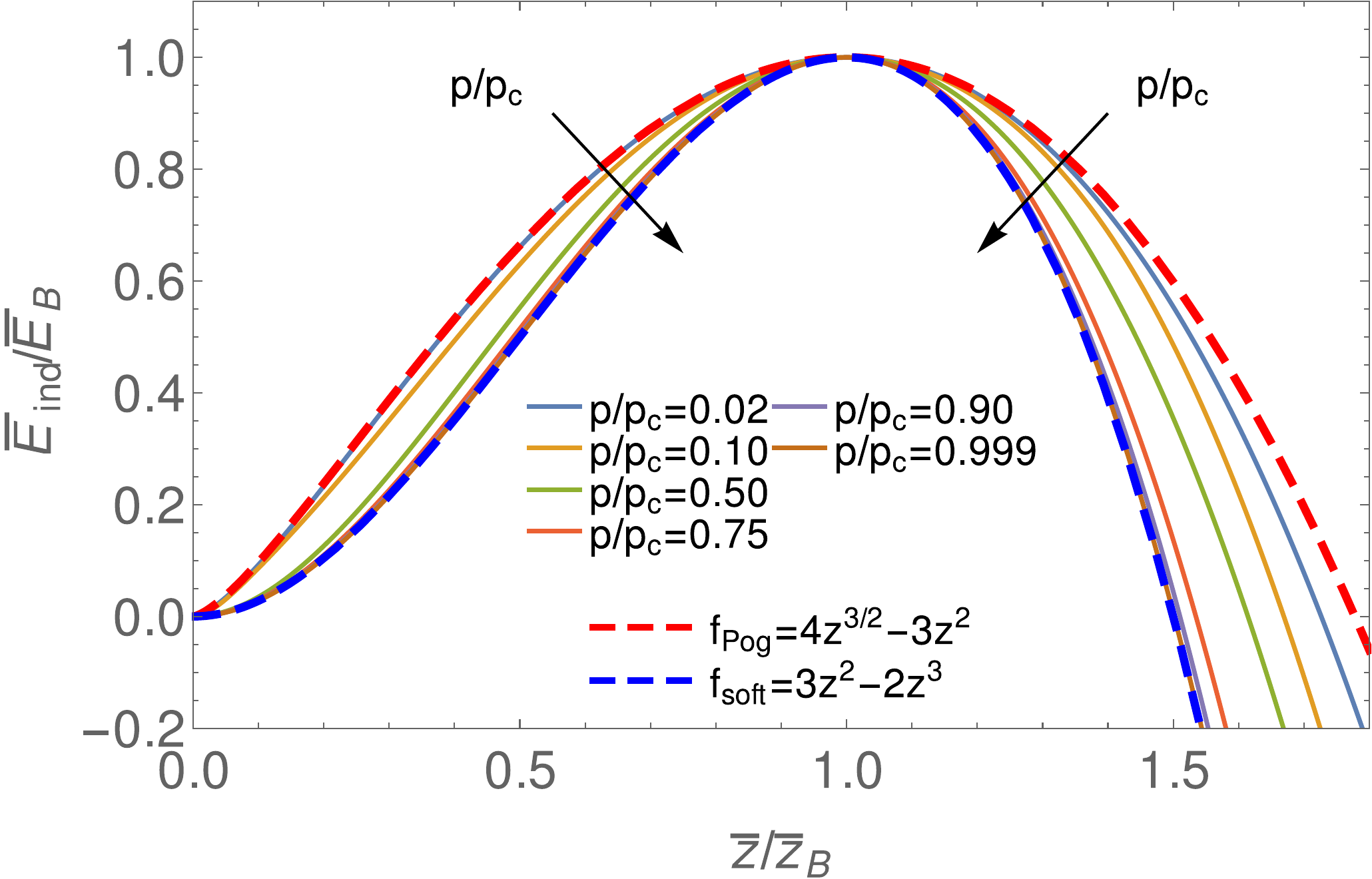}
\caption{
  Solid lines: Numerical results for the 
   rescaled energy landscape  $\bar{E}_{\rm ind}/\bar{E}_B$
   as a function of rescaled indentation $z\equiv \bar{z}/\bar{z}_B$  for 
  various pressures $p/p_c$ according to Eq.\ (\ref{eq:landscape_scaling}).
  Arrows indicate increasing pressure.
  All rescaled landscapes lie between the two limiting 
  scaling curves $f_{\rm Pog}(z)$ and $f_{\rm soft}(z)$, 
   which are known analytically 
  [dashed lines; see Eq.\ (\ref{eq:landscapescaling})] and approached 
for $p/ p_c\ll 1$ or $p/p_c$ close to unity, respectively. 
}
\label{fig:landscape_scaling}
\end{center}
\end{figure}

Both in the Pogorelov limit and the softening limit close to $p_c$,
we obtained the exact asymptotics of the full 
energy  landscape, (\ref{eq:landscapePog}) and 
(\ref{eq:landscapelin}),
which also contains all information about the energy barrier 
[by maximizing $\bar{E}_{\rm ind}(\bar{z})$ with respect to $\bar{z}$] 
and the force-indentation 
curves $\bar{F}(\bar{z})/2\pi = d\bar{E}_{\rm ind}/d\bar{z}$.
Both limits 
 can be written in a scaling form 
\begin{align}
 \bar{E}_{\rm ind} &= 
 \bar{E}_B f\left(\frac{\bar{z}}{\bar{z}_B} \right),
\label{eq:landscape_scaling}
\\
 \frac{\bar{F}(\bar{z})}{2\pi} &= 
 \frac{\bar{E}_B}{\bar{z}_B} f'\left(\frac{\bar{z}}{\bar{z}_B} \right)
\label{eq:Fz_scaling}
\end{align}
  with  two characteristic scaling functions
\begin{equation}
\begin{split}
   f_{\rm Pog}(z) &=  4 z^{3/2} - 3 z^2, 
 \\
 f_{\rm soft}(z) &= 3z^2 - 2 z^3 
\end{split}
\label{eq:landscapescaling}
\end{equation}
determining the shape of the energy landscape, which is 
in good agreement  with our numerical results 
in Fig.\ \ref{fig:landscape_scaling}.
We note that the Pogorelov result applies for $\bar{z}\gg 1$ and $p\ll p_c$, 
whereas the softening regime applies to $\bar{z}\ll 1$ and $p$ close to $p_c$.
In particular, it is not possible to calculate the linear shell 
stiffness $k=\left. d^2\bar{E}_{\rm ind}/d\bar{z}^2\right|_{\bar{z}=0}$
from the Pogorelov result for $p\ll p_c$. For this, one has to resort
to the linear response result (\ref{eq:k}) as has been discussed
in Ref.\ \cite{Vella2012b}.

\subsection{Volume and area change during indentation}

\begin{figure}
\begin{center}
\includegraphics[width=0.4\textwidth]{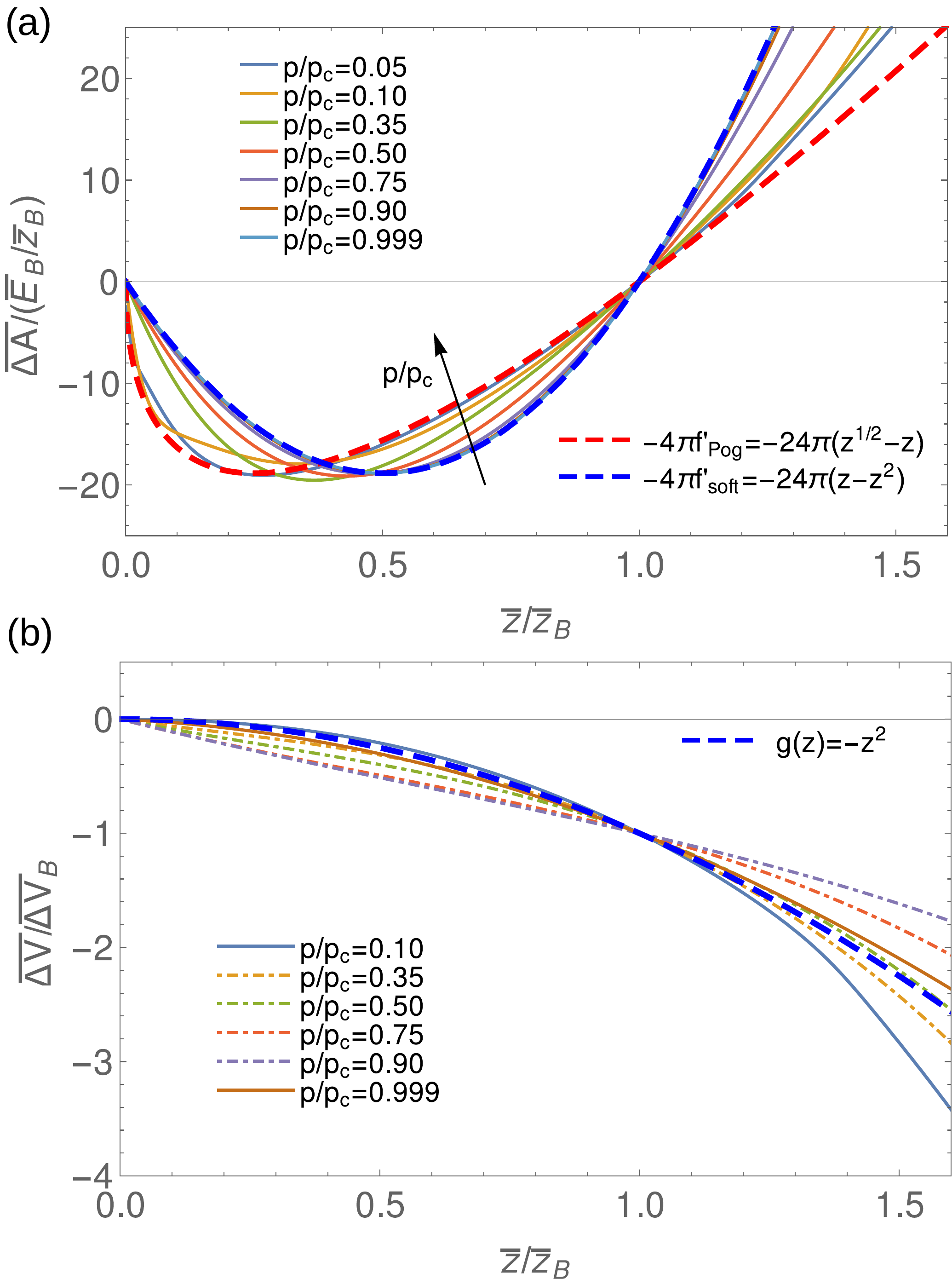}
\caption{
  Numerical results for (a) the
  rescaled area $\overline{\Delta A}/(\bar{E}_B/\bar{z}_B)$  and (b) rescaled
  volume  $\overline{\Delta V}/\overline{\Delta V}_B$
  change as a function of $z\equiv \bar{z}/\bar{z}_B$
   for 
   various pressures $p/p_c$ according to Eqs.\ (\ref{eq:Fz_scaling})
   and (\ref{eq:volume_scaling}). The dashed lines indicated the limiting 
   scaling functions   according to Eqs.\ (\ref{eq:landscapescaling})
   and (\ref{eq:volume_scaling}).
}
\label{fig:areavol_scaling}
\end{center}
\end{figure}

During point force indentation, area and volume change according
to Eqs.\ (\ref{eq:DeltaV}) and (\ref{eq:DeltaA}). 
According to relation (\ref{eq:FVA}), the indentation area change 
is given by  the force-indentation relation 
via $\overline{\Delta A}(\bar{z})   = -2\bar{F}(\bar{z})$.
The above scaling relation (\ref{eq:Fz_scaling}) for
the force-indentation relation is 
in good agreement  with our numerical results  for the area change,
as Fig.\ \ref{fig:areavol_scaling}(a) shows.
The area change is a non-monotonous function of $\bar{z}$
with a minimum at the maximal point force $-\bar{F}_{\rm max} =
\frac{1}{2}\overline{\Delta A}_{\rm min}$.

In the linear reponse regime $\bar{z}\ll 1$, also 
the volume change $\overline{\Delta V}(\bar{z})$ is
given by the point force via
$\overline{\Delta V}(\bar{z}) =
2\pi \int_0^{\infty} d\rho \rho \bar{w}_{\rm lin} = -\bar{k}(p)\bar{z}
  = -\bar{F}$;
  see Eqs.\ (\ref{eq:Fwlin}) and (\ref{eq:k}).
  For deeper indentations, we can use the results
  (\ref{eq:intwlin})  close to $p_c$ and the
  mirror-inverted dimple result $\overline{\Delta V} =  -\pi \bar{z}^2/2$
  to find
\begin{align}
  -\overline{\Delta V}(\bar{z}) &\approx \begin{cases}
    \overline{k}(p)\bar{z}
              +\frac{\bar{z}^2}{2\sqrt{2}}  \left(1- p/p_c  \right)^{-1/2}
              &\mbox{for}~ p\approx p_c
  \\
         \overline{k}(p)\bar{z}
                                 +\pi \bar{z}^2/2~~
                               &\mbox{for}~ p \ll p_c
                             \end{cases}.
   \label{eq:Vz}
\end{align}
Rescaling with the barrier indentation $\bar{z}_B$ [see Eqs.\ (\ref{eq:zBlin})
and (\ref{eq:zBPog})] and the indentation volume
$\overline{\Delta V}_B$ at the barrier  [see Eqs.\ (\ref{eq:Vindlin})
and (\ref{eq:VindPog})], we find that both for $p\approx p_c$ and
$p\ll p_c$ a scaling relation
\begin{align}
  -\overline{\Delta V}(\bar{z}) &= 
 \overline{\Delta V}_B\, g\left(\frac{\bar{z}}{\bar{z}_B} \right)
                                  \nonumber\\
  \mbox{with}~~g(z) &= \pi \bar{z}^2/2
                      \label{eq:volume_scaling}\\
\end{align}
holds over a wide range  $\bar{z}/\bar{z}_B \gg (1-p/p_c)^{1/2}$  and 
$\bar{z}/\bar{z}_B \gg  (p/p_c)^{2} $, respectively.
For intermediate pressures,
however,
an additional  linear regime emerges; see Fig.\ \ref{fig:areavol_scaling}(b).
Volume and area change during point force indentation
combine to a reduced volume
\begin{equation}
  v= V/(4\pi/3)(A/4\pi)^{3/2} = \frac{1+\gamma^{-1}\frac{3}{4\pi}
    \overline{\Delta V}}
  {\left(1+\gamma^{-1}\frac{1}{2\pi}\overline{\Delta A}\right)^{3/2}}
  <1,
\end{equation}
which is monotonously decreasing and reduced to  $v = 1- O(\gamma^{-1})$
up to the barrier. It becomes
significantly reduced and finally vanishes only close to the 
snap through where $z\propto R_0$ or $\overline{\Delta V}\propto \bar{z}^2
\propto \gamma$.

\subsection{Quantitative interpolation for energy barrier and indentation}

The scaling forms (\ref{eq:landscape_scaling}) and (\ref{eq:Fz_scaling})
give quantitatively  accurate energy barrier shapes and force-indentation 
relations if accurate interpolation results for the energy barrier height
$\bar{E}_B$ and the barrier indentation $\bar{z}_B$ are available. 
As non-dimensionalization of the 
shallow shell Eqs.\ (\ref{eq:forcebal3}) and (\ref{eq:comp3})  showed,
the dimensionless energy barrier $\bar{E}_B$ at $\bar{F}=0$ can only
depend on  $p/p_c$,
\begin{align}
   \bar{E}_B &= f_p(p/p_c).
\label{eq:interpol1}
\end{align}
Based on our analytic asymptotic results 
(\ref{eq:EBpPog}) and (\ref{eq:EBplin}) for the function $f_p(x)$,
 we can give a new interpolation 
formula, significantly improving the 
interpolation formula for the function $f_p(x)$ 
proposed in Ref.\ \cite{Baumgarten2018}
(which was based on scaling results in the Pogorelov limit and 
numerical results only):
\begin{align}
    f_p(x) &= 
   \left(a_3 x^{-3} + a_2x^{-2}+ a_1x^{-1}+a_0+(1-\sum_{n=0}^3a_n)x\right)
\nonumber\\
    &\times\left[b_{3/2}(1-x)^{3/2}+b_2(1-x)^2\right.
      \nonumber\\
  &~\left. +(1-b_{3/2}-b_2)(1-x)^{5/2}\right],
\nonumber\\
   &a_3 = \frac{p_1^4}{3}\simeq 0.161,~a_2=-0.0168,
\nonumber\\
   &a_1=1.653,~a_0=1.951,
\nonumber\\
   &b_{3/2} = \frac{128\sqrt{2}}{81\pi}\simeq 0.711,~b_2=3.794.
\label{eq:interpol}
\end{align}
Our analytical asymptotic results constrain the values $a_3$ and $b_{3/2}$.
The remaining fit parameters are determined by a Levenberg-Marquardt fit 
on numerical shallow shell data for the energy barrier that are equally 
distributed between $p=0$ and $p=1$ in steps of $\Delta p=0.01$. 
The fact that the fit gives 
$|a_2|\ll 1$ is consistent with our numerical finding 
that the next non-vanishing 
term in Eq.\  (\ref{eq:EBpPog}) is
 actually  $O((p/p_c)^{-1})$ rather
 than $O((p/p_c)^{-2})$.
Relative 
deviations from the numerical shallow shell data are smaller than 5\% for 
$p/p_c<0.6$ and smaller then 20\% over the whole range of 
 $p$; see Fig.\ \ref{fig:compareEB}.

 A similar interpolation can be given for the barrier indentation:
\begin{align}
   &\bar{z}_B = g_p(p/p_c)   ~~\mbox{with}
\nonumber\\
    g_p(x) &= 
   \left(a_2 x^{-2} +a_0+ a_{-1}x + (1-\sum_{n=-1}^2a_n)x^2\right)
\nonumber\\
   &\times\left[b_{1/2}(1-x)^{1/2}+b_1(1-x)\right.
     \nonumber\\
   &~\left.+(1-b_{1/2}-b_1)(1-x)^{3/2}\right],
\nonumber\\
   &a_2 ={p_1^2}\simeq 0.695,~a_0=4.779,~a_{-1}=-3.144,
\nonumber\\
   &b_{1/2} = \frac{8\sqrt{2}}{3\sqrt{3}}\simeq 2.177,~b_1=-1.377,
\label{eq:zBinterpol}
\end{align}
see Fig.\ \ref{fig:zp}, where 
our analytical results constrain the values $a_2$,  $b_{1/2}$, and 
 $a_1=0$ [the 
leading non-vanishing correction in the Pogorelov limit 
is  $O((p/p_c)^{0})$
because of $p_2=0$; see Eqs.\ (\ref{eq:p2}) and (\ref{eq:zBPog})]. 
Relative deviations from the numerical results are smaller than 2\% for 
$p/p_c<0.6$ and smaller than 15\% over the whole range of 
 $p$.
 
We can also employ the strategy of Evkin {\it et al.} \cite{Evkin2016}
and use a Pad{\'e} interpolation of  $p/p_c$ as a function of
$\bar{z}_B$. 
With an ansatz 
\begin{align}
   \frac{p}{p_c} = \frac{\sum_{n=0}^4 \alpha_n \varepsilon^n}
    {1+ \sum_{n=1}^4 \beta_n \varepsilon^n}    ~~~\mbox{with}~\varepsilon\equiv
    2\bar{z}_B^{-1/2},
\label{eq:Pade}
\end{align}
we can incorporate the analytical 
constraints $\alpha_0=0$, $\alpha_1=p_1/2\simeq 0.417$  
from the Pogorelov limit 
$\varepsilon \ll 1$ with (\ref{eq:pzPog})  and 
constraints $\alpha_n=\beta_n$ for $n=1,2,3,4$ and $\alpha_4=8/27\simeq 0.296$ 
from the shallow dimple regime $\varepsilon \gg 1$ with (\ref{eq:zBlin2}). 
Furthermore, the constraint $p_2=0$ from the  Pogorelov limit 
$\varepsilon \ll 1$ [see Eq.\ (\ref{eq:p2})]
 gives $\alpha_2=p_1^2/4\simeq 0.174$. 
The only unconstrained coefficient $\alpha_3$ can be used for 
a   Levenberg-Marquardt fit of
the numerical shallow shell data, which gives $\alpha_3\simeq 0.476$. 
Our Pad{\'e} interpolation of  $p/p_c$
differs significantly from the one given in Ref.\ \cite{Evkin2016}
because of the corrected behavior in the  limit $p\approx p_c$
[see Fig.\ \ref{fig:zp}(a)];
relative 
deviations from the numerical results for $p/p_c$ are smaller than 5\% for 
 over the whole range of $z_B$.

\section{Maxwell and  unbuckling pressure}

We can  use the energy  
landscape (\ref{eq:landscapePog}) in the Pogorelov limit $\bar{z}\gg 1$
to  calculate the critical unbuckling pressure 
$p_\text{cu}$ and the Maxwell pressure $p_{c1}$.
When we  reduce the  compressive pressure to $p<p_\text{cu}<p_c$, the buckled
state can no longer be stabilized, and both the buckled state and the 
unstable barrier  transition state merge and vanish in a saddle-node
bifurcation at $p=p_\text{cu}$. The unbuckling pressure 
 is the smallest compressive 
pressure at which a metastable buckled state still exists and is 
thus also called  minimum buckling load
\cite{Karman1939,Koiter1969,Evkin2001,Evkin2017}.
The stable buckled states assumed after buckling at 
 $p_c$ are snap-through buckled states ($z\approx 2R_0$)
and can no longer be described in shallow shell approximation,
which assumes small slopes
 $|w'|\sim z/R_0 \ll 1$, i.e., small rotation 
angles of shell elements. 
By numerical solution of the full shape equations \cite{Knoche2011},
which are valid  beyond the shallow
shell approximation, we find that 
at unbuckling at the  pressure $p_\text{cu}$ the shell is 
 no longer fully 
snapped-through but the indentation is still  deep and proportional 
to $R_0$, i.e., $z= \alpha R_0$ with $\alpha \sim 1.4-1.5$ 
(in agreement with Refs.\ \cite{Evkin2001,Evkin2017}).

If we assume  that  also the unbuckling state
has  maximal  indentation $z=2R_0$ or $\bar{z} = 2\gamma^{1/2}$ 
with $\bar{z}\gg 1$ for thin shells $R_0\gg h$ or large 
F\"oppl-von K\'arm\'an numbers  $\gamma \gg 1$, 
this state  
can only be a  boundary minimum of the energy landscape
(\ref{eq:landscapePog}) with $\bar{F}=0$
if the maximum of the energy landscape at $z_B$ has smaller 
indentation $z_B<2R_0$ or $\bar{z}_B < 2\gamma^{1/2}$. 
This is the case above the critical 
unbuckling pressure 
\begin{equation}
   p_\text{cu} =  \frac{p_1}{\sqrt{2}} p_c \gamma^{-1/4}.
\label{eq:pcuss}
\end{equation}
For $p<p_\text{cu}$, the precompressed spherical state is the 
{\it only} accessible energy minimum and, thus, spherical shells have to 
unbuckle. 
The parameter dependence $p_\text{cu} \propto p_c \gamma^{-1/4}$ 
has been previously observed \cite{Evkin2001,Knoche,Paulose2012,Evkin2017}.
In Refs.\ \cite{Evkin2001,Evkin2017},
 $p_\text{cu} \simeq 2.65\, p_c \gamma^{-1/4}$ has been 
found using a shell theory that allows for large deflections and rotation
angles. Solving 
 the full shape equations from Ref.\ \cite{Knoche2011} numerically 
we find 
$p_\text{cu} \simeq 1.2\,p_c \gamma^{-1/4}$, which is by a factor of 2 larger
than the shallow shell result (\ref{eq:pcuss})  but also shows the 
same parameter dependence. 
We conclude that in 
shallow shell theory  the unbuckling state is not properly accessible 
but can be approximately 
regarded as a boundary minimum at $z=2R_0$ corresponding to 
a snap-through state.
The   simple condition $z_B=2R_0$ 
for the barrier state can predict the unbuckling pressure $p_\text{cu}$ 
up to a numerical factor of about 2.

We can  calculate the Maxwell pressure $p_{c1}$, at which the 
snap-through buckled state and the unbuckled precompressed 
spherical state  have equal energies,  from calculating 
the zero $\bar{E}_{\rm ind}(\bar{z}_1)=0$ of the  energy landscape
(\ref{eq:landscapePog}) (see also Fig.\ \ref{fig:landscape})
\begin{equation}
  \bar{z}_1 = \frac{16 p_1^2}{9}   (p/p_c)^{-2} = 
     \frac{16}{9} \bar{z}_B.
\label{eq:z1}
\end{equation}
The buckled state with equal energy  is the snap-through state 
if  $z_1=2R_0$ or $\bar{z}_1 = 2\gamma^{1/2}$.
This condition determines 
 the  Maxwell  pressure 
\begin{equation}
   p_{c1} =  \frac{4p_1}{3\sqrt{2}} p_c \gamma^{-1/4} = \frac{4}{3}p_\text{cu},
\label{eq:pcu}
\end{equation}
 confirming the parameter dependence 
$p_{c1} \sim  p_{c} \gamma^{-1/4}$ \cite{Knoche2014o}. 
Using the Pogorelov theory $p_{c1} \simeq 0.901
 (1-\nu^2)^{-1/4} p_c \gamma^{-1/4}$ 
has been obtained \cite{Knoche}, which slightly deviates 
from our above result $p_{c1} \simeq 0.786\, p_c\gamma^{-1/4}$
from shallow shell theory. 
The relation $p_{c1} = 4p_\text{cu}/3$ 
is obtained identically using Pogorelov theory  \cite{Knoche}.

\section{Soft spots}

The critical unbuckling pressure can be interpreted as a finite-size effect 
that leads to spontaneous unbuckling if the critical 
indentation at the barrier does no longer ``fits'' 
into the capsule in normal direction, i.e., if $z_B>2R_0$ or 
$\bar{z}_B > 2\gamma^{1/2}$.
Because $\rho_B \propto \bar{z}_B^{1/2}\ll \bar{z}_B$ for mirror-inverted 
dimples the lateral extent $\rho_B$ of the critical 
barrier state does not conflict 
with the finite size $R_0$ of the shell.
This can happen, however, for  spherical caps 
under pressure \cite{Huang1964} or  soft spots 
on a sphere under pressure \cite{Paulose2012} if 
their lateral size $L$ (or opening angle $\alpha = L/R_0$ for a spherical cap) 
are small.
Then the finite lateral size $L$ can trigger unbuckling of 
the cap or the soft spot. 
The important parameter governing the buckling of a finite 
spherical cap is 
$\lambda \equiv L/l_{\rm el} = \bar{L}$ 
\cite{Huang1964,Paulose2012}.
Although the boundary conditions play an important role and 
differ from those of a complete spherical shell 
both  for clamped and free  caps or soft spots,
 we expect that unbuckling is triggered if 
$\lambda=\bar{L}<\rho_B(p)$
 because a  fully buckled state of extent $\bar{L}$ becomes an
unstable  boundary energy maximum then.
According to Eqs.\ (\ref{eq:rhoBPog}) and (\ref{eq:rhoBlin}),
it will unbuckle if  $p/p_c < p_1/\bar{L}$ for  $\bar{L} \gg 1$ and 
for {\it all} $\bar{L} <0.70$ for $p$ close to $p_c$. 
Therefore, soft spots sufficiently small compared to the 
the elastic length ($L<0.70 l_{\rm el}$)
will immediately  unbuckle for $p$ smaller than  $p_c$. 
This will suppress the existence of the subcritical barrier state and 
hysteresis in the buckling of  sufficiently  small 
soft spots: The soft spot buckles {\it and} unbuckles at the 
same threshold pressure $p_c$.

\section{Bifurcation behavior as a function of pressure}

\begin{figure}
\begin{center}
\includegraphics[width=0.4\textwidth]{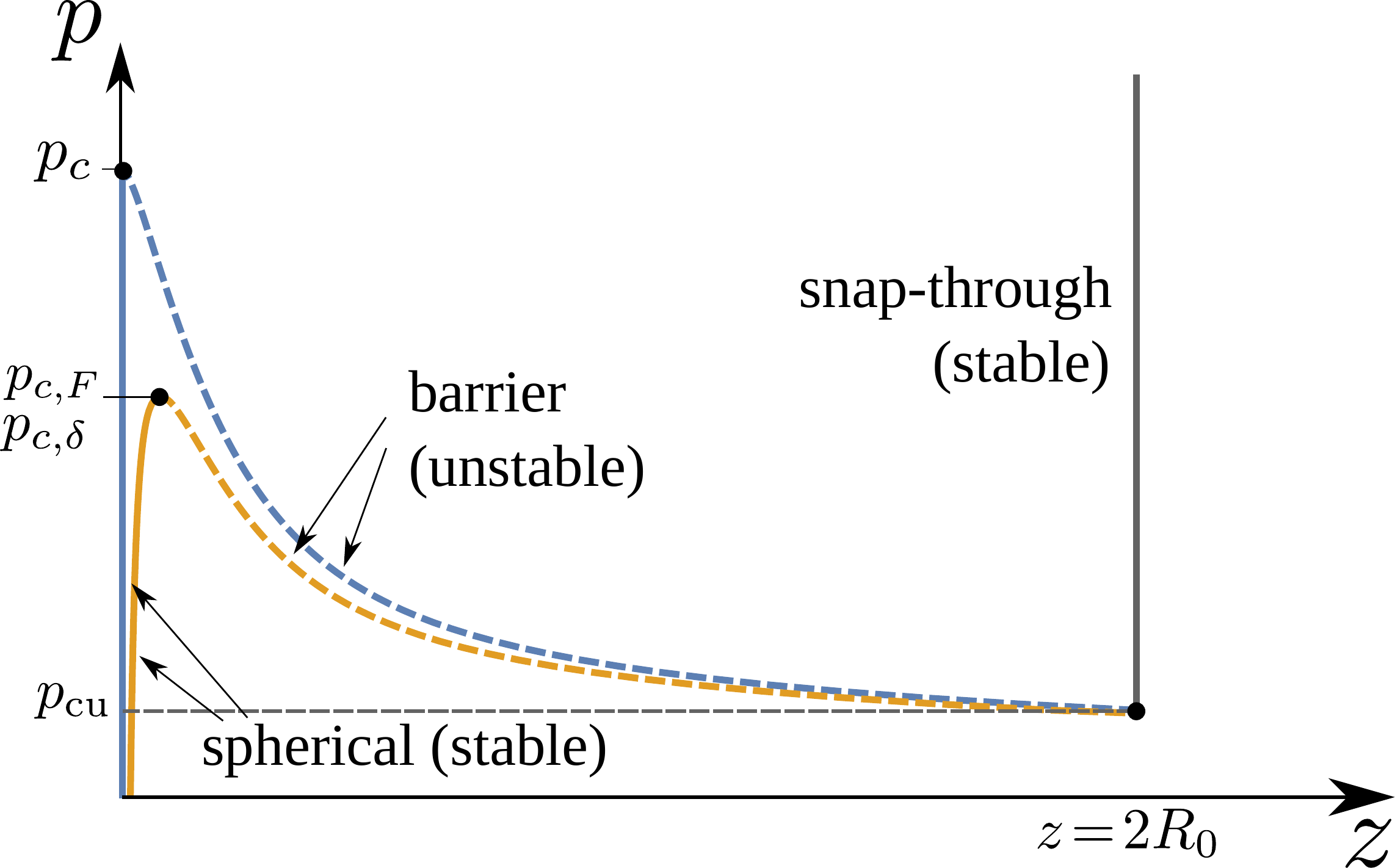}
\caption{
Bifurcation behavior of the indentation $z$ 
with  
pressure $p$  as control parameter for a perfect shell (blue) or in the
presence of a preindenting force or for an imperfect shell (yellow)
in shallow shell theory. 
There are three types of stationary states: a stable spherical state 
for $p<p_c$ (or $p<p_{c,F}$, $p<p_{c,\delta}$) (solid lines), an unstable 
barrier state corresponding to an energy maximum (dashed lines),
and a stable 
snap-through state above the unbuckling pressure $p>p_\text{cu}$ (solid gray
line).
In shallow shell theory, the unbuckling and  snap-through states are
 not properly accessible 
but can be regarded as boundary minimum at $z=2R_0$. 
}
\label{fig:bifurcation}
\end{center}
\end{figure}

Our results on the buckling barrier of a perfect spherical shell 
allow us to classify the buckling bifurcation as a function 
of the control parameter $p$ in more detail, as schematically 
shown in Fig.\ \ref{fig:bifurcation}. 
A suitable order parameter to trace  the bifurcation is the indentation $z$. 
The functional 
form of the energy landscape (\ref{eq:landscapelin}) close to $p_c$
(containing $z^2$- and $z^3$-terms)
suggests a transcritical 
bifurcation at $p=p_c$, but the unstable barrier state does not continue 
as a stable equilibrium state into the buckled phase $p>p_c$, where 
Eq.\ (\ref{eq:landscapelin}) is no longer applicable. 
Clearly, the  bifurcation at $p=p_c$ is 
subcritical as the barrier states represent
a subcritical branch of unstable stationary points, which are 
already present for $p<p_c$ (and $p>p_\text{cu}$). 
From this fixed point structure with an unstable barrier state 
with $\bar{z}_B \propto  (1-p/p_c)^{1/2}$ [Eq.\ (\ref{eq:zBlin})] and 
the stable spherical state joining at $p=p_c$ and resulting in 
an unstable spherical state for $p>p_c$, the bifurcation at $p=p_c$ 
is similar to a subcritical pitchfork, in which the spherical state $z=0$ 
becomes unstable.
We have, however, only  a ``one-sided'' pitchfork  because  we only 
consider compressed states $z<0$ and there is no inversion symmetry 
$z \to -z$ between compression and deflation of a sphere.
The buckled snap-through state $z=2R_0$ is 
stabilized as  a boundary minimum  for $p>p_\text{cu}$ and becomes 
the only remaining  minimum after the bifurcation at $p=p_c$.
It becomes the global energy minimum at the Maxwell pressure $p_{c1}$
between $p_\text{cu}$ and $p_c$. 
The appearance of the buckled  state together with  the energy barrier 
state at $p=p_\text{cu}$ is a saddle-node or fold bifurcation
as results in Refs.\ \cite{Evkin2001,Knoche2011,Knoche2014o} suggest. 
For pressures slightly above $p_\text{cu}$ the buckled state snaps through.
Bifurcations at $p_c$ and  $p_\text{cu}$ result
 in hysteresis   for  $p_\text{cu} <p<p_c$
between saddle-node and subcritical bifurcation.
Upon approaching the buckling instability 
from $p<p_c$, the shell indentation at the energy barrier maximum
 vanishes $\propto (1-p/p_c)^{1/2}$ and the linear restoring 
force vanishes as the linear stiffness $\propto (1-p/p_c)^{1/2}$;
see also Eq.\ (\ref{eq:Fz}). 
Right at the instability, the linear restoring force vanishes and 
the force-indentation curve 
$\bar{F} \propto -\bar{z}^2$ misses linear terms 
such that we can also expect  a 
``critical slowing down''  of  the buckling dynamics \cite{Gomez2017}.

\section{Bifurcation behaviour and softening 
in the presence of a preindenting point force}

One view of the point force $F$  is to consider $F$ as a 
probe of the buckling barrier for a fixed subcritical pressure $p<p_c$,
which gives access to the bifurcation behavior as a function of $p$.
An alternative view is to 
consider the point force as an additional control parameter 
and consider its  effect on the buckling bifurcation, i.e., 
to consider how the buckling bifurcation as a function of $p$ 
is modified if the sphere is preloaded by a small point force $F_{\rm pre}$. 
 Then, $F_{\rm pre}$ 
acts analogously to  an additional  ordering field in a phase
transition bifurcation favoring one phase (here the buckled state),
 gives rise to an {\it avoided} or {\it perturbed bifurcation}
at a reduced critical pressure $p_{c,F}<p_c$, and turns the bifurcation 
at $p_{c,F}$  into a saddle-node bifurcation (see Fig.\ \ref{fig:bifurcation}). 
The bifurcation at $p_\text{cu}$ remains essentially
unchanged as long as $\bar{F}_{\rm pre}$ is small. 
Interestingly, imperfections will have a very similar effect 
as we will  show below.

In the presence of a preloading 
 point force $F_{\rm pre}>0$, the critical buckling pressure  is reduced 
to $p_{c,F}<p_c$ as it is easier to buckle a preloaded shell. 
The additional force $F_{\rm pre}$ tilts the energy landscape to 
$\bar{E}_{\rm ind} - \bar{F}_{\rm pre}\bar{z}/2\pi$, 
resulting in  the modified
force-indentation relation $\bar{F}(\bar{z}) = 
   2\pi d\bar{E}_{\rm ind}/d\bar{z}- \bar{F}_{\rm pre}$.
Equivalently, we can say that $\bar{F}$ is replaced by 
the total force $\bar{F}+\bar{F}_{\rm pre}$ in the 
original force-indentation relation.
This 
turns the bifurcation at $p_{c,F}$ into a saddle-node 
bifurcation, at which both the stationarity 
condition $\bar{F}_{\rm pre} =  2\pi d\bar{E}_{\rm ind}/d\bar{z}$
for the tilted energy landscape (i.e., $\bar{F}(\bar{z})=0$ for 
the modified force-indentation relation) 
and the 
saddle condition  $0=d^2\bar{E}_{\rm ind}/d\bar{z}^2 
   = d\bar{F}(\bar{z})/d\bar{z}$
have to be fulfilled. 

For small forces $\bar{F}_{\rm pre}$, the bifurcation still occurs 
close to $p_c$ and for 
small indentations $\bar{z}$, such that we 
 can use the asymptotic energy landscape
(\ref{eq:landscapelin}) and the 
 force-indentation relation  (\ref{eq:Fz}). 
Stationarity and saddle conditions
then  result in a reduced critical pressure 
with a knockdown factor
\begin{equation}
   \frac{p_{c,F}}{p_c} = 1-   \frac{3\sqrt{3}}{16} \bar{F}_{\rm pre}.
\label{eq:knockdown1}
\end{equation}
For  $p<p_{c,F}$,   two stationary states emerge in the saddle-node 
bifurcation (see Fig.\ \ref{fig:bifurcation}):
 a stable preindented spherical state 
and the unstable barrier state. The stable preindented state is no longer 
a perfect precompressed sphere with $\bar{z}=0$ but has 
a finite indentation $\bar{z}_{\rm sph}\approx  F_{\rm pre}/k$, 
which is very well described by the linear stiffness $k$ from Eq.\ 
(\ref{eq:k}). 
Solving the stationarity condition  
$\bar{F}_{\rm pre} = 2\pi d\bar{E}_{\rm ind}/d\bar{z}$, i.e., solving the 
force-indentation relation  (\ref{eq:Fz}) for $\bar{z}$, 
we obtain the indentation for both branches,
\begin{align*}
\bar{z}_{\rm B,sph} &= \bar{z}_{+,-} = 
    \frac{4\sqrt{2}}{3\sqrt{3}}\left[ \left(1-\frac{p}{p_c}\right)^{1/2} \pm 
              \left(\frac{p_{c,F}-p}{p_c}\right)^{1/2}\right],
\end{align*}
which meet for $p=p_{c,F}$ at 
$z_{\rm B}=z_{\rm sph}\sim \sqrt{\bar{F}_{\rm pre}}$. 
Interestingly, we find the same softening behavior as in the 
absence of the preloading force if the pressure approaches 
the critical value $p_{c,F}$.
The  indentation difference from spherical to barrier state is 
\begin{align}
   \bar{z}_{B,F} &=  \bar{z}_{B} - \bar{z}_{\rm sph} 
  =  \frac{8\sqrt{2}}{3\sqrt{3}}
     \left(\frac{p_{c,F}-p}{p_c}\right)^{1/2}
\label{eq:zBlinF}
\end{align}
 and the corresponding  energy barrier  is
 \begin{align} 
   \bar{E}_{B,F} &= \bar{E}_B-\bar{E}_{\rm sph} = 
   \frac{1}{2\pi}\int_{z_{\rm sph}}^{\bar{z}_B} \bar{F}(z) dz 
\nonumber\\
   &= \frac{128\sqrt{2}}{81\pi}  \left(\frac{p_{c,F}-p}{p_c}\right)^{3/2}
   =\frac{\sqrt{3}}{8\pi}  \bar{z}_{B,F}^3.
\label{eq:EBplinF}
\end{align}
Both results are completely analogous to Eqs.\ (\ref{eq:zBlin}), 
(\ref{eq:EBzlin}), and 
(\ref{eq:EBplin}) for $\bar{F}_{\rm pre}=0$, with 
$\Delta \bar{p} \equiv ({p_{c,F}-p})/{p_c}$ replacing  $(1-p/p_c)$. 
They result in a linear stiffness
\begin{equation*}
  k_F =
  \left.\frac{d^2\bar{E}_{B,F}}{d\bar{z}_{B,F}^2}\right|_{\bar{z}_{B,F}=0}
 =  4\sqrt{2}  \left(\frac{p_{c,F}-p}{p_c}\right)^{1/2},
\end{equation*}
giving rise to the same softening behavior close to $p_{c,F}$ 
as in Eq.\ (\ref{eq:k}) in the absence of the preloading force.
The  properties of the subcritical barrier such as the scaling
$\bar{z}_{B,F} \propto \Delta \bar{p} ^{1/2}$ and 
$\bar{E}_{B,F} \propto \Delta \bar{p} ^{3/2}$, which
characterize the softening of the shell close to the critical pressure
$p_{c,F}$   are {\it universal}
and independent of the applied point force $\bar{F}_{\rm pre}$.
For a saddle-node bifurcation, where two branches 
of fixed points ($\bar{z}_{B}$ and  $\bar{z}_{\rm sph}$) smoothly  merge 
at the critical value $p=p_{c,F}$ of the control parameter, 
the behaviors $\bar{z}_{B,F} \propto \Delta \bar{p} ^{1/2}$ and 
$\bar{E}_{B,F} \propto \Delta \bar{p} ^{3/2}$ are actually generic.

\begin{figure*}
\begin{center}
\includegraphics[width=0.99\textwidth]{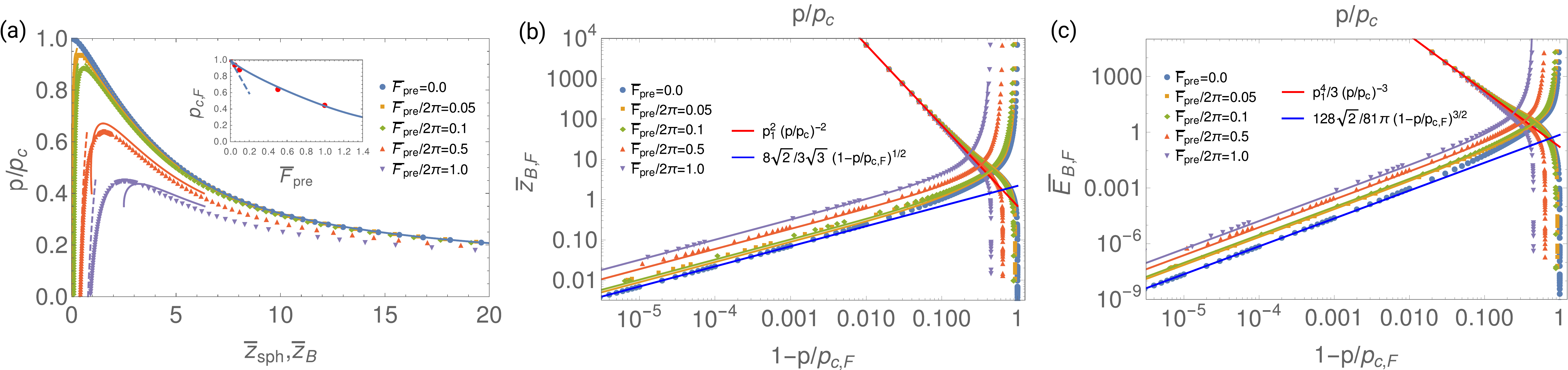}
\caption{
 (a) Bifurcation of the indentation as a function of the pressure
  for a shell preindented with force $\bar{F}_{\rm pre}$.
  Below a critical pressure $p_{c,F}$, two branches of 
  shapes emerge in a saddle-node bifurcation, an indented  spherical 
  branch with $\bar{z}_{\rm sph}$ and an unstable barrier state 
  with $\bar{z}_B$. Dashed lines: Linear 
   response  $\bar{z}_{\rm sph}\approx  F_{\rm pre}/k$ of spherical shell. 
  Solid lines: Approximation (\ref{eq:zBsphF}). 
   Inset: Knockdown factor $p_{c,F}/p_c$ as a function of preindenting force 
   $\bar{F}_{\rm pre}$ as compared  to Eq.\ (\ref{eq:knockdown2}) (solid line)
  and small $\bar{F}_{\rm pre}$ approximation 
    (\ref{eq:knockdown1}) (dashed line). 
 (b) Indentation  difference 
   $\bar{z}_{B,F} =  \bar{z}_{B} - \bar{z}_{\rm sph}$ 
   as a function of pressure  $p/p_c$ (upper curves and upper horizontal scale)
 together with analytical result 
      (\ref{eq:zBPog}) for $F_{\rm pre}=0$ (red line)
 and as a function of  $1-p/p_{c,F}$  (lower curves and lower horizontal scale)
together with analytical result   (\ref{eq:zBlin}) 
   for $F_{\rm pre}=0$ (blue line). 
   Lines are approximation (\ref{eq:zBpF}). 
(c)
   Energy barrier $\bar{E}_{B,F}$ 
  as a function of pressure  $p/p_c$ (upper curves and upper horizontal scale)
 together with the analytical result (\ref{eq:EBpPog}) 
   for  $F_{\rm pre}=0$ (red  line)
 and as a function of  $1-p/p_{c,F}$  (lower curves and lower horizontal scale)
together with the analytical result (\ref{eq:EBplin}) for $F_{\rm pre}=0$
(blue line). Lines  are approximation (\ref{eq:EBpF}).
}
\label{fig:Force}
\end{center}
\end{figure*}

 Figure \ref{fig:Force} shows numerical results 
for the knockdown factor, the bifurcation of the indentation,
 the indentation difference, and the energy barrier between 
barrier state and preindented spherical state. 
The numerical method is unchanged, in principle. 
In the presence of a preindenting force,
the total force $\bar{F}+\bar{F}_{\rm pre}$  is acting  on the 
shell, and 
$\bar{F}$ is replaced by 
the total force $\bar{F}+\bar{F}_{\rm pre}$ in the 
shallow shell equations that are solved numerically.
 Both at the barrier state and at the preindented 
spherical state we have an applied force $\bar{F}=0$, but the total force 
acting on the shell is not vanishing but equals 
the preindenting force $\bar{F}_{\rm pre}$. 
In order to calculate the barrier energy 
$\bar{E}_{B,F} = \bar{E}_B-\bar{E}_{\rm sph}$ directly without numerically 
integrating
the full force-indentation relation,
we now employ a generalized version of relation 
(\ref{eq:Etot}), which gives direct numerical access to the energy difference 
$\Delta \bar{E}_{\rm tot,F}$ between 
a state indented with a force  $\bar{F}_{\rm pre}$ and 
the precompressed unindented spherical state (with the same 
pressure but with $\bar{z}=0$) as
\begin{align}
   \Delta \bar{E}_{\rm tot,F} &= -\frac{1}{4} \int_0^{\infty}d\rho  \bar{\psi}
   (\partial_\rho \bar{w})^2 -\frac{\bar{F}_{\rm pre}}{4\pi} \bar{z} 
\nonumber\\
  &~~~ - 2(1+\nu) \frac{p}{p_c}\frac{\bar{F}_{\rm pre}}{2\pi}.
\label{eq:EtotF}
\end{align} 
A derivation is given 
 in Appendix \ref{app:barrierenergy}.
The energy 
barrier $\bar{E}_{B,F} =  \Delta \bar{E}_{\rm tot,B}-\bar{E}_{\rm tot,sph}$
can then be  obtained as difference of the values of $\Delta \bar{E}_{\rm tot,F}$
between  the barrier state  and the preindented spherical state, 
both of which can  be obtained via continuation of   
solutions of the shallow shell
equations (\ref{eq:forcebal3}) 
and (\ref{eq:comp3}) with force $\bar{F}_{\rm pre}$.

The  numerical results in Fig.\ \ref{fig:Force} show
that the above results 
(\ref{eq:zBlinF}) and (\ref{eq:EBplinF}) are quantitatively 
correct only for very small $\bar{F}_{\rm pre} \ll 1$, where the knockdown 
factor (\ref{eq:knockdown1}) is close to unity and we can use the 
asymptotic result (\ref{eq:Fz}) for the  force-indentation relation
[see inset in  Fig.\ \ref{fig:Force}(a)].
For larger $\bar{F}_{\rm pre}$, 
it turns out that prefactors in Eqs.\ (\ref{eq:zBlinF}) and (\ref{eq:EBplinF})
depend weakly on the preindentation force $\bar{F}_{\rm pre}$, 
and the knockdown factor deviates from Eq.\  (\ref{eq:knockdown1}).

In order to explain these results 
quantitatively, we use analytical estimates
based on the scaling form (\ref{eq:landscape_scaling})
of the energy landscape for $\bar{F}=0$ employed in conjunction with the 
interpolation formulas (\ref{eq:interpol}) and (\ref{eq:zBinterpol})
for the pressure dependence of $\bar{E}_{B,0}$ and $\bar{z}_{B,0}$
for $F_{\rm pre}=0$. 
The knockdown factor  $p_{c,F}/p_c$ is then determined by the solution of 
\begin{align}
   \frac{3}{2} \frac{\bar{E}_{B,0}(p_{c,F}/p_c)}{\bar{z}_{B,0}(p_{c,F}/p_c)} &= 
\frac{\bar{F}_{\rm  pre} }{2\pi}
\label{eq:knockdown2}
\end{align}
and 
\begin{align}
\bar{z}_{\rm B,sph} &= 
    \frac{1}{2}\bar{z}_{B,0} \left( 1 \pm \left(1 - 
     \frac{\bar{F}_{\rm  pre}}{2\pi}\frac{2}{3} 
     \frac{\bar{z}_{B,0}(p)}{\bar{E}_{B,0}(p)} \right)^{1/2} \right),
\label{eq:zBsphF}\\
 \bar{z}_{B,F} &=  \bar{z}_{B} - \bar{z}_{\rm sph} 
  = \bar{z}_{B,0}\left(1 - 
     \frac{\bar{F}_{\rm  pre}}{2\pi}\frac{2}{3} 
     \frac{\bar{z}_{B,0}(p/p_c)}{\bar{E}_{B,0}(p/p_c)} \right)^{1/2}, 
\label{eq:zBpF} \\
 \bar{E}_{B,F} &= \bar{E}_{B,0}  \left(1 - 
     \frac{\bar{F}_{\rm  pre}}{2\pi}\frac{2}{3} 
     \frac{\bar{z}_{B,0}(p/p_c)}{\bar{E}_{B,0}(p/p_c)} \right)^{3/2},
\label{eq:EBpF} 
\end{align}
where we used the scaling function $f_{\rm soft}(z)$ appropriate 
for $p/p_c \gtrsim 0.75$. 
This describes our  numerical data for the 
modified bifurcation behavior in Fig.\ \ref{fig:Force}
well for $\bar{F}_{\rm pre} \lesssim 1$. 
We recover again the universal 
 properties of the subcritical barrier 
$\bar{z}_{B,F} \propto \Delta \bar{p} ^{1/2}$ and 
$\bar{E}_{B,F} \propto \Delta \bar{p} ^{3/2}$,
but prefactors in these scaling laws now depend 
on the applied point force $\bar{F}_{\rm pre}$.

Effects of the preindenting force $F_{\rm pre}$ are 
negligible  for the barrier state 
in the Pogorelov limit 
 $\bar{z}_B\gg 1$ and $p\ll p_c$. Then 
the barrier state is deeply indented, and the preindenting force 
$\bar{F}_{\rm pre}$  can be neglected versus 
the elastic and pressure terms in the force-indentation relation 
 (\ref{eq:FzPog}) for $\bar{z}_B \gg \bar{F}_{\rm pre}^2$
or $p/p_c \ll 1/\bar{F}_{\rm pre}$.
The linear shell stiffness  $\bar{k} = \frac{d\bar{F}}{d\bar{z}}$
[see Eq.\ (\ref{eq:k})], 
on the other hand, is independent of an additional constant 
force $\bar{F}_{\rm pre}$ in the force-indentation relation. 
Thus,
the preindented spherical state approaches 
$\bar{z}_{\rm sph}\approx  \bar{F}_{\rm pre}/k \simeq 0.8\,\bar{F}_{\rm pre}$
for $p/p_c\approx 0$, which is also negligible 
versus $\bar{z}_B \sim (p/p_c)^{-2}$ for $p/p_c \ll 1/F_{\rm pre}^{1/2}$. 
Therefore, the 
 indentation difference from spherical to barrier state  
$\bar{z}_{B,F}\approx \bar{z}_B$ and the barrier energy 
$\bar{E}_{B,F}\approx \bar{E}_B$  
approach the Pogorelov asymptotics (\ref{eq:zBPog}) and 
 (\ref{eq:EBpPog}) in the Pogorelov limit for 
 $p/p_c \ll  1/F_{\rm pre}^{1/2}$; see also Fig.\ \ref{fig:Force}. 
Therefore, the bifurcation at $p_\text{cu}/p_c \sim \gamma^{-1/4}$ 
[see Eq.\ (\ref{eq:pcuss})] remains essentially
unchanged as long as $\bar{F}_{\rm pre}\ll \gamma^{2}$, which is a
rather weak condition.

\section{Imperfections}

For applications, 
another important class of ``quenched'' defects are imperfections in form of a
normal axisymmetric displacement field $w_I(r)$, which
is  already present in the strain-free state of the shell
\cite{Hutchinson1967,Hutchinson2016,Lee2016,Hutchinson2018}.
Then the strain is defined  relative to the configuration of a sphere 
with radius $R(r)=R_0+w_I(r)$
containing already  normal displacements $w_I(r)$.
Similar to the preindenting force, also imperfections  are known to affect 
the nature of the bifurcation at $p_c$  and cause a pronounced 
reduction of the critical buckling pressure $p_c$ 
\cite{Hutchinson1967, Koiter1969,Lee2016}.
We will show that the effect of localized axisymmetric imperfections is 
very similar  to the effect of a preindenting force.

\subsection{Shallow shell theory in the presence of imperfections}

We consider here axisymmetric imperfections and demonstrate
that they  can be 
 incorporated  in an exact manner  into  our 
analytical barrier calculation based on the shallow shell equations 
in the regime of  small $\bar{z}_B\ll 1$  from 
Sec.\ \ref{sec:theory2}. A detailed derivation of the 
nonlinear shallow shell equations in the presence of imperfections 
is given in Appendix \ref{app:shshell}.

In-plane  strains $u_{ij}$ are defined relative to the imperfect initial shape
and depend on the imperfection field $w_I$ via 
 nonlinear terms in the normal displacement $w$. 
Changes in the curvature tensor (curvature strains) $k_{ij}$, on the other
hand, are independent of $w_I$. 
The Hookean stress-strain relations giving in-plane stresses $\sigma_{ij}$ 
and bending moments in terms of the in-plane strains $u_{ij}$ 
and curvature strains $k_{ij}$, respectively, are not modified 
by imperfections, as well as  the Hookean elastic 
energy of the shell in terms of strains 
 [see Eq.\ (\ref{eq:Eel2}) in 
 Appendix \ref{app:shshell}]. 
Variation with respect to the additional normal displacements $w(r)$ 
(and in-plane displacements) finally gives the modified shallow shell 
equations 
\begin{align}
  & \kappa \nabla^4 w
   + \frac{1}{R_0} \frac{1}{r} \partial_r (r\psi)
      - \frac{1}{r} \partial_r \left( \psi \partial_r(w+w_I) \right) 
\nonumber\\
 & ~~~~~~~~~= -p - \frac{F}{2\pi} \frac{\delta(r)}{r},
  \label{eq:forcebalI} \\
 &\frac{1}{Y} r  \partial_r \left[ 
     \frac{1}{r} \partial_r (r\psi) \right] 
  = \frac{r}{R_0} \partial_r w  - 
   \frac{1}{2} \left( \partial_r w\right)^2 - 
   (\partial_r w)(\partial_r w_I),
  \label{eq:compI}
\end{align}
which generalize Eqs.\ (\ref{eq:forcebal}) and (\ref{eq:comp}) 
in the presence of imperfections. 

As in Eq.\ (\ref{eq:forcebal2}), we can absorb the effect of the 
pressure $p$ in Eq.\ (\ref{eq:forcebalI}) into
 a uniform precompression 
 $\psi(r) = \psi_0(r)  = -pR_0r /2$ (corresponding to stresses 
 $\sigma_{rr} = \sigma_{\phi\phi} = \sigma_0 = -pR_0/2$) and  
 consider changes with respect to this prestress  by substituting
$\psi(r) \to \psi_0(r) + \psi(r)$  [see Eq.\ (\ref{eq:forcebal2_I_app} 
in  Appendix \ref{app:shshell}].
Also 
nondimensionalization  (\ref{eq:non-dim})
proceeds as before. Integrating on both 
sides from $\rho$ to infinity finally gives 
\begin{align}
  & -\rho \partial_\rho (\nabla_\rho^2 \bar{w}) - \rho \bar{\psi} 
    + \bar{\psi} \partial_\rho (\bar{w}+\bar{w}_I) 
  -2\frac{p}{p_c} \rho \partial_\rho\bar{w}
\nonumber\\
   &~~~~~~~~~~~~~  = \frac{\bar{F}}{2\pi} 
  +2\frac{p}{p_c} \rho \partial_\rho\bar{w}_I,
\label{eq:exact1_I}
\end{align}
which has to be solved together with the 
compatibility condition 
\begin{align}
& \rho  \partial_\rho \left[ 
     \frac{1}{\rho} \partial_\rho (\rho\bar{\psi}) \right] 
  = \rho \partial_\rho \bar{w}  - 
   \frac{1}{2} \left( \partial_\rho \bar{w}\right)^2 
  -  \left( \partial_\rho \bar{w}\right)\left( \partial_\rho \bar{w}_I\right).
  \label{eq:comp3_I}
\end{align}
In comparison with our original Eqs.\  (\ref{eq:exact1}) for $\bar{w}$ and 
the compatibility condition (\ref{eq:comp3}) for $\bar{\psi}$,
there are three additional terms coupling to the imperfection displacement. 
The first term in Eq.\ (\ref{eq:exact1_I}) and the last term
in Eq.\ (\ref{eq:comp3_I}) are couplings caused by  the non-linearities of the 
shallow shell equations and are of order $O(\bar{z}\bar{z}_I)$ if
$\bar{z}_I$ is the amplitude of the imperfection displacement.
They correct the homogeneous part of the shallow shell equations. 
The  term $-2(p/p_c) \rho\bar{w}_I' = O(\bar{z}_I)$ 
in Eq.\  (\ref{eq:exact1_I}), on the other hand,  is
 of lower order and corrects the inhomogeneity of the shallow 
shell equations. It can be interpreted as an 
additional effective pointlike force, which is  localized to the extent
of the imperfection $\bar{w}_I$ and
 is caused by the response of the imperfection displacement to additional 
pressure. 
The effective force term  gives the leading order effect of 
imperfections as long as $\bar{z}\lesssim 1$. 
This is a first hint that the combination of pressure and imperfection
displacement leads to similar indentation behavior as a
preindenting point force.

\subsection{The avoided buckling bifurcation}

In the following, we consider axially symmetric imperfections
in the form of a localized indentation of Gaussian shape 
\cite{Hutchinson2016,  Lee2016}
\begin{equation}
\bar{w}_I(\rho) = -\delta e^{-\rho^2/\rho_I^2}
\label{eq:wIrho}
\end{equation}
with dimensionless depth $\delta$ (measured in the same units as normal
 displacements $w$) and dimensionless size $\rho_I$ (measured in the same 
units as radial distances $r$). 
In the presence of such imperfections, the critical pressure 
is reduced to $p_{c,\delta}<p_c$ because it is easier to buckle 
the preindented shell, in which an indentation imperfection 
displacement $\bar{w}_I$ with  $\bar{w}_I'>0$ 
leads  to  additional compressive strains 
upon  compression by   pressure.

\begin{figure*}
\begin{center}
\includegraphics[width=0.99\textwidth]{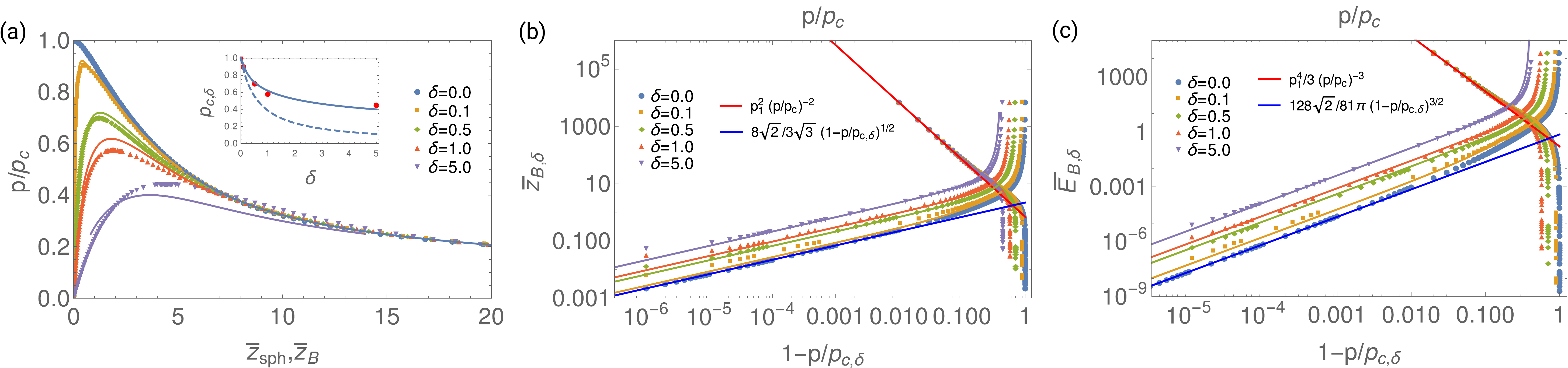}
\caption{
 (a) Bifurcation of the indentation as a function of the pressure
  for an imperfect shell with different imperfection 
  depths $\delta$ (and size $\rho_I=1$). 
  Below a critical pressure $p_{c,\delta}$, two branches of 
  shapes emerge in a saddle-node bifurcation, an indented  spherical 
  branch with $\bar{z}_{\rm sph}$, and an unstable barrier state 
  with $\bar{z}_B$.  Lines: Approximation with effective 
   preindenting force (\ref{eq:FpreAI}) and (\ref{eq:AIPog}). 
   Inset: Knockdown factor $p_{c,\delta}/p_c$ 
  as a function of $\delta$
    as compared  to approximations with effective 
   preindenting force (\ref{eq:FpreAI}) and (\ref{eq:AIPog}) (solid line)
  and small $\delta$ approximation (\ref{eq:AIlin2}) together with 
     (\ref{eq:knockdown1})  (dashed line). 
(b)  Indentation  difference 
   $\bar{z}_{B,\delta} =  \bar{z}_{B} - \bar{z}_{\rm sph}$ 
   as a function of pressure  $p/p_c$ (upper curves and upper horizontal scale)
 together with the analytical result (\ref{eq:zBPog}) for $\delta=0$ 
 (red line)
 and as a function of  $1-p/p_c$  (lower curves and lower horizontal scale)
together with the analytical result (\ref{eq:zBlin}) for $\delta=0$ 
  (blue  line).
   Lines are approximations with effective 
   preindenting force (\ref{eq:FpreAI}) and (\ref{eq:AIPog}). 
(c)
   Energy barrier $\bar{E}_{B,\delta}$ 
  as a function of pressure  $p/p_c$ (upper curves and upper horizontal scale)
 together with the analytical result (\ref{eq:EBpPog}) for $\delta=0$
  (red  line)
 and as a function of  $1-p/p_{c,\delta}$  (lower curves and lower 
   horizontal scale)
together with the analytical result (\ref{eq:EBplin}) for $\delta=0$ 
 (blue  line). Lines are approximations with effective 
   preindenting force (\ref{eq:FpreAI}) and (\ref{eq:AIPog}).
}
\label{fig:imp}
\end{center}
\end{figure*}

The numerical results in Fig.\ \ref{fig:imp} show 
that the effect of the preindentation imperfection is indeed 
qualitatively 
very similar to the effect of a preindenting point force.
For a localized imperfection field $\bar{w}_I$, 
 the additional inhomogeneous term in 
Eq.\  (\ref{eq:exact1_I}), which is the leading order 
effect for $\bar{z}\lesssim 1$,  becomes a localized effective 
force, which acts essentially in the same way as a 
preindenting point force.
Therefore, we find 
the same bifurcation behavior as discussed in the 
previous section for a  preindenting point force.
The imperfection gives rise to an 
 {\it avoided} or {\it perturbed bifurcation}
at a reduced critical pressure $p_{c,\delta}<p_c$, and the 
bifurcation at $p_{c,\delta}$
becomes a saddle-node bifurcation, in which 
 two stationary states emerge (see also Fig.\ \ref{fig:bifurcation}):
 a stable preindented spherical state with energy $\bar{E}_{\rm sph}$ and 
indentation $\bar{z}_{\rm sph}$ and 
 the unstable barrier state with  energy $\bar{E}_{B}$ and 
indentation $\bar{z}_{B}$ ($>\bar{z}_{\rm sph}$).

In the numerical approach, we solve the modified shallow shell 
equations as given explicitly in Eqs.\
(\ref{eq:forcebal3_I_app}) and  (\ref{eq:comp3_I_app}) in  Appendix
\ref{app:shshell}. 
In the presence of imperfections, 
the total energy difference 
in Eq.\ (\ref{eq:Etot}) is no longer the barrier energy 
but measures the energy difference between
 the barrier state or the preindented spherical state 
with $\bar{F}=0$ and  the unindented  and
precompressed state (with the same pressure but 
$\bar{F}=0$ and $\bar{w}=0$),  which is 
no longer a stationary state satisfying the force balance 
 (\ref{eq:forcebalI}) but which is
still an admissible and well-defined shell state satisfying the compatibility
condition (\ref{eq:comp3_I}). 
The total energy difference $\Delta \bar{E}_{\rm tot, imp}$ 
to this state can be obtained analytically;
see Eq.\ (\ref{eq:EtotIapp}) in Appendix \ref{app:shshell}.
In order to  calculate the barrier energy 
$\bar{E}_{B,\delta} = \bar{E}_B-\bar{E}_{\rm sph}$, we calculate 
this total energy difference for the barrier and the preindented 
spherical state numerically and use
$\bar{E}_{B,\delta} =  \Delta \bar{E}_{\rm tot, imp, B}-\bar{E}_{\rm tot, imp,
  sph}$  
to directly access the energy barrier  by continuation methods 
without the need to numerically integrate 
the force-indentation relation.
Our numerical shallow shell theory results for 
 the indentation as a function of the pressure
agree well with results from moderate rotation theory 
\cite{Hutchinson2016,Lee2016}.

For small imperfection indentations $\delta\ll 1$,
 the bifurcation still occurs 
close to $p_c$ and for 
small indentations $\bar{z}$. Then 
we employ the same expansion  (\ref{eq:expansionlin}) as in Sec.\
\ref{sec:theory2} for shallow dimples and find in the leading order
\begin{align*}
  & \hat{M} 
\begin{pmatrix} \bar{w}_{\rm lin,0}' \\ \bar{\psi}_{\rm lin,0} \end{pmatrix}
+\begin{pmatrix} 2(1-p/p_c)\bar{w}_{\rm lin,0}' 
    + \bar{\psi}_{\rm lin,0}\bar{w}_I' \\  \bar{w}_{\rm lin,0}'\bar{w}_I'
   \end{pmatrix}
\\
 & ~~~~=  \begin{pmatrix}   F_0 +\bar{z}^{-1} 2(p/p_c)\rho \bar{w}_I' 
 \\
  0 
\end{pmatrix}.
\end{align*}
Scalar multiplication with 
$(\bar{w}_0' , \bar{\psi}_0)$ on both sides gives the solvability 
condition 
\begin{align*}
   F_0 &=  \frac{2\sqrt{2}}{\pi} (1-p/p_c)^{1/2}   
    +2 \int d\rho \bar{w}_I' (\bar{\psi}_0 \bar{w}_0')  
\\
    &- 
      \bar{z}^{-1}  2(p/p_c)\int d\rho \rho \bar{w}_I' \bar{w}_0',
\end{align*}
with  an additional  $\bar{z}^{-1}$ contribution to the leading order,
which means an additional imperfection force 
and an additional imperfection contribution to the 
linear stiffness. 
Both contributions are linear in the imperfection displacement 
$\bar{w}_I$. 
We evaluated integrals approximately
 in the limit $p\to p_c$ 
as previously done in Eq.\ (\ref{eq:intwo2}).
The  force-displacement relation becomes
\begin{subequations}
\begin{align}
&\bar{F}+(p/p_c)\bar{A}_I =  
  \left( {4\sqrt{2}} (1-p/p_c)^{1/2}+ {\bar{k}_I}\right) \bar{z} 
\nonumber\\
    &~~~~~~~~~~~ - \frac{3\sqrt{3}}{2}  \bar{z}^2  + O(\bar{z}^3)
~~~\mbox{with}
\label{eq:Fzimp}\\
 &\bar{A}_I \equiv 4\pi \int d\rho \rho \bar{w}_I' \bar{w}_0',
\label{eq:AIlin}\\
 & \bar{k}_I \equiv 4\pi \int d\rho \bar{w}_I' (\bar{\psi}_0 \bar{w}_0')  
     = 4\pi \int d\rho \rho  \bar{w}_I' \bar{w}_0' \bar{\sigma}_{rr,0}.
\label{eq:kI}
\end{align}
\end{subequations}
Both imperfection modifications have a transparent physical 
interpretation. 
The strength of the imperfection force  $(p/p_c)\bar{A}_I$ 
 is proportional 
to the pressure and 
 the amplitude of the imperfection displacement but also depends 
on the ``overlap area'' of the imperfection field $\bar{w}_i(\rho)$ 
and the indentation mode $\bar{w}_0$ for  $p=p_c$ and $\bar{F}=0$
centered at the pole. 
Imperfections localized away from the pole will greatly reduce 
the imperfection force because $\bar{w}_I'$ hardly overlaps with the 
indentation mode $\bar{w}_0'$. 
The imperfection contribution $\bar{k}_I$ to the linear 
stiffness  is the  overlap area weighted by 
the radial stress $\sigma_{rr,0}$ profile at $p=p_c$. 

\begin{figure}
\begin{center}
\includegraphics[width=0.4\textwidth]{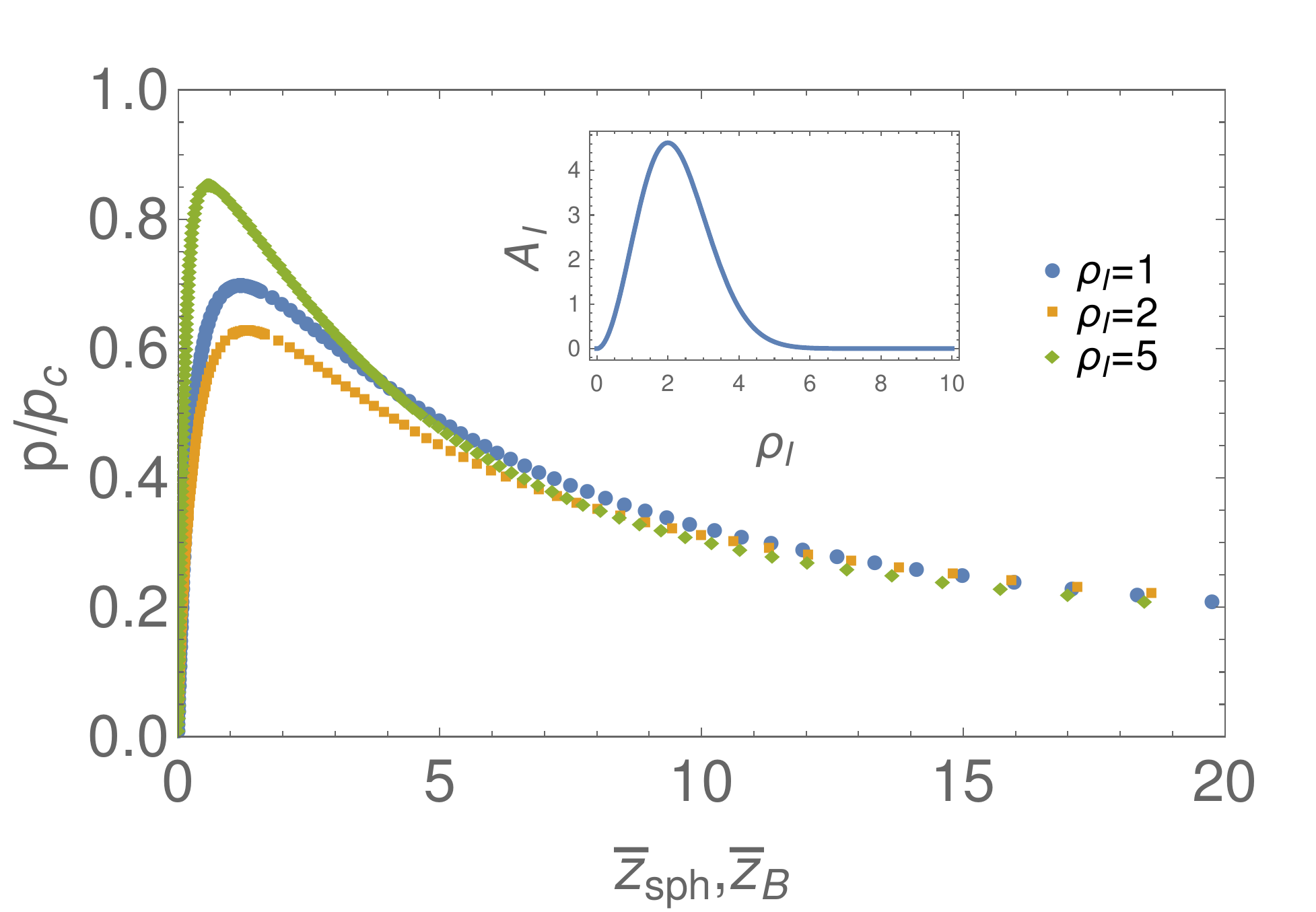}
\caption{
 Indentation as a function of the pressure
  for an imperfect shell with fixed  imperfection 
  depth $\delta=0.5$ as a function of the imperfection size $\rho_I=1$. 
  The knockdown factor  $p_{c,\delta}/p_c$  is maximal for $\rho_I\sim 2$. 
   Inset: Overlap area $A_I$ as a function of $\rho_I$ for $\delta=0.5$ 
   according to Eq.\ (\ref{eq:AIlin2}) with a maximum at $\rho_I=2$. 
}
\label{fig:zpri}
\end{center}
\end{figure}

The buckling bifurcation is now governed by the presence of the additional 
$z$-independent imperfection force 
in the force-displacement relation, which acts analogously to a
preindentation force $\bar{F}_{\rm pre}=(p/p_c)\bar{A}_I$. 
For the imperfections with shape (\ref{eq:wIrho}) we can explicitly
evaluate $\bar{A}_I$ in Eq.\ (\ref{eq:AIlin}) as 
\begin{equation}
   \bar{A}_I =2\pi \delta \rho_I^2\left( 1 - \frac{\sqrt{\pi}}{2}
   \rho_I e^{-\rho_I^2/8} I_{1/2}\left(\frac{\rho_I^2}{8}\right)\right),
\label{eq:AIlin2}
\end{equation}
where $I_\nu(x)$ is the modified Bessel function.
This reveals that, for fixed indentation depth $\delta$, the 
overlap area $\bar{A}_I$ and thus the imperfection force $\bar{F}_{\rm
  pre}=(p/p_c)\bar{A}_I$ become maximal for an imperfection 
size $\rho_I\approx 2$, i.e., if the imperfection has twice
the size of the elastic length $l_{\rm el} = (hR_0)^{1/2}k^{-1}$
(see inset of Fig.\ \ref{fig:zpri}).
We conclude that this
 imperfection size is most effective in reducing the 
buckling threshold. This is confirmed by 
our numerical data in  Fig.\ \ref{fig:zpri}.

The description by an effective 
preindentation force $\bar{F}_{\rm pre}=(p/p_c)\bar{A}_I$ 
also implies that the softening behavior 
of the shell close to the critical pressure
$p_{c,\delta}$    is {\it universal} and the indentation  difference 
   $\bar{z}_{B,\delta} =  \bar{z}_{B} - \bar{z}_{\rm sph}$ 
 from spherical to barrier state and the 
energy barrier $\bar{z}_{B,\delta} =  \bar{E}_{B} - \bar{E}_{\rm sph}$ 
are governed by the 
same exponents $\bar{z}_{B,\delta} \propto \Delta \bar{p}^{1/2}$ and 
$\bar{E}_{B,\delta} \propto \Delta \bar{p}^{3/2}$ 
[$\Delta \bar{p} \equiv ({p_{c,\delta}-p})/{p_c}$] as for 
a preindenting  point force $\bar{F}_{\rm pre}$ or as in the absence 
of a preindenting point force.

We can use the results of the previous section with 
an effective  preindenting force $\bar{F}_{\rm pre}=(p/p_c)\bar{A}_I$
 given by Eqs.\  (\ref{eq:AIlin}) or (\ref{eq:AIlin2})
to describe the 
avoided bifurcation in the presence of imperfections. 
This is  quantitatively correct only for small $\delta$ 
[$\delta < 0.1$, see also inset in  Fig.\ \ref{fig:imp}(a)]  
such that $p_{c,\delta}$ remains close to $p_c$. 
Our numerical results show, however, that 
the concept of an 
effective preindenting force $\bar{F}_{\rm pre}=(p/p_c)\bar{A}_I$,
which is given by an ``overlap area''  area $\bar{A}_I$ can be 
generalized for larger $\delta$  and smaller $p$,  if 
 we replace the shallow shell solution in the 
linear approximation 
$\bar{w}_0=  \lim_{p\to p_c} \bar{w}_{\rm lin}/\bar{z}$ [see Eqs.\ 
(\ref{eq:w0}) and (\ref{eq:expansionlin})] in Eq.\ (\ref{eq:AIlin}) 
by the general solution $\bar{w}/\bar{z}$.
The resulting effective preindenting force
\begin{equation}
   \bar{F}_{\rm pre} = (p/p_c)\bar{A}_I =  
  (p/p_c) 4\pi  \int d\rho \rho \bar{w}_I' \frac{\bar{w}'}{\bar{z}}
 \label{eq:FpreAI}
\end{equation}  
can describe our numerical data  for the 
modified bifurcation behavior in Fig.\ \ref{fig:imp} well 
for  $0.05 \lesssim \delta\lesssim 1.0$ 
when we evaluate $\bar{A}_I$ using the  mirror-inverted Pogorelov dimple
with $\bar{w}(\rho) = -\bar{z}+ \rho^2$, which gives
\begin{equation}
   \bar{A}_I =8\pi \delta  \frac{\rho_I^2}{\bar{z}}
   \left( 1 - e^{-\bar{z}/\rho_I^2} \left(\frac{\bar{z}}{\rho_I^2} +1\right)
   \right),
\label{eq:AIPog}
\end{equation}
and  when we use  $\bar{z}= \bar{z}_{B,0}$ as given by the 
 interpolation formula (\ref{eq:zBinterpol}).

Also analogously to a preindenting force, 
effects of imperfections are negligible  for the deeply-indented 
 barrier states 
in the Pogorelov limit  $\bar{z}_B\gg 1$ and $p\ll p_c$,
where elastic and pressure terms dominate the  force-indentation relation 
 (\ref{eq:FzPog}). 
The indentation difference  from spherical to barrier state  
$\bar{z}_{B,\delta}\approx \bar{z}_B$ and the barrier energy 
$\bar{E}_{B,\delta}\approx \bar{E}_B$  
approach the Pogorelov asymptotics (\ref{eq:zBPog}) and 
 (\ref{eq:EBpPog}) in the Pogorelov limit for 
 $p/p_c \ll  1/F_{\rm pre}^{1/2}$ with the effective 
$\bar{F}_{\rm pre}$ from Eq.\ (\ref{eq:FpreAI}), which is 
 also seen in the numerical data in  Fig.\ \ref{fig:imp}.

\section{Discussion and conclusion}

We characterized the buckling bifurcation  of closed 
spherical shells in the framework of continuum elastic theory under the 
combined action of pressure and point forces.
Spherical shells have numerous realizations on the micro- and macroscale
to which our theory applies.

 Typical 
 artificial  micrometer-sized capsules have 
shell thicknesses of  $h\sim 10{\rm nm}$ and are made from 
 soft materials with  bulk Young moduli $E \sim 0.1\,{\rm GPa}$. 
This results in 2D Young's moduli $Y = Eh \sim 1 {\rm N/m}$ and 
bending moduli $\kappa \sim  Eh^3
 \sim 10^{-16} {\rm Nm} \sim  2\times 10^4 k_BT$ 
in agreement with elastometry measurements \cite{Hegemann2018}. 
For $R_0 = 10{\rm \mu m}$,  typical  F\"oppl-von K\'arm\'an numbers
(\ref{eq:FvK}) are 
$\gamma\sim 10^6-10^7$.

Related but distinct systems are red blood cells, viruses, and biological 
cells. 
Red blood cells and viruses also have stretching, shear 
and bending elasticity featuring elastic moduli similar to 
 artificial spherical microcapsules.
Red blood cells are also micrometer-sized with a somewhat 
smaller shear modulus 
$\mu \sim 10^{-5}{\rm N/m}$  and 
bending moduli $\kappa \sim  10^{-16} {\rm Nm}$
\cite{evans1980,Iglic1997}.
One important difference from  artificial microcapsules is the
local inextensibility of the lipid bilayer membrane of the red blood cell,
which enforces a local area constraint
such that lipid membranes and red blood cells are even less extensible 
than shells \cite{evans1980}. 
Furthermore, there are differences in the rest shape.
The buckling of spherical shells is governed 
by their spherical rest shape, which has minimal area at given volume. 
Thus, any deformation stretches the shell, which leaves the 
mirror-inverted Pogorelov dimple as preferred buckled shape
under pressure.  Red blood cells have a fixed  volume $V_0$ and
fixed area $A_0$ combining to a  reduced volume 
$v_{\rm RBC}=V_0/(4\pi/3)(A_0/4\pi)^{3/2} \sim 0.6$
and an oblate spheroidal rest shape under physiological conditions,
while the reduced volume of a sphere reaches the maximal value
$v_{\rm sph} =1$. Under point-force indentation, the reduced volume
of a spherical shell only  reduces  to $v = 1- O(\gamma^{-1})$
up to the barrier.
The additional area that is available in a nonspheroidal rest shape
with $v_{\rm RBC}<1$  contributes  to the rich 
deformation behavior of red blood cells
\cite{evans1980,Iglic1997,LimHW2002,RBC_Book}.
Another  crucial difference is the 
 role of spontaneous curvature (or the conjugate 
integrated mean curvature or area difference) as 
control parameter. For red blood cells, area difference is an important 
control parameters of shape sequences \cite{Iglic1997,LimHW2002,RBC_Book}
whereas spherical shells are treated with a fixed spontaneous curvature
given by  the spherical  rest shape (only recently, Pezulla {\it et al.}
started to address the role of spontaneous curvature in buckling of 
spherical shells \cite{Pezzulla2018}).
Viruses range in size from $15$ to $500 {\rm nm}$ and 
have elastic moduli $Y = Eh \sim 0.1-1 ~{\rm N/m}$ and 
bending moduli $\kappa \sim  Eh^3
\sim 5\times 10^{-19}~{\rm to}~ 5\times 10^{-15} {\rm Nm}$ 
\cite{Buenemann2008},
similar to artificial microcapsules.
Virsues  are, however,  crystalline spherical shells consisting of
discrete protein building blocks. On a sphere, the crystal structure must
contain 
 topologically unavoidable triangulation defects. 
This results  in faceted 
equilibrium shapes of large viruses with
F\"oppl-von K\'arm\'an numbers $\gamma > 150$, while 
only small viruses maintain a  spherical equilibrium shape
\cite{Lidmar2003}. Therefore,
our results regarding, for example, the linear stiffness $k$ apply only
to small viruses \cite{Buenemann2008}.

For pressurized spherical shells, 
we showed that  nonlinear shallow shell theory can
quantify many aspects of the energy barrier, which
has  to be overcome if buckling is triggered by  ``poking'' 
with a point force $F$ while the pressure is still subcritical, 
i.e., below the classical buckling pressure $p_c$. 
In particular, we could derive the exact asymptotics of the 
energy barrier properties (including the numerical prefactors)
 in two relevant limiting regimes, namely in the Pogorelov limit
at small pressures $p\ll p_c$, where the 
indentation at the transition state is 
 much deeper than shell thickness,  $z_B \gg h$,  and in the 
opposite limit for pressures very close to $p_c$, where the 
indentation at the transition state is shallow, $z_B \ll h$,
and develops oscillations. 
 We developed a very accurate 
 numerical approach, which is based on the 
closed analytical expression (\ref{eq:Etot}) 
for the energy barrier height and allows 
us to trace the energy barrier height by numerical continuation techniques. 
The numerical study reveals that there are  only these two regimes 
for all barrier properties, such as the barrier 
energy $E_B$,  the indentation depth $z_B$, the indentation volume 
$\overline{\Delta V}_B $, and the indentation width $\rho_B$, 
with a clear crossover if the indentation $z_B\sim h$ or 
$\bar{z}_B \sim 1$ in dimensionless units (\ref{eq:non-dim}). 

Using  systematic expansions of the nonlinear shallow shell 
equations about the 
Pogorelov mirror-inverted dimple for $p/p_c\ll 1$
 and the linear response state for $(1-p/p_c)\ll 1$,
we obtained a complete analytical 
characterization of the energy barrier 
in both limits; see Eqs.\ (\ref{eq:barrierPog}) for the 
Pogorelov limit and Eqs.\ (\ref{eq:barrierlin}) for 
pressures  close to $p_c$. 
The analytical results agree with our numerical shallow shell 
data over several decades of control parameters $p/p_c$ or $1-p/p_c$, 
respectively (see Figs.\ \ref{fig:zp} and \ref{fig:EB})
 and allow us to formulate quantitatively correct 
interpolation formulas for the barrier energy $\bar{E}_B$ and the 
corresponding indentation $\bar{z}_B$ [Eqs.\ (\ref{eq:interpol}) and 
(\ref{eq:zBinterpol}), respectively], which incorporate all 
analytical constraints.

Reverting the  nondimensionalization  (\ref{eq:non-dim}) 
gives
$E_B = YR_0^2 \gamma^{-3/2} 2\pi f_p(p/p_c)$, and 
 we can estimate  values for the
energy barrier in applications. 
The  typical barrier energy scale  is 
$E_B \propto YR_0^2 \gamma^{-3/2} = (\kappa^{3}/YR_0^2)^{1/2}=
 [12(1-\nu^2)]^{-3/2} EhR_0^2 (h/R_0)^3 $ with a 
pressure-dependent numerical prefactor $2\pi f_p(p/p_c)$, which is given 
by the interpolating function (\ref{eq:interpol}). 
For $p=p_c/2$ at half the buckling pressure, $2\pi f_p(1/2) \simeq 18$. 
For a typical artificial 
thin microscale 
capsule with $R_0 = 10{\rm \mu m}$, $h=10{\rm nm}$ made from 
a soft material with $E = 0.1\,{\rm GPa}$ with $\nu=1/2$ this gives 
an energy  barrier
$E_B \sim 17 k_BT$ at room temperature or $E_B/U_c =2.4 \times 10^{-4}$,
where $U_c = \frac{1}{2} p_c \Delta V_c = 16\pi (1-\nu)  \kappa$
is the elastic energy stored in the spherical shape just before 
it buckles. These estimates demonstrate that the buckling energy barrier 
is rather small: It is comparable to the thermal energy and much smaller 
than the total elastic energy stored in the capsule. 
Therefore, capsules become very susceptible to thermally induced buckling 
\cite{Kosmrlj2017,Baumgarten2018}, small point forces, or 
 imperfection effects  already at moderate compressive pressures. 
The typical normal indentation at the barrier is $z_B \propto
R_0 \gamma^{-1/2} = h/[12(1-\nu^2)]^{1/2}$, i.e., the shell thickness. 
The pressure-dependent numerical prefactor $g_p(p)$ is given by the 
interpolation function (\ref{eq:zBinterpol}) with $g_p(1/2)\simeq 1.7$
at  $p=p_c/2$. This demonstrates that only small indentations have to 
be achieved to overcome the energy barrier.

We also obtain a complete picture of the buckling energy landscapes
 (\ref{eq:landscapePog}) and  (\ref{eq:landscapelin}) in both regimes, 
i.e., the total indentation energy $\bar{E}_{\rm ind}$
as a function of the indentation $\bar{z}$,   from which also the 
force-displacement curves can be calculated in both 
regimes. 
In the Pogorelov limit 
we could show the equivalence of our  
systematic expansion of the nonlinear shallow shell 
equations about the 
Pogorelov mirror-inverted dimple to a  boundary layer 
approach by  Evkin {\it et al.} 
\cite{Evkin2016} and establish the connection to and generalize 
 recent work 
of  Gomez {\it et al.} \cite{Gomez2016} on the $p=0$ case. 
While Gomez  {\it et al.}  addressed the case $F>0$ and $p=0$, our 
analytical calculation  
focused on the barrier state, where $F=0$, for arbitrary $0\le p \le p_c$. 
We could draw conclusions for the general case by showing that 
both results are consistent if the 
 elastic part of the indentation energy  is 
{\it independent of the pressure}.
Future work should try to develop a systematic expansion covering the 
full behavior for $F>0$ and $0\le p \le p_c$.

The  regime $\bar{z}_B \ll 1$ for $p$ close to $p_c$ 
 is particularly interesting as it 
characterizes the ``critical properties'' if 
the buckling instability is approached from below. 
In this regime close to $p_c$, we find a 
 softening of the spherical shell, which is 
characterized by three critical exponents:
(i) 
the stiffness $k$ for the linear response to point forces vanishes 
 $\propto (1-p/p_c)^{1/2}$; (ii)  the buckling 
energy barrier maximum
vanishes $\propto (1-p/p_c)^{3/2}$; and 
(iii)  the shell indentation at the energy barrier maximum
 vanishes $\propto (1-p/p_c)^{1/2}$.
These results are shown analytically and  agree 
with our numerical shallow shell data. They extend and correct 
previous findings in Ref.\ \cite{Baumgarten2018}, which
were based on less accurate SURFACE EVOLVER simulations.

The linear stiffness $k$ with respect to a point force 
is experimentally accessible 
in  mechanical compression tests such as plate compression
\cite{carin_compression_2003,fery_mechanical_2007} or 
compression by microscopy tips \cite{Fery2004,fery_mechanical_2007}
because all of these compression devices 
 effectively act as point forces in the initial regime. 
Therefore our result (\ref{eq:k}) for the softening of the shell 
can be directly tested  
if such compression tests are combined 
with external pressure $p$. For microcapsules in liquids
external pressure can be generated as osmotic pressure \cite{Knoche2014o},
and for macroscopic capsules as mechanical air or liquid pressure. 
For compressive pressure  $p=p_c/2$ at half the buckling pressure,
Eq.\ (\ref{eq:k}) predicts that the stiffness 
reduces to only 
$64\%$ of its pressure-free value $k  = 8  Y \gamma^{-1/2}$. 
Measurements  of the linearized stiffness 
at two different pressures 
 can also  be used to infer two unknown 
quantities, for example, the  shell's Young's modulus  
and  capsule pressure via Eq.\ (\ref{eq:k}) and, thus,
determine elastic properties of the shell. 
While the linear stiffness increases for pressurized shells \cite{Vella2012b},
it decreases for shells under compressive pressures. 
In Ref.\ \cite{Wischnewski2018}, it has been shown that a positive interfacial
tension also acts as a stretching pressure and gives rise to linear
stiffening. Softening can therefore also 
 be induced by a negative interfacial 
tension. This shows that active expansion or stretching tensions generated 
in biological cells by molecular motors in the cell cortex, which will 
give rise to negative interfacial 
tension in the cortex, will  effectively soften 
the cell. 

Knowledge of the buckling energy landscape also  enables us to 
calculate the Maxwell pressure $p_{c1}\sim  p_{c} \gamma^{-1/4}$ 
and the critical unbuckling pressure 
$p_\text{cu}\sim 3p_{c1}/4$ from 
shallow shell theory. 
The buckling energy landscape also suggests that  soft spots, 
which are  small compared to 
the elastic length $l_{\rm el} = (R_0h)^{1/2}$ will  
immediately  unbuckle for $p$ smaller than  $p_c$ and thus 
exhibit no hysteresis in buckling and unbuckling.

Our results shed light on the nature of the buckling bifurcation
as schematically shown in Fig.\ \ref{fig:bifurcation}.
The bifurcation at $p=p_c$ is a subcritical bifurcation 
which has a fixed point structure similar to a 
subcritical pitchfork bifurcation. 
The barrier states are a 
subcritical branch of unstable stationary points, which appear 
together with the buckled snap-through state 
  for $p>p_\text{cu}$ in a  
type of  saddle-node or fold bifurcation.
The snap-through state  becomes 
the only  minimum after the bifurcation at $p=p_c$.
Within the pressure window $p_\text{cu} <p<p_c$ there is 
bistability between buckled and unbuckled solutions with the 
barrier state separating these two stable branches
resulting in hysteresis. 
Upon approaching the instability 
from $p<p_c$, the softening of the 
shell with vanishing  linear stiffness and energy barrier 
can give rise to 
important dynamical consequences such as  a 
``critical slowing down''  of  the buckling dynamics \cite{Gomez2017}.
A  complete  theory  of the  buckling dynamics 
of a shell close to $p_c$  remains to be developed.

This bifurcation behavior is modified if  a preindenting
point force $F_{\rm pre}$ is applied or in the presence of localized 
axisymmetric imperfections of indentation depth $\delta$ and size $r_I$.
Both the numerical shallow shell  analysis and the 
analytical approach could be extended to these situations. 
Interestingly, a localized preindentation imperfection's effect 
 on the buckling instability is very similar to that of 
 a preindenting point force,
as can be immediately recognized by comparing Figs. \ref{fig:Force}
and \ref{fig:imp}. For both cases, the buckling bifurcation 
becomes an  avoided or  perturbed bifurcation
at a reduced critical pressure $p_{c,F}<p_c$ or $p_{c,\delta}<p_c$. 
Below this critical pressure, two stationary states emerge
in a saddle-node bifurcation, namely a stable quasi-spherical state and an
unstable barrier state. 
Interestingly, the softening behavior sketched above 
with 
(ii) the buckling 
energy barrier between quasispherical and barrier state 
vanishing as $\bar{E}_B \propto  \Delta \bar{p}^{3/2}$ and 
(iii)  the shell indentation difference between quasispherical and 
barrier state vanishing  $\delta \bar{z}_B\propto \Delta \bar{p}^{1/2}$
remains universally valid both in the presence of a preindenting force
and imperfections. 
This also suggests that (i) remains valid, and the linear stiffness 
vanishes as $k\propto \bar{E}_B/\delta \bar{z}_B^2 \propto \Delta
\bar{p}^{1/2}$.

We were able to make the equivalence between preindenting point force 
and imperfections quantitative with Eq.\ (\ref{eq:FpreAI})
and found that imperfections 
effectively act as a point force proportional to the pressure $p$ and 
an ``overlap area'' $A_I$, which depends on the shape of the imperfection 
$w_I(r)$  and the indentation $w(r)$. 
This allowed us to conclude that there exists an ``optimal'' size 
for an imperfection of $r_I = 2l_{\rm el}$, where it maximizes the 
knockdown factor for the buckling pressure. 
The quantitative prediction  (\ref{eq:FpreAI}) may be useful 
in designing spherical shells with specific buckling thresholds.

\appendix

\section{Derivation of total  energy difference}
\label{app:barrierenergy}

The difference in total energy 
$\Delta E_{\rm tot} = \Delta E_s + \Delta E_b +p\Delta V$ 
(i.e., the sum of stretching and  bending energy 
 and mechanical work by pressure)
between the indented barrier state ($F=0$) 
and the precompressed unindented spherical state (with the 
same pressure but $\bar{z}=0$) is given by 
the simple, explicit formula (\ref{eq:Etot}):
\begin{align}
  \Delta \bar{E}_{\rm tot}= \bar{E}_B 
   = -\frac{1}{4} \int_0^{\infty}d\rho  \bar{\psi}
   (\bar{w}')^2,
\label{eq:Etotapp}
\end{align}
where   $\bar{\psi}$ and $\bar{w}$ are solutions of the 
 shallow shell equations (\ref{eq:forcebal3}) and (\ref{eq:comp3}) 
in the barrier state with  $\bar{F}=0$.

More generally, for an arbitrary indented state  with a point force $F$  
and a corresponding indentation $z$,
the total energy difference 
$\Delta E_{\rm tot,F} = \Delta E_s + \Delta E_b +p\Delta V -Fz$
between the indented state and the 
 precompressed unindented spherical state (with the same pressure 
and $z=0$) is given by 
\begin{align}
   \Delta \bar{E}_{\rm tot,F} = -\frac{1}{4} \int_0^{\infty}d\rho  \bar{\psi}
   (\bar{w}')^2 -\frac{\bar{F}}{4\pi} \bar{z} 
   - 2(1+\nu) \frac{p}{p_c}\frac{\bar{F}}{2\pi},
\label{eq:EtotappF}
\end{align}
where $\bar{\psi}$ and $\bar{w}$ are solutions of the 
 shallow shell equations (\ref{eq:forcebal3}) and (\ref{eq:comp3}) 
for indentation $\bar{z}=-\bar{w}(0)$ and 
corresponding point force $\bar{F}$.

In order to derive these formulas, we start 
with the dimensionless elastic energy, i.e., stretching and bending energy 
of an axisymmetric  state with stress function $\bar{\psi}$ and 
normal displacement $\bar{w}$ in shallow shell theory
[see also Eq.\ (\ref{eq:Eel2}) below] \cite{Ventsel,Paulose2012}, 
\begin{align}
  E_{\rm el} &= E_s+E_b\nonumber\\
  &=  \frac{1}{2}\int d\rho \rho 
  \left[\left( \bar{\psi}' + \frac{\bar{\psi}}{\rho}\right)^2
    -2(1+\nu) \bar{\psi}'\frac{\bar{\psi}}{\rho} \right]
\nonumber\\
  & +  \frac{1}{2}\int d\rho \rho 
   \left[\left( \bar{w}'' + \frac{\bar{w}'}{\rho}\right)^2
    -2(1-\nu) \bar{w}'\frac{\bar{w}}{\rho}    \right].
\label{eq:Eel}
\end{align}
The stretching energy difference $\Delta E_s$ is the difference between 
the stretching energy due to the total stress function 
$\psi+\psi_0$ and the stretching energy due to the prestress
$\psi_0= -pR_0r /2$ [or $\bar{\psi}_0(\rho) = -2\rho p/p_c$],
\begin{align}
  \Delta\bar{E}_{s} &=  \frac{1}{2} \int d\rho \rho \left[ 
    \left( (\bar{\psi}_0+ \psi)' + \frac{\bar{\psi}_0+ \psi}{\rho}\right)^2
    \right.
    \nonumber\\
  & -2(1+\nu) (\bar{\psi}_0+ \psi)' \frac{\bar{\psi}_0+ \psi}{\rho}
   \nonumber\\
  &\left. - \left( \bar{\psi}_0' + \frac{\bar{\psi}_0}{\rho}\right)^2
    +2(1+\nu) \bar{\psi}_0'\frac{\bar{\psi}_0}{\rho} \right]
  \nonumber \\
  &=  \frac{1}{2}\int d\rho \rho \left[
    \left( \bar{\psi}' + \frac{\bar{\psi}}{\rho}\right)^2\right] +
     2(1-\nu) \frac{\bar{F}}{2\pi} \frac{p}{p_c},
\label{eq:Esapp}
\end{align}
where we used $\bar{\psi}=0$ for $\rho=0$ and $\rho=\infty$ 
and the asymptotics (\ref{eq:psiasy}). 
The last term vanishes at the barrier state, where $\bar{F}=0$. 
For the bending energy difference, we find 
(note that $w$ was already defined as the normal displacement relative to the 
precompressed state)
\begin{align}
\Delta \bar{E}_b &=  \frac{1}{2}\int d\rho \rho \left[
    \left( \bar{w}'' + \frac{\bar{w}'}{\rho}\right)^2\right],
\label{eq:Ebapp}
\end{align}
where we used $\bar{w}'=0$  for $\rho=0$ and $\rho=\infty$. 
The dimensionless mechanical work is 
\begin{align}
   \overline{p\Delta V} &= 4\frac{p}{p_c} \int d\rho \rho \bar{w}
     = \frac{2}{\pi} \frac{p}{p_c}  \overline{\Delta V}
\label{eq:pVapp}
\end{align}
(again, note that $w$ was defined as the normal displacement relative to the 
precompressed state). 

Using   partial integration [with $\bar{w}(\infty)=0$ and $\bar{w}'(0)=0$],
the relation 
 (\ref{eq:exact1}), and $\int d\rho \bar{w}' = -\bar{w}(0)=\bar{z}$,
we  can re-write $\Delta\bar{E}_b$ as  
\begin{align*}
  \Delta \bar{E}_b &= \frac{1}{2} \int d\rho \left[ 
  \rho \bar{\psi}\bar{w}'  -\bar{\psi}(\bar{w}')^2 + 2\frac{p}{p_c}
  \rho\bar{w}'\right] + \frac{\bar{F}}{2\pi} \frac{1}{2}\bar{z}.
\end{align*}
With Eq.\ (\ref{eq:exact2}),
we obtain
\begin{align*}
  \Delta \bar{E}_b &= -\overline{p\Delta V} + \frac{1}{2} \int d\rho \rho
    \bar{\psi}\bar{w}' \left( 1 - \frac{\bar{w}'}{\rho} \right)
   \\
  &~~~~
    +\frac{\bar{F}}{2\pi}\left(-4\frac{p}{p_c} +  \frac{1}{2}\bar{z}\right).
\end{align*}
This results in a total energy  difference
\begin{align*}
  \Delta \bar{E}_{\rm tot,F} &=   \Delta \bar{E}_s+\Delta \bar{E}_b+
   \overline{p\Delta V} -\frac{\bar{F}}{2\pi} \bar{z} 
\nonumber\\ 
&= \frac{1}{2} \int d\rho \rho 
    \left[
    \left( \bar{\psi}' + \frac{\bar{\psi}}{\rho}\right)^2+
   \bar{\psi}\bar{w}' \left( 1 - \frac{\bar{w}'}{\rho} \right) \right]
\nonumber\\
  &~~~ +\frac{\bar{F}}{2\pi} \left(  2(1-\nu)  \frac{p}{p_c} - 4 \frac{p}{p_c} 
     + \frac{1}{2}\bar{z} - \bar{z} \right) 
\nonumber\\
&= -\frac{1}{4} \int_0^{\infty}d\rho  \bar{\psi}
   (\bar{w}')^2-\frac{\bar{F}}{4\pi} \bar{z} 
   - 2(1+\nu) \frac{p}{p_c}\frac{\bar{F}}{2\pi},
\end{align*}
where we used the compatibility condition (\ref{eq:comp3}).
This is Eq.\ (\ref{eq:EtotappF}) and specializes to Eq.\ 
 (\ref{eq:Etotapp}) 
at the barrier where $\bar{F}=0$. 

A further generalization to shells containing  imperfections
is given in Appendix \ref{app:shshell}. 

\begin{table*}
 	\caption{Units for nondimensionalization and 
    dimensionless quantities.}
  \label{tab:dimensionless}
 \begin{ruledtabular}
  \begin{tabular}{l|l|l|l|l} 
    & Here    & Gomez {\it et al.} \cite{Gomez2016}  & 
     Hutchinson {\it et al.} \cite{Hutchinson2017b,Marthelot2017} &
    Evkin {\it et al.} \cite{Evkin2016,Evkin2017}
    \\\hline
   Norm.\ displacement unit
    &   $(\kappa/Y)^{1/2} = R_0 \gamma^{-1/2}$ &  &   
   $\Delta w_R\equiv \sqrt{12}hk^{-2}$  &
\\
 & $~~~~=  hk^{-2}$   &   $h$     &   
   $~~~~=h(1-\nu^2)^{-1/2}$ & 
   \\
   Norm.\ displacement    &    $\bar{w}\equiv wk^2/h$     &  $W =\bar{w}k^{-2}$    &   
     ${\Delta w}/{\Delta w_R} = \xi g(\tilde{s},\xi)$   &
\\
   &&&   $~~~~= -\bar{w}/\sqrt{12}$ & 
   \\
 Indentation depth      & $\bar{z}\equiv -\!\left.\bar{w}\right|_{\rho=0}$  
   &    $\Delta \equiv -\!\left.W\right|_{\rho=0}= \bar{z}k^{-2}$     
 &    $\xi \equiv \left.{\Delta w}/{\Delta w_R}\right|_{\tilde{s}=0} = 
      \bar{z}/\sqrt{12}$ &    $\varepsilon^{-2} = \bar{z}/4$
\\
  Pressure & $p/p_c$    &    $p=0$   &   $p/p_c = f(\xi)$    &  $\bar{q}=p/p_c$
\\
   Radial distance unit & 
    $(\kappa R_0^2/Y)^{1/4} = R_0 \gamma^{-1/4} $ &&&\\
 &  $~~~~~~~~~= (hR_0)^{1/2} k^{-1} $ 
    & $(hR_0)^{1/2}$    &   $12^{1/4}(hR_0)^{1/2}$ & 
   \\
  Radial distance
       & $\rho\equiv rk/(hR_0)^{1/2} $     &    $\rho_{\rm Gomez} = \rho k^{-1}$    & 
      $\tilde{s}\simeq \rho/12^{1/4}$ & \\
  Indentation volume & 
  $\overline{\Delta V}_B \equiv
                       - 2\pi\int_0^{\infty} d\rho \rho \bar{w}$ & -- & 
  $\overline{\Delta V}_B  = 24\pi h(\xi)$ &
   \\
    Force unit & $YR_0/\gamma= \kappa/R_0$ & & &\\
   & $~~~~~ = (Yh^2/R_0)k^{-4}$   & $Y h^2/R_0$  
   &   $2\pi \kappa/R_0$ \cite{Marthelot2017}&
  \\
  Force    &  $\bar{F}$   &   $F = \bar{F} k^{-4}$ & 
   $\bar{F}_{\rm Marthelot} = \bar{F}/2\pi$ \cite{Marthelot2017} &
    $\bar{Q} = \bar{F}/12\pi$
 \\
    Energy unit    & $e\equiv 2\pi(\kappa^{3}/YR_0^2)^{1/2}$ & --   &
     $W_c=\frac{1}{2} p_c\Delta V_c C h/R_0=48e$ &
   $U_A = \frac{1}{2} (p/p_c)^2 p_c\Delta V_c$ 
\\
   &&& $~~~~= 96\pi(\kappa^{3}/YR_0^2)^{1/2}$ &
\\
   &&& $C\equiv \frac{\sqrt{3}}{(1-\nu)\sqrt{1-\nu^2}}$ &  \\
 Barrier energy  & $\bar{E}_B$       & -- & $W/W_c = \bar{E}_B/48$ & 
    $\bar{E}= \frac{\bar{E}_B}{2(p/p_c)^2 \lambda^2(1-\nu)}$ 
\\
   &&& $~~~~~~~~= q(\xi) - f(\xi) h(\xi)$&  $\lambda^2 \equiv 4k^2R_0/h$
 \\ 
\end{tabular}
\end{ruledtabular}
\end{table*}

\section{Nonlinear shallow shell equations with imperfections}
\label{app:shshell}

\subsection{Derivation of shallow shell equations}

We follow Refs.\  \cite{Hutchinson1967,Hutchinson2016,Lee2016,Hutchinson2018}
 and consider imperfections 
in  form of a
normal axisymmetric displacement field $w_I$,
which is already present in the strain-free state of the shell. 
With a displacement field 
${\bf u} =  u{\bf e}_x + v{\bf e}_y + w {\bf e}_z$,
where $x$ and $y$ are Cartesian directions in  the two-dimensional 
reference plane over which shallow shell configurations are 
parametrized and ${\bf e}_z$ is the outward pointing normal,
the in-plane strain tensor in the presence of the imperfection field is 
given by $u_{ij} = u_{ij}(w+w_I) - u_{ij}(w_I)$, resulting 
in 
\begin{align*}
   u_{xx} &= \partial_xu + \frac{1}{2} (\partial_x w)^2 + \frac{w}{R_0}
   + \partial_xw \partial_xw_I
\\
 u_{yy} &= \partial_yv + \frac{1}{2} (\partial_y w)^2 + \frac{w}{R_0} 
  + \partial_yw \partial_yw_I
   \\
u_{xy} &= \frac{1}{2}\left( \partial_yu + \partial_xv+ \partial_xw\partial_yw + 
   \partial_xw \partial_y w_I + \partial_xw_I \partial_yw\right).
\end{align*}
These modified strains are used in the linear  Hookean 
 stress-strain relation 
\begin{align*}
  \sigma_{xx} &= \frac{Y}{1-\nu^2} (u_{xx}+ \nu u_{yy}), \nonumber\\
  \sigma_{yy} &= \frac{Y}{1-\nu^2} (u_{yy}+ \nu u_{xx}), \nonumber\\
  \sigma_{xy} &= \frac{Y}{1-\nu} u_{xy},
\end{align*}
which is unaffected by imperfections. 
The changes in the curvature tensor  $k_{ij}= k_{ij}(w+w_I) - k_{ij}(w_I)$
due to normal displacement are independent of $w_I$,
\begin{equation*}
    k_{ij} = \partial_i\partial_j w,
\end{equation*}
in linear order. Imperfections thus only affect the 
in-plane strain tensor $u_{ij}$ via the nonlinear term in the 
normal displacement $w$. 

Also, in the presence of imperfections the Hookean elastic energy
is given by the sum of stretching and bending energies: 
\begin{align}
  E_{el} &= \int dxdy( e_s + e_b) 
 \label{eq:Eel2} \\
e_s&=\frac{Y}{2(1-\nu^2)}\!\left[(u_{xx}\!+\!u_{yy})^2
    \!-2(1-\nu)(u_{xx}u_{yy}\!-u_{xy}^2)  \right],
\nonumber\\
e_B &=  \frac{\kappa}{2} \left[(\partial_x^2w + \partial_y^2w) \!- 
    2(1-\nu) (\partial_x^2w\partial_y^2w-\!(\partial_x\partial_yw)^2) \right].
\nonumber
\end{align}
Variation with respect to $u$ and $v$ gives 
\begin{align*}
   \partial_x \sigma_{xx} + \partial_y \sigma_{xy} &=0~,~~
 \partial_y \sigma_{yy} + \partial_x \sigma_{xy} = 0,
\end{align*}
which is automatically satisfied by the introduction of the Airy 
stress function 
\begin{align}
   \sigma_{xx} &= -\partial_y^2\Phi ~,~~
   \sigma_{yy} = -\partial_x^2 \Phi ~,~~
   \sigma_{xy} =  \partial_x\partial_y \Phi.
\label{eq:Phi}
\end{align}
These relations are unchanged in the presence of the imperfection field 
$w_I$. 

The first nonlinear shallow shell equation [cf.\ Eq.\ (\ref{eq:forcebal})]
for $\Phi$ and $w$ 
is obtained from extremizing 
$E_{el}$ with respect to $w$ (and expressing stresses by the Airy stress
function $\Phi$) to get the elastic force in normal direction, 
\begin{align}
 & \kappa \nabla^4 w - \frac{1}{R_0} \nabla^2\Phi + [\Phi,w+w_I] = 
   -p - \frac{F}{2\pi} \frac{\delta(r)}{r},
  \label{eq:shshell1} \\ 
 &\mbox{where}~  [f,g]\equiv \partial_x^2f \partial_y^2g + 
      \partial_y^2f \partial_x^2g 
  - 2(\partial_x\partial_y f)(\partial_x\partial_y g).
\nonumber
\end{align}
By eliminating the displacement fields $u$ and $v$ from the 
stress-strain relation we obtain the additional  
compatibility condition [cf.\ Eq.\ (\ref{eq:comp})]. 
In the presence of imperfections we obtain
\begin{align}
  -\frac{1}{Y} \nabla^4\Phi &= \partial_y^2 u_{xx} + \partial_x^2 u_{yy}
     -2\partial_x\partial_y u_{xy}
\nonumber\\
 &= \frac{1}{R_0} \nabla^2 w -\frac{1}{2} [w,w] - [w,w_I].
\label{eq:shshell2}
\end{align}

Both Eqs.\ (\ref{eq:shshell1}) and (\ref{eq:shshell1}) can be brought 
into a simpler form for axisymmetric shapes if coordinates $r$, $\phi$ 
are used. For axisymmetric functions $w$, $w_I$, $\Phi$ that only depend 
on $r$, we can use
$\nabla^2... =(\frac{1}{r}\partial_r +\partial_r^2)... 
 = \frac{1}{r} \partial_r r \partial_r... $
and 
\begin{equation*}
   [f,g] = \frac{1}{r} \partial_r (\partial_rf \partial_rg).
\end{equation*}
If we also use the derivative of the stress function 
$\psi = -\partial_r \Phi$ (with 
$\sigma_{\phi\phi} = \psi'$ and $\sigma_{rr} = \psi/r$) and integrate 
eq.\ (\ref{eq:shshell2}) once, we find 
\begin{align}
  & \kappa \nabla^4 w
   + \frac{1}{R_0} \frac{1}{r} \partial_r (r\psi)
      - \frac{1}{r} \partial_r \left( \psi \partial_r(w+w_I) \right) 
\nonumber\\
 & ~~~~~~~~~= -p - \frac{F}{2\pi} \frac{\delta(r)}{r},
  \label{eq:forcebalI_app} \\
 &\frac{1}{Y} r  \partial_r \left[ 
     \frac{1}{r} \partial_r (r\psi) \right] 
  = \frac{r}{R_0} \partial_r w  - 
   \frac{1}{2} \left( \partial_r w\right)^2 - 
   (\partial_r w)(\partial_r w_I). 
  \label{eq:compI_app}
\end{align}
These are Eqs.\ (\ref{eq:forcebalI}) and (\ref{eq:compI}) in the main 
text, which  generalize the nonlinear shell equations 
 (\ref{eq:forcebal}) and (\ref{eq:comp}) 
in the presence of imperfections. 
As in Eq.\ (\ref{eq:forcebal2}), we can absorb the effect of the 
pressure $p$ in Eq.\ (\ref{eq:forcebalI_app}) into 
 a uniform precompression   with $w(r) =w_0 <0$ and
 $\psi(r) = \psi_0(r)  = -pR_0r /2$ (corresponding to stresses 
 $\sigma_{rr} = \sigma_{\phi\phi} = \sigma_0 = -pR_0/2$) and  
 consider changes with respect to this prestress  by substituting
 $w(r) \to w_0 + w(r)$ and 
$\psi(r) \to \psi_0(r) + \psi(r)$. This gives
\begin{align}
   & \kappa \nabla^4 w
    + \frac{1}{R_0} \frac{1}{r} \partial_r (r\psi)
       -\sigma_0  \nabla^2 (w+w_I)
 \nonumber\\
 & ~~~ - \frac{1}{r} \partial_r \left( \psi
    \partial_r (w+w_I) \right) 
   =  - \frac{F}{2\pi} \frac{\delta(r)}{r}
   \label{eq:forcebal2_I_app} 
 \end{align}
 in the presence of imperfections. 
We note that, in the presence of imperfections, a precompressed 
state  with $w(r) =w_0 <0$ and
 $\psi(r) = \psi_0(r)  = -pR_0r /2$  is no longer a stationary state 
as it does {\em not} satisfy the force balance (\ref{eq:forcebalI_app}). 
It is, however, an admissible shell state which satisfies the 
compatibility conditions (\ref{eq:compI_app}). Therefore, 
we can still  consider all quantities relative to this state as 
for an ideal shell. 

Non-dimensionalization proceeds as before using  (\ref{eq:non-dim})
and we obtain 
\begin{align}
  &  \nabla_\rho^4 \bar{w}
   + \frac{1}{\rho} \partial_\rho (\rho\bar{\psi})
      +2\frac{p}{p_c} \nabla_\rho^2 (\bar{w}+\bar{w}_I)
\nonumber\\
& ~~~- \frac{1}{\rho} \partial_\rho \left( \bar{\psi}
\partial_\rho(\bar{w}+\bar{w}_I)\right) 
 =  - \frac{\bar{F}}{2\pi} \frac{\delta(\rho)}{\rho}
  \label{eq:forcebal3_I_app}\\
& \rho  \partial_\rho \left[ 
     \frac{1}{\rho} \partial_\rho (\rho\bar{\psi}) \right] 
  = \rho \partial_\rho \bar{w}  - 
   \frac{1}{2} \left( \partial_\rho \bar{w}\right)^2 
  -  \left( \partial_\rho \bar{w}\right)\left( \partial_\rho \bar{w}_I\right),
  \label{eq:comp3_I_app}
\end{align}
with 
$\nabla_\rho^2... = \frac{1}{\rho}\partial_\rho(\rho \partial_\rho
...)$. 
In this form,  the shallow shell equations are solved numerically 
in the presence of imperfections analogously to the 
ideal case.

We can also generalize the exact relations (\ref{eq:exact1}) and
(\ref{eq:exact2}) for the imperfect shell and find 
the relation (\ref{eq:exact1_I}) in the main text and 
\begin{align}
  & -\frac{\bar{F}}{2\pi} = \int_0^{\infty} d\rho \rho \bar{w}
    + \int_0^\infty d\rho \rho \left[\frac{1}{4} (\bar{w}')^2
    + \frac{1}{2} \bar{w}'\bar{w}_I'\right].
\label{eq:exact2_I_app}
\end{align}

\subsection{Derivation of total  energy difference}

In the form (\ref{eq:forcebal3_I_app}) and (\ref{eq:comp3_I_app}), 
 the shallow shell equations are solved numerically  analogously to the 
ideal case. One numerical complication is the correct calculation 
of barrier energies without the need to numerically 
integrate force-indentation relations by a suitable generalization 
of the exact result (\ref{eq:Etotapp}) for the total 
energy difference between  the barrier state ($\bar{F}=0$) 
and  the precompressed unindented 
state (with $\bar{F}=0$ and $\bar{w}=0$, but which is no longer a stationary
state in the presence of imperfections as discussed above) to the 
imperfect situation:
\begin{align}
  \Delta \bar{E}_{\rm tot, imp}= 
  -\frac{1}{4} \int_0^{\infty}d\rho  \bar{\psi}
   (\bar{w}')^2  
    -\frac{p}{p_c} \int d\rho \rho \bar{w}'\bar{w}_I'.
\label{eq:EtotIapp}
\end{align}
We note that the last term can be written as 
$-[(p/p_c) \bar{A}_I/4\pi] \bar{z}$ with 
$\bar{A}_I \equiv   4\pi \int d\rho \rho \bar{w}_I' (\bar{w}'/\bar{z})$,
which is analogously defined to the imperfection area occurring in the 
force-indentation relation (\ref{eq:Fzimp}). Then the imperfection 
force $(p/p_c)\bar{A}_I$ in Eq.\ (\ref{eq:EtotIapp}) 
is analogous to the point force $\bar{F}$ in the second 
term in  Eq.\ (\ref{eq:EtotappF}), which also corroborates the 
use of an effective preindenting force (\ref{eq:FpreAI}) 
in the presence of imperfections.

The derivation  follows the same lines as in 
Appendix \ref{app:barrierenergy}  without the imperfection field. 
We consider the total energy difference 
$\Delta E_{\rm tot} = \Delta E_s + \Delta E_b +p\Delta V$ 
and expressions (\ref{eq:Esapp}) for the stretching energy 
 $\Delta \bar{E}_s$, 
 (\ref{eq:Ebapp}) for the bending energy $\Delta \bar{E}_b$,
and (\ref{eq:pVapp}) for the mechanical work $\overline{p\Delta V}$
remain valid also in the presence of imperfections (we consider 
the case $\bar{F}=0$ here). 

Using partial integration and Eqs.\ (\ref{eq:exact1_I})  and 
(\ref{eq:exact2_I_app})  we can  rewrite $\Delta\bar{E}_b$ for $\bar{F}=0$ 
as  
\begin{align*}
  \Delta \bar{E}_b &= -\overline{p\Delta V} + \frac{1}{2} \int d\rho \rho
    \bar{\psi}\bar{w}' \left( 1 - \frac{\bar{w}'}{\rho} \right)
   \\
  &~~~~
  -\frac{p}{p_c} \int d\rho \rho \bar{w}' \bar{w}_I' - 
   \frac{1}{2} \int d\rho  \bar{\psi}\bar{w}'\bar{w}_I'.
\end{align*}
Using (\ref{eq:comp3_I_app}),  this leads to  the total energy 
difference (\ref{eq:EtotIapp}).

\section{Dimensionless quantities}
\label{app:dim}

We provide a conversion table (Table \ref{tab:dimensionless})
for the different
dimensionless quantities used here [see Eq.\ (\ref{eq:non-dim})], 
by Gomez {\it et al.} in  Ref.\ \cite{Gomez2016}, 
and by Hutchinson {\it et al.} in Refs.\ 
\cite{Hutchinson2017b,Hutchinson2018,Marthelot2017}.

The shell thickness is called $h$. We define $k\equiv [12(1-\nu^2)]^{1/4}$
and the  F\"oppl-von K\'arm\'an number
$\gamma \equiv YR_0^2/\kappa = (R_0/h)^2k^4$ [see Eq.\ (\ref{eq:FvK})], 
and use the critical buckling pressure
  $p_c =  4 (Y\kappa)^{1/2}/R_0^2 = 4 (Yh/R_0^2) k^{-2}$ 
[see Eq.\ (\ref{eq:pcb})]. 
Note that in the shallow shell approximation the arc length $s$
used in  Refs.\ \cite{Hutchinson2017b,Hutchinson2018} approaches 
the radial coordinate $r$. 
In  Refs.\ \cite{Hutchinson2017b,Hutchinson2018}, 
 $C\equiv \sqrt{3}/(1-\nu)\sqrt{1-\nu^2}$  is used.

\bibliography{references}

\end{document}